\documentclass[12pt,a4paper,notitlepage]{article}

\usepackage{amsfonts}
\usepackage{amsmath}
\usepackage{amssymb}
\usepackage{amsthm}
\usepackage{color}
\usepackage[T1]{fontenc}
\usepackage{graphicx}
\usepackage{hyperref}
\usepackage{mathrsfs}
\usepackage{multirow}
\usepackage[numbers]{natbib}
\usepackage[format=hang]{subcaption}
\usepackage{times}
\usepackage{upgreek}
\usepackage[utf8]{inputenc}
\usepackage{pgfplots}
\usepackage{longtable}
\usepackage{booktabs}
\usepackage{algorithm, algpseudocode}
\usepackage{adjustbox}
\usepackage{siunitx}
\usepackage{svg}
\usepackage{tikz}
\usepackage{import}
\usetikzlibrary{patterns,shapes.arrows}
\pgfplotsset{compat=newest}
\usepackage{float}
\pgfplotsset{compat=newest}
\usetikzlibrary{plotmarks}
\usetikzlibrary{arrows.meta}
\usetikzlibrary{calc}
\usepackage{grffile}
\AtBeginDocument{}

\date{}

\graphicspath{{./images/}}

\newcounter{remark}

\begin{document}

\author{ \large {M. Sesa${}^{ *, \,\dag}$, H. Holthusen${}^{\,\dag}$, L. Lamm${}^{\,\dag}$, C. B\"ohm${}^{\,\ddag}$,} \\ {T. Brepols${}^{\,\dag}$, S. Jockenh\"ovel${}^{\,\ddag}$, S. Reese${}^{\,\dag}$}\\[0.5cm]
\hspace*{-0.1cm}
\normalsize{\em ${}^{\dag}$ Institute of Applied Mechanics, RWTH Aachen
  University,}\\
\normalsize{\em Mies-van-der-Rohe-Str.\ 1, 52074 Aachen, Germany}\\[0.25cm]\
\normalsize{\em ${}^{\ddag}$ Biohybrid \& Medical Textiles, Institute of Applied Medical Engineering,} \\ \normalsize{\em RWTH Aachen University, Forckenbeckstr. \ 55, 52074 Aachen, Germany}\\
%\normalsize{\em  }
\normalsize{\em ${}^{*}$ Corresponding author: mahmoud.sesa@ifam.rwth-aachen.de }
%\normalsize{\em swu@tf.uni-kiel.de}
}
\title{\LARGE Mechanical modeling of the maturation process for tissue-engineered implants: application to biohybrid heart valves}
\maketitle

\small
{\bf Abstract.}
The development of tissue-engineered cardiovascular implants can improve the lives of large segments of our society who suffer from cardiovascular diseases. Regenerative tissues are fabricated using a process called tissue maturation. Furthermore, it is highly challenging to produce cardiovascular regenerative implants with sufficient mechanical strength to withstand the loading conditions within the human body. Therefore, biohybrid implants for which the regenerative tissue is reinforced by standard reinforcement material (e.g.\ textile or 3d printed scaffold) can be an interesting solution. In silico models can significantly contribute to characterizing, designing, and optimizing biohybrid implants. The first step towards this goal is to develop a computational model for the maturation process of tissue-engineered implants. This paper focuses on the mechanical modeling of textile-reinforced tissue-engineered cardiovascular implants. First, we propose an energy-based approach to compute the collagen evolution during the maturation process. Then, we apply the concept of structural tensors to model the anisotropic behavior of the extracellular matrix and the textile scaffold. Next, the newly developed material model is embedded into a special solid-shell finite element formulation with reduced integration. Finally, we use our framework to compute two structural problems: a pressurized shell construct and a tubular-shaped heart valve. The results show the ability of the model to predict collagen growth in response to the boundary conditions applied during the maturation process. Consequently, we can predict the implant's mechanical response, such as the deformation and stresses of the implant.

\vspace*{0.3cm}

\normalsize

%%%%%%%%%%%%%%%%%%%%%%%%%%%%%%%%%%%%%%%%%%%%%%%%%%%%%%%%%%%%%%%%%%%%%%%%%%%%%%%%%%%%%%%%%%%%%%%%%%%%%%%%%

\section{Introduction}
\label{sec:1}

In modern industrialized societies, a significant portion of the population suffers from valvular heart disease \cite{Lung_Vahanian_2011}. The number in the US is estimated by Nkoma et al.\ \cite{Nkomo_EtAl_2006} to be $2.5 \%$ of the population. Valvular heart defects are not restricted to a specific age group. They affect young and old people since many are born with congenital heart defects \cite{Hoffman_Kaplan_2002}. A typical treatment approach is to replace the defective native valve with a biomedical heart valve implant. Finding the optimal type and design of heart valve implants has been investigated for decades. Heart valve implants can be categorized into three groups: (i) mechanical valves, (ii) bioprosthetic valves and (iii) tissue-engineered heart valves \cite{Mela_2019}.   

The essential structural requirement of a heart valve implant is to withstand the cyclic loading conditions ($\approx$ 90000 loading cycles/day) for decades. In this regard, mechanical valves offer the most reliable solution. However, it is necessary for patients using mechanical valves to take anti-coagulation medications for the rest of their life, adversely affecting their lifestyle. This severe drawback leads to the increasing adaptation of bioprosthetic valves produced out of decellularized bovine or porcine heart valves. Bioprosthetic valves improve the patient's lifestyle since they do not need to take anti-coagulation medicines \cite{Mol_2009}. However, the lifetime of bioprosthetic valves is often shorter than the patient's expected, lifespan making it more suited for older patients. The short lifetime is caused by damage due to calcification and the inability of the implant to grow or remodel. 

Studies have shown that the heart valve's ability to adapt to hemodynamic conditions through tissue growth and remodeling is essential for its long-term durability and integrity \cite{Chester_etal_2014}. Adaptation is crucial in the case of children born with congenital valvular defects since hemodynamic conditions are altered significantly during their lifespan (e.g.\ change in blood pressure and heart rate) \cite{Oomen_etal_2016}. The drawbacks of mechanical and bioprosthetic valves were the main impetus for many researchers to investigate the implantation of living biological valves. One of the earliest developments in this field was the Ross procedure performed in 1967 \cite{Ross_1967}, which transplanted damaged aortic and mitral heart valves with pulmonary autografts. Tissue engineering aims to avoid the transplantation procedure through the fabrication and implantation of functional living valves \cite{Yacoub_Takkenberg_2005}.   

One of the main deterrents to using tissue-engineered heart valves (TEHVs) is their low mechanical strength. Improving the mechanical properties can be achieved by impeding a reinforcement material with the desired mechanical properties. These reinforced valves are called biohybrid heart valves. Biohybrid valve constituents can be decomposed into mechanobiologically active and passive materials. The reinforcement materials (e.g.\ scaffold) act as a passive constituent, while the extracellular matrix (ECM) actively reacts to mechanobiological stimulation. The scaffold is a porous material that supports the ECM during maturation. Among the widely used scaffold types are textile and 3D printed scaffolds \cite{Vukicevic_etal_2017}. Textile scaffolds are produced from biocompatible polymers using special techniques such as electrospinning \cite{Hinderer_etal_2017} or knitting \cite{Liberski_etal_2016}. Biomimetic non-woven textile scaffolds have been used by Moreira et al.\ \cite{Moreira_etal_2016} to fabricate TEHVs. A macroporous textile scaffold provides structural support to the implant and guides the ECM growth during the maturation process. The long-term goal is to produce valves in which the ECM develops enough mechanical strength to sustain physiological loads while the scaffold degrades after the implantation procedure \cite{Mendelson_2006}.

An essential aspect of developing TEHVs is tailoring their mechanical properties to resemble the behavior of native human heart valves. From a structural mechanics point of view, a native heart valve is a soft, thin and highly anisotropic shell structure. The density and orientation of collagen fibers highly influence the valve's mechanical properties. In the case of a biohybrid heart valve we are considering in this work, the textile scaffold is the main load-carrying constituent in the implant. Consequently, it is necessary to tailor the anisotropic behavior of the scaffold to mimic collagenous native valves. A native-like mechanical behavior can be achieved through (i) producing scaffolds with anisotropic microstructure and (ii) embedding fiber reinforcements to induce anisotropic mechanical behavior \cite{Moreira_etal_2016}.

% literature review
There is extensive literature on the mechanical modeling of heart valves. Among them is the work of Driessen et al.\ \cite{Driessen_etal_2007, Driessen_etal_2008} on modeling tissue-engineered heart valves and remodeling of angular collagen fiber distribution. With the help of these finite element models, Loerakker et al.\ \cite{Loerakker_etal_2013} showed that tissue anisotropy could significantly influence the hemodynamics of the valves. On another front, Driessen et al.\ \cite{Driessen_etal_2005} developed an approach to consider the angular collagen fiber distribution in the constitutive model. Later, Gasser et al.\ \cite{Gasser_etal_2006} introduced a new approach based on generalized structural tensors to address a similar issue in modeling arteries.

One of the main challenges in manufacturing heart valves is their complex geometry. Native valves have a semi-lunar shape with curved cusps. Fabricating and suturing an artificial valve with a semilunar shape is a complex and challenging process which suffers from many unpredictable conditions. Reliability can be improved by using simpler designs that mimic the hemodynamics of native valves. Therefore, Weber et al.\ \cite{Weber_etal_2014} proposed a tubular valve design. Mechanical properties can be enhanced by attaching fiber reinforcement. Stapleton et al.\ \cite{Stapleton_etal_2015} studied the influence of fiber reinforcement density and orientation on the mechanical behavior of the valves. Furthermore, Sodhani et al.\ \cite{Sodhani_etal_2017, Sodhani_etal_2018a} developed a multiscale modeling approach to simulate valves made of knitted textile scaffolds. The model was then extended to consider a coupled fluid-structural interaction \cite{Sodhani_etal_2018b}. Although these models are highly accurate in predicting the valve's hemodynamics, their computational cost is enormous. This makes them ill-suited for exploring and optimizing the design of TEHVs. Furthermore, the heart valve models from Stapleton et al.\ \cite{Stapleton_etal_2015} and Sodhani et al.\ \cite{Sodhani_etal_2017, Sodhani_etal_2018a, Sodhani_etal_2018b} only considered the passive mechanical behavior of the implant without considering mechanobiologically induced tissue growth and remodeling during the maturation process.

ECM is secreted during the maturation process. It surrounds the cells and provides structural support to the tissue through networks of protein fibers that provide mechanical strength. Biological tissues undergo various types of growth \cite{Kuhl_2014, Eskandari_2015}, such as volumetric growth, mass growth or cross-sectional area growth. Among the pioneering work on modeling finite volumetric growth is the article from Rodriguez et al.\ \cite{Rodriguez_1994}, in which they applied the multiplicative split of the deformation gradient to describe the material's inelastic response during the growth process. Later, Humphrey \& Rajagopal \cite{Humphrey_2002} introduced the constrained mixture model which considers the kinetics of the production and degradation processes of the tissue's constituents. Volumetric changes in soft biological tissue's can be either isotropic or anisotropic. Anisotropic growth can be implemented by defining the growth direction \cite{Ambrosi_2019}. However, this approach is only valid for simple geometries. More generalized approach based on the homogenized constrained mixture models was introduced by Braeu et al.\ \cite{Braeu_2017, Braeu_2019}. Based on the multiplicative split concept from Rodriguez et al.\ \cite{Rodriguez_1994}, Lamm et al.\ \cite{Lamm_2021, Lamm_2022} introduced a general approach to model volumetric growth without pre-defining a growth tensor. This approach is based on defining a homeostatic stress surface. The model was extended by Holthusen et al.\ \cite{Holthusen_2023} to consider collagen fiber reorientation using a novel framework based on a co-rotated intermediate configuration. Another interesting application for growth models is in-stent restenosis which was investigated by Manjunatha et al.\ \cite{Manjunatha_2022}.

During the in-vitro maturation process of soft tissues, volumetric contraction and mass growth due to the production of collagen fibers are observed. Volumetric contraction is undesirable during the fabrication of heart valves as it makes them unsuitable for clinical use.  On the other hand, high collagen content improves the mechanical properties of the tissue. Moreira et al.\ \cite{Moreira_etal_2015} showed that volumetric contraction can be prevented by using textile reinforcement and choosing a tubular instead of semi-lunar heart valve design. Since modeling textile-reinforced tubular valves is the topic of this paper, we will focus on the evolution of collagen density (collagen mass growth) during the maturation process.

Collagen fiber bundles are responsible for most of the load-carrying capacity in soft biological tissues. Therefore, changes in collagen density and orientation significantly alter the mechanical properties of the tissue. Growth drivers in living tissues differ from one tissue type to another depending on the environment and the tissue microstructure \cite{Kuhl_2014, Eskandari_2015}. Finding the relevant growth driving factors is essential to optimize the design of tissue-engineering processes. Various studies found that the enzymatic degradation rate is influenced by the strain applied to the tissue \cite{Huang_Yannas_1977, Wyatt_etal_2009, Siadat_Ruberti_2023}. This makes it possible to control the collagen density in the tissue \cite{Ruberti_Hallab_2005}. With regards to heart valves, Ku et al.\ \cite{Ku_etal_2009} found that collagen synthesis by valvular interstitial cells (VICs) from the aortic valve can be controlled by the level and duration of the applied tissue stretching. Furthermore, Oomen et al.\ \cite{Oomen_etal_2016} studied the mechanical properties of an extensive data set of post-mortem healthy human heart valves for various age groups. The study showed that native human heart valves maintain the same level of stretch along the circumferential direction for different age groups. However, the circumferential stress varies considerably from one age group to another. They concluded that human heart valves maintain stretch homeostasis. These results were later applied to study the influence of growth and remodeling on the evolution of heart valves \cite{Oomen_etal_2018}. With regards to textile-reinforced biohybrid implants, Hermans et al.\ \cite{Hermans_etal_2022} investigated the influence of the scaffold microstructure on the collagen evolution during the maturation process.  

Modeling the evolution of collagen fibers in soft and hard biological tissues has been widely investigated. Baek et al.\ \cite{Baek_etal_2006} introduced a model for collagen fiber deposition during intracranial fusiform aneurysms, where fiber stretching drives collagen evolution. Martufi \& Gasser \cite{Martufi_Gasser_2012} applied a similar approach to model aortic aneurysms. Furthermore, Hadi et al.\ \cite{Hadi_etal_2012} modeled collagen fiber's enzymatic degradation and growth as a function of collagen fiber strain. A similar assumption was applied to study wound healing by Gierig et al.\ \cite{Gierig_etal_2021}. Also, Topol et al.\ \cite{Topol_etal_2021} introduced a model that examines the influence of tissue stretch on collagen fiber density. However, modeling the evolution of collagen fibers for cardiovascular tissue-engineering applications needs to be better investigated. Among the limited work in this area, Szafron et al.\ \cite{Szafron_etal_2019} applied a constrained mixture model to optimize the micro-structural design of tissue-engineered vascular grafts with 3d printed scaffold. In their model, collagen growth is driven by the graft's wall-shear stress. However, in the case of heart valves, experimental investigations from \cite{Oomen_etal_2016, Oomen_etal_2018} showed that growth is driven by the strain of the collagen fibers. Furthermore, in \cite{Szafron_etal_2019}, the collagen evolution depends explicitly on time, making the model not general enough to consider a different process setup.

In this paper, we introduce a simple approach to model the in-vitro maturation of textile-reinforced biohybrid implants. In Section \ref{sec:2}, we introduce an energy-based approach to model the evolution of collagen content during tissue maturation. Although energy-based models were investigated for other biomechanics applications, to our knowledge, a similar approach has not been yet developed to model the maturation of regenerative tissues. A special challenge in applying an energy-based approach to tissue maturation is that at the beginning of the cultivation process, the initial collagen density is zero and hence the corresponding Helmholtz free energy is zero. Also, our experimental investigation showed that collagen growth occurs for unloaded samples. We find the idea of decomposing the collagen growth into a mechanically-driven and biologically-driven part by Szafron et al.\ \cite{Szafron_etal_2019} to be very beneficial in this regard. Then, we apply the concept of structural tensors to model the anisotropic mechanical behavior of the scaffold and collagen fibers. Then Section \ref{sec:3} discusses the finite element implementation where we embed the material model into a special solid-shell finite element formulation with reduced integration. We also show the necessary steps to compute the collagen density. In Section \ref{sec:4}, we experimentally measure the stress-stretch behavior of collagenous regenerative tissues and their corresponding collagen fiber density. The results are used to validate our choice for the Helmholtz free energy function of the collagen fibers and identify the material parameters concerning this part. Furthermore, we perform biaxial tensile testing experiments on the textile scaffold and use the results to identify the values of the material parameters for the textile scaffold. Finally in Section \ref{sec:5}, we use our finite element framework to compute two structural problems. The first problem concerns a pressurized shell construct. In this example, we explore the validity of our model and study sensitivity of the results to the newly introduced modeling parameters. The second problem, we compute the collagen evolution and stress distribution for a tubular-shaped heart valve.

\section{Continuum mechanical model}
\label{sec:2}

During the maturation process of tissue-engineered implants, collagen fibers are secreted by VICs, which infuse the textile scaffold. In this section, we introduce a continuum mechanical approach to model the material during the maturation process. The model is described by defining the main constituents of the material and their corresponding constitutive equations. Then we sum up each part's contribution to describe the total response of the material. The main constituents of biohybrid implants are the regenerative tissue and the textile scaffold. The regenerative tissue contains VICs and the ECM. VICs lack mechanical strength; therefore, only the ECM is considered in our constitutive model. The ECM mechanical response is highly influenced by the mass growth of collagen fibers. 

The section is organized in the following way. We start by explaining the kinematic relations. Then, we explain the thermodynamic basis of our model. Afterwards, the collagen density evolution equations are introduced. Finally, define show the constitutive equations used for each constituent.  

\subsection{Kinematics}
To define the kinematic relations of our model, we consider a three-dimensional body that occupies the domain $\Omega_{0}$ with the boundary $\partial \Omega_{0}$ in the reference configuration at a reference time $t = 0$. A material particle within the domain is defined in the reference configuration by the position vector $\mathbf{X}$. The current position vector is denoted by $\mathbf{x} = \phi(\mathbf{X}, t) $. This gives us the deformation gradient tensor
\begin{equation}
\label{eq:2-1}
\mathbf{F} = \frac{\partial \mathbf{x}}{\partial \mathbf{X}}, 
\end{equation}
and the right Cauchy-Green tensor
\begin{equation}
\label{eq:2-2}
\mathbf{C} = \mathbf{F}^{\mathrm{T}} \, \mathbf{F} .
\end{equation}

\subsection{Thermodynamic basis}

An essential aspect of fabricating functional tissue-engineered implants is the production of collagen fibers. At the beginning of the maturation process, the tissue does not contain collagen fibers; then, the valvular interstitial cells (VIC) produce collagen during the maturation. This process leads to a change in the mass of collagen fibers or, in other words, mass growth. The mass growth can be expressed in a continuum model as a change in the material density. The in-vitro maturation process spans several weeks, while the loading and unloading of cardiovascular tissues, such as heart valves, takes less than a second. As a result, mass growth occurs over a significantly longer time scale than the time scale for loading the material. This makes it possible to simplify the balance equations by applying the slow-growth assumption \cite{Goriely_2017}.

From a thermodynamic perspective, mass growth turns our problem into an open system. In our model, we consider mass change due to the evolution of the material density in the reference configuration $\rho^{0}$. To formulate the balance of mass equation for an open system, we introduce the mass flux vector $\mathbf R$ and a local mass source term $R_{0}$. The mass balance equation becomes
\begin{equation}
\label{}
\dot{\rho}^{0} = \mathrm{Div}({\mathbf R} ) + R_{\mathrm{0}},
\end{equation} 
with the shorthand notation $({\bullet})$ denotes the material time derivative. From the slow-growth assumption \cite{Goriely_2017}, the mass flux term becomes $\mathrm{Div}({\mathbf R} ) = 0$. Consequently, the mass balance equation is reduced to $\dot{\rho}^{0} = R_{\mathrm{0}}$, where the density change rate equals the mass source $R_{\mathrm{0}}$. Furthermore, the application of the slow-growth assumption lead to the standard balance of linear momentum equation
\begin{equation}
\label{eq:2-3}
\mathrm{Div}({\mathbf F} \, {\mathbf S}) + {\mathbf b}_{0} = {\mathbf 0},
\end{equation}
where ${\mathbf S}$ is the second Piola-Kirchhoff stress and ${\mathbf b}_{0}$ is the referential body force vector.

To ensure that the second law of thermodynamics is satisfied, an entropy term $S_{0}$ is introduced by Kuhl \& Steinmann \cite{Kuhl_Steinmann_2003}. The Clausius-Duhem inequality becomes 
\begin{equation}
\label{eq:2-4}
\frac{1}{2} \, {\mathbf S} \cdot \dot{{\mathbf C}} - \dot{\psi} + S_{\mathrm{0}} \geqslant 0,
\end{equation}
where $S_{0}$ accounts for both the local entropy production and entropy fluxes through the boundaries of our system. The concept was successfully used in various problems, such as bone remodeling \cite{Kuhl_2003} or volumetric growth of soft tissue \cite{Lamm_2021, Lamm_2022}. Then, we can formulate the Clausius-Duhem inequality as 
\begin{equation}
\label{eq:2-8}
\left( {\mathbf S} - 2 \, \frac{\partial \psi}{\partial \mathbf{C}} \right) \cdot \, \frac{1}{2} \, \dot{{\mathbf C}} + S_{\mathrm{0}} \geqslant 0 .
\end{equation}
By applying the Coleman-Noll procedure \cite{Coleman_Noll_1963}, we get the following term for the second Piola-Kirchhoff stress
\begin{equation}
\label{}
{\mathbf S} = 2 \, \frac{\partial {\psi}}{\partial \mathbf{C}},
\end{equation}

where the total Helmholtz free energy function ${\psi}$ is the sum of the energy contribution from the textile reinforcement ${\psi}_{\mathrm{tex}}$ and the ECM ${\psi}_{\mathrm{ECM}}$ as described by the equation
\begin{equation}
\label{}
{\psi} =  {\psi}_{\mathrm{tex}} + {\psi}_{\mathrm{ECM}} .
\end{equation}

The extracellular matrix contains protein fibers such as collagen and elastin fibers as well as proteoglycans and glycosaminoglycans. Collagen bundles are stiff and highly anisotropic, while other constituents show isotropic mechanical behavior. Collagen bundles are responsible for most cardiovascular tissues' mechanical strength, as in the case of heart valve tissue. Therefore, the extracellular matrix is modeled as an isotropic ground matrix with anisotropic collagen reinforcement. The energetic part ${\psi}_{\mathrm{ECM}}$ is defined as the sum of the contribution from the isotropic matrix ${\psi}_{\mathrm{matrix}}$ and anisotropically oriented collagen bundles ${\psi}_{\mathrm{co}}$. This gives us the relation
\begin{equation}
\label{}
{\psi}_{\mathrm{ECM}} = {\psi}_{\mathrm{matrix}} + {\psi}_{\mathrm{co}},
\end{equation}
for the Helmholtz free energy function ${\psi}_{\mathrm{ECM}}$. Therefore, the total Helmholtz free energy can be described as the sum of the contributions from the textile reinforcement, matrix, and collagen parts as described by the equation
\begin{equation}
\label{eq:2-30}
{\psi} =  {\psi}_{\mathrm{matrix}}(\mathbf{C}) + {\psi}_{\mathrm{co}}(\mathbf{C}, \mathbf{H}, {\rho}^{0}_{\mathrm{co}}) + {\psi}_{\mathrm{tex}}(\mathbf{C}, \mathbf{M}_{1}, \mathbf{M}_{2}) 
\end{equation}
where ${\rho}^{0}_{\mathrm{co}}$ is the collagen density in the reference configuration. It is important to mention that in this model only the density of the collagen changes while the density of other constituents remain constant. The evolution equations for ${\rho}^{0}_{\mathrm{co}}$ are introduced in Section \ref{subsec:2-3}. Furthermore, we introduce the generalized structural tensor $\mathbf{H}$, and the structural tensors $\mathbf{M}_{1}$ and $\mathbf{M}_{2}$, which are defined in the reference configuration. 

The anisotropic behavior of the collagen fibers is introduced into the constitutive model using the structural tensor $\mathbf{H}$. One unique aspect to consider while modeling the anisotropic behavior of collagen fibers is that the fibers are not uni-oriented, but they show dispersed orientation over a range of angles. The level of fiber dispersity needs to be considered in the constitutive model. Here, we apply the concept of generalized structural tensors introduced by Gasser et al.\ \cite{Gasser_etal_2006} to model collagen fibers in arteries. To account for fiber dispersity, an additional material parameter $\kappa$ is introduced. We define the mean orientation of the collagen fibers in the reference configuration by the vector ${\mathbf a}$. This leads to the generalized structural tensor
\begin{equation}
\label{eq:2-18}
\mathbf{H} = \kappa{\bf I} + (1 - 3 \kappa)({\mathbf a} \otimes {\mathbf a}),
\end{equation}
where the value of $\kappa$ varies within the range $ 0 \le \kappa \le \frac{1}{3} $, with $\kappa = 0$ in the case of uni-oriented collagen fibers, and $\kappa = \frac{1}{3}$ in the case of isotropically dispersed fibers.

From the structural tensor $\mathbf{H}$ and the right Cauchy-Green tensor $\mathbf{C}$, the square of collagen stretching $\lambda_{\mathrm{co}}$ is computed by
\begin{equation}
\label{eq:2-19}
\lambda_{\mathrm{co}}^{2} = \mathrm{tr}(\mathbf{C} \, \mathbf{H}),
\end{equation}
and the strain along the fiber direction reads
\begin{equation}
\label{ }
E_{\mathrm{co}} = \lambda_{\mathrm{co}}^{2} - 1.
\end{equation}

The scaffold is an orthotropic material made out of non-woven textiles. The orthotropic behavior is modeled using structural tensors defined in the reference configuration. In the case of orthotropic materials, there are two anisotropic material orientations to consider. These orientations are defined by the vectors ${\bf n}_{1}$ and ${\bf n}_{2}$. Then, we use the dyadic product of the directional vectors to compute the corresponding structural tensors
\begin{equation}
\label{eq:2-24}
\mathbf{M}_{1} = {\bf n}_{1} \otimes {\bf n}_{1} \; \; \text{and} \; \; \mathbf{M}_{2} = {\bf n}_{2} \otimes {\bf n}_{2} .
\end{equation}

The strain energy function ${\psi}_{\mathrm{tex}}$ is defined in terms of material invariants \cite{Reese_etal_2001, Reese_etal_2021}. In this model, ${\psi}_{\mathrm{tex}}$ is a function of the following invariants
\begin{equation}
\label{eq:2-25}
\begin{gathered}
\mathit{I}_{\mathrm{1}} = \mathrm{tr}(\mathbf{C}),  \\
\mathit{I}^{\mathrm{tex}}_{\mathrm{2}} = \mathrm{tr}(\mathbf{C} \, \mathbf{M}_{1}), \;  \mathit{I}^{\mathrm{tex}}_{\mathrm{3}} = \mathrm{tr}(\mathbf{C}^{2} \, \mathbf{M}_{1}), \\
\mathit{I}^{\mathrm{tex}}_{\mathrm{4}} = \mathrm{tr}(\mathbf{C} \, \mathbf{M}_{2}) \; \text{and} \; 
\mathit{I}^{\mathrm{tex}}_{\mathrm{5}} = \mathrm{tr}(\mathbf{C}^{2} \, \mathbf{M}_{2}) .
\end{gathered}
\end{equation}

The next step is to compute the second Piola-Kirchhoff stresses. Similar to the energy term in Eq.\ \ref{eq:2-30}, the total second Piola-Kirchhoff stress $\mathbf{S}$ can be computed by summing up the stresses of each constituent which leads to the relation
\begin{equation}
\label{eq:2-27}
\mathbf{S} =  \mathbf{S}_{\mathrm{matrix}} + \mathbf{S}_{\mathrm{co}} + \mathbf{S}_{\mathrm{tex}}.
\end{equation}

The second Piola-Kirchhoff stress of the textile part $\mathbf{S}_{\mathrm{tex}}$ is computed by taking the partial derivative of the Helmholtz free energy ${\psi}_{\mathrm{tex}}$ with respect to the right Cauchy-Green tensor $\mathbf{C}$ which can be written as 
\begin{equation}
\label{eq:2-28}
\mathbf{S}_{\mathrm{tex}} = 2 \, \frac{\partial {\psi}_{\mathrm{tex}}}{\partial \mathbf{C}},
\end{equation}
whereas the second Piola-Kirchhoff stress for the matrix part reads
\begin{equation}
\label{eq:2-29}
\mathbf{S}_{\mathrm{matrix}} = 2 \, \frac{\partial {\psi}_{\mathrm{matrix}}}{\partial \mathbf{C}}.
\end{equation}

The evaluation of the second Piola-Kirchhoff stress of the collagen part ${\mathbf S}_{\mathrm{co}}$ requires considering the evolution of the referential collagen density ${\rho}^{0}_{\mathrm{co}}$. The Helmholtz free energy of the collagen fibers ${\psi}_{\mathrm{co}}$ per unit volume can be expressed in terms of the collagen density in the reference configuration ${\rho}^{0}_{\mathrm{co}}$ and the corresponding free energy per unit mass ${\psi}_{\mathrm{co, m}}$, i.e.
\begin{equation}
\label{eq:2-5}
{\psi}_{\mathrm{co}} =  {\rho}^{0}_{\mathrm{co}} \, {\psi}_{\mathrm{co, m}} .
\end{equation}

By taking the time derivative of ${\psi}_{\mathrm{co}}$ in Eq.\ (\ref{eq:2-5}), we obtain the expression for the energy rate
\begin{equation}
\label{eq:2-6}
\dot{\psi}_{\mathrm{co}} =  \dot{\rho}^{0}_{\mathrm{co}} \, {\psi}_{\mathrm{co, m}} \, + \, {\rho}^{0}_{\mathrm{co}} \, \dot{\psi}_{\mathrm{co, m}}.
\end{equation}

Then with applying the chain rule of differentiation, the energy rate $\dot{\psi}_{\mathrm{co}}$ can be rewritten in the following way:
\begin{equation}
\label{eq:2-7}
\dot{\psi}_{\mathrm{co}} =  \frac{\partial {\rho}^{0}_{\mathrm{co}}}{\partial \mathbf{C}} \cdot \dot{{\mathbf C}} \, {\psi}_{\mathrm{co, m}} + {\rho}^{0}_{\mathrm{co}} \, \frac{\partial {\psi}_{\mathrm{co, m}}}{\partial \mathbf{C}} \cdot \dot{{\mathbf C}}.
\end{equation}
Later, we insert Eq.\ \ref{eq:2-7} into Eq.\ \ref{eq:2-8} to get the following expression for the second Piola-Kirchhoff stress of the collagen part
\begin{equation}
\label{eq:2-9}
{\mathbf S}_{\mathrm{co}} = 2 \, \left(\frac{\partial {\rho}^{0}_{\mathrm{co}}}{\partial \mathbf{C}} \, {\psi}_{\mathrm{co, m}} + {\rho}^{0}_{\mathrm{co}} \, \frac{\partial {\psi}_{\mathrm{co, m}}}{\partial \mathbf{C}} \right).
\end{equation}

The second Piola-Kirchhoff stress does not have a clear physical interpretation. This makes it necessary to transform it into physically meaningful stress measures. The stress measures that we will use in this paper are the first Piola-Kirchhoff stress $\mathbf{P} = \mathbf{F} \, \mathbf{S}$ and the Cauchy stress $\boldsymbol{\sigma} = J^{-1} \, \mathbf{P} \, \mathbf{F}^{T}$, where $J = \mathrm{det}(\mathbf{F})$, and it represents the volumetric change.

\subsection{Collagen evolution equations}
\label{subsec:2-3}

In our experiments, we observed that during the maturation, collagen fibers are secreted even for unloaded tissues. Furthermore, work in the literature over decades shows that mechanobiological stimulation plays a major role in collagen growth. A reasonable way to consider these factors is to decompose the total rate of collagen density change $\dot{\rho}^{0}_{\mathrm{co}}$ into two parts. The first part considers collagen growth due to bio-chemical factors $\dot{\rho}^{0}_{\mathrm{bio}}$, and the second part considers the growth due to mechanobiological stimulation $\dot{\rho}^{0}_{\mathrm{mech}}$ \cite{Szafron_etal_2019}. This gives us the following expression for the collagen density change rate
\begin{equation}
\label{eq:2-10}
\dot{\rho}^{0}_{\mathrm{co}} = \dot{\rho^{0}}_{\mathrm{bio}} + \dot{\rho}^{0}_{\mathrm{mech}}, 
\end{equation}
where the quantities $\dot{\rho}^{0}_{\mathrm{co}}$, $\dot{\rho}^{0}_{\mathrm{bio}}$ and $\dot{\rho}^{0}_{\mathrm{mech}}$ are defined in the reference configuration.

Experimental measurements of collagen density for unloaded samples showed that a collagen density increase with respect to maturation time follows an S-shaped curve. This can be described by the following Weibull cumulative distribution function
\begin{equation}
\label{eq:2-11}
{\alpha}_{\mathrm{bio}}(t) = 1 - e^{-(t / \tau)^{h}} ,
\end{equation}
where $\tau$ is the half-time and $h$ is a parameter that controls the curve's steepness. Compared to other types of S-shaped curves, Weibull cumulative distribution function is more suited to our model because: (i) it satisfies the condition that ${\alpha}_{\mathrm{bio}} = 0$ when $t = 0$ and (ii) it can describe curves with positively skewed first derivative. By taking the time derivative of ${\alpha}_{\mathrm{bio}}$, we get the following expression
\begin{equation}
\label{eq:2-12}
\dot{\alpha}_{\mathrm{bio}}  = \frac{h}{\tau}\, e^{-(t / \tau)^{h}}  \, \left(\frac{t}{\tau}\right)^{h - 1}.
\end{equation}

From Eq.\ \ref{eq:2-12} we can describe the collagen density evolution for the biologically-driven part by the following equation
\begin{equation}
\label{eq:2-13}
\dot{\rho}^{0}_{\mathrm{bio}} = a_{1} \, c_{\mathrm{cell}} \, \dot{\alpha}_{\mathrm{bio}} .
\end{equation}
where $c_{\mathrm{cell}}$ is the valvular interstitial cell density and $a_{1}$ ($\mathrm{\SI{}{\micro\gram} / cells}$) is an additional material parameter.

In the second part, we consider collagen growth rate is driven by the stretching of the collagen fibers. To formulate this in a thermodynamically consistent  manner, we choose the collagen fiber strain energy per unit mass ${\psi}_{\mathrm{co, m}}$ to be the driving factor for the density change.  In several works, the strain energy was assumed to be the driving factor for collagen mass growth. See \cite{Waffenschmidt_etal_2012} for hard tissue such as bone and \cite{Saez_etal_2013} for hypertension in soft tissue. However, the strain energy of the collagen fibers is a function of the fiber density, which means that in cases where the initial collagen density is zero, the growth rate will remain zero. Therefore splitting the growth rate as we described in Eq.\ (\ref{eq:2-10}) provides a straightforward and physically motivated remedy for this particular problem. 

Here we assume that mechanical stimulation occurs when the energy per mass for collagen fibers ${\psi}_{\mathrm{co, m}}$ exceeds the threshold value ${\psi}_{\mathrm{crit}}$, leading to the case-dependent evolution equation
\begin{equation}
\label{eq:2-14}
\dot{\rho}^{0}_{\mathrm{mech}}  = 
\begin{cases}
a_{2} \, c_{\mathrm{cell}} \, f_{\mathrm{mech}}  \, {\rho}^{0}_{\mathrm{co}} \, \frac{({\psi}_{\mathrm{co, m}} - {\psi}_{\mathrm{crit}})}{{\psi}_{\mathrm{crit}}} ,
&        {\psi}_{\mathrm{co, m}} \ge  {\psi}_{\mathrm{crit}},\\
0  , &       {\psi}_{\mathrm{co, m}} <  {\psi}_{\mathrm{crit}},
\end{cases} 
\end{equation}
where we introduce the material parameter $a_{2}$ ($\mathrm{mm^{3}/cells/day}$).

The collagen density evolution depends on the local density of collagen fibers. To describe an S-shaped evolution of the collagen density, we introduce the exponential decay function 
\begin{equation}
\label{eq:2-15}
f_{\mathrm{mech}} =  e^{-({\rho}^{0}_{\mathrm{co}} / \rho_{\mathrm{th}})},
\end{equation}
where $\rho_{\mathrm{th}}$ is the parameter that controls the saturation level of the collagen density. The value of $f_{\mathrm{mech}}$ is higher when collagen density is significantly lower than $\rho_{\mathrm{th}}$, while for collagen densities significantly higher than $\rho_{\mathrm{th}}$, the value of $f_{\mathrm{mech}}$ is close to zero.

\subsection{Particular choice of the Helmholtz free energy}
The next step is to define the Helmholtz energy per unit volume for each part in the Eq.\ \ref{eq:2-30}. The first part is for the isotropic matrix is described by a compressible Neo-Hookean material law, which is expressed by
\begin{equation}
\label{eq:2-16}
{\psi}_{\mathrm{matrix}} = \frac{\mu}{2}  (\mathrm{tr}({\mathbf C})  -3) - \mu \mathrm{ln}(J) + \frac{\lambda}{4}(J^{2} - 1 - 2 \mathrm{ln}(J)) .
\end{equation}

The second part in the Eq.\ \ref{eq:2-30} is ${\psi}_{\mathrm{co}}$. Collagen bundles in soft biological tissues show exponential stress-strain behavior. This hyperelastic material behavior is prevalent in cardiovascular tissues such as arteries and heart valves. We model this behavior using a Fung-type material model with an exponential strain function \cite{Fung_1990, Holzapfel_etal_2000}. Here, we choose the following strain energy function introduced by Holzapfel et al.\ \cite{Holzapfel_etal_2000}  which is defined in terms of the fiber strain $E$, i.e. 
\begin{equation}
\label{eq:2-20}
{\psi}_{\mathrm{co}} = \frac{\rho^{0}_{\mathrm{co}}}{\rho_{\mathrm{co, f}}} \begin{cases} \frac{k_{1}}{2 \, k_{2}} \, [e^{\mathrm{k_{2}} \, E_{\mathrm{co}}^2} - 1], \: \; \; \lambda_{\mathrm{co}} \ge 1,  \\ 0, \; \; \; \; \; \; \; \; \; \; \; \; \; \; \: \; \; \: \; \; \: \:\; \; \lambda_{\mathrm{co}} < 1, \end{cases}
\end{equation}
where $\rho^{0}_{\mathrm{co}}$ is the density of the collagen fibers during the maturation process and $\rho_{\mathrm{co, f}}$ is the final density of the collagen fiber at the end of the process. The ratio $\frac{\rho^{0}_{\mathrm{co}}}{\rho_{\mathrm{co, f}}}$ is the relative density of the collagen fibers during maturation. The densities $\rho^{0}_{\mathrm{co}}$ and $\rho_{\mathrm{co, f}}$ are defined in the reference configuration. Furthermore, $k_{1}$ is a stiffness-like parameter and $k_{2}$ is a dimensionless quantity.

The last energy in Eq.\ \ref{eq:2-30} is ${\psi}_{\mathrm{tex}}$. Here we use the constitutive equation introduced by Reese et al.\ \cite{Reese_etal_2001} which reads
\begin{equation}
\label{eq:2-26}
\begin{gathered}
{\psi}_{\mathrm{tex}} = K^\mathrm{tex,1}_\mathrm{1} (\mathit{I}^{\mathrm{tex}}_{\mathrm{2}} - 1)^{\beta_1} + K^\mathrm{tex,1}_\mathrm{2} (\mathit{I}^{\mathrm{tex}}_{\mathrm{3}} - 1)^{\beta_2} \\ + K^\mathrm{tex,2}_\mathrm{1} (\mathit{I}^{\mathrm{tex}}_{\mathrm{4}} - 1)^{\gamma_1}  + 
K^\mathrm{tex,2}_\mathrm{2} (\mathit{I}^{\mathrm{tex}}_{\mathrm{5}} - 1)^{\gamma_2}   
+  K^\mathrm{tex}_\mathrm{coup,1} (\mathit{I}_{\mathrm{1}} - 3)^{\delta_1}  (\mathit{I}^{\mathrm{tex}}_{\mathrm{2}} - 1)^{\delta_1} 
\\  +  K^\mathrm{tex}_\mathrm{coup,2} (\mathit{I}_{\mathrm{1}} - 3)^{\delta_2}  (\mathit{I}^{\mathrm{tex}}_{\mathrm{4}} - 1)^{\delta_2} 
+ K^\mathrm{tex}_\mathrm{coup,ani} (\mathit{I}^{\mathrm{tex}}_{\mathrm{2}} - 1)^{\xi}  (\mathit{I}^{\mathrm{tex}}_{\mathrm{4}} - 1)^{\xi} ,
\end{gathered}
\end{equation} 
where $K^\mathrm{tex,1}_\mathrm{1}$, $K^\mathrm{tex,1}_\mathrm{2}$, $K^\mathrm{tex,2}_\mathrm{1}$, $K^\mathrm{tex,2}_\mathrm{2}$, $K^\mathrm{tex}_\mathrm{coup,1}$, $K^\mathrm{tex}_\mathrm{coup,2}$ and $ K^\mathrm{tex}_\mathrm{coup,ani}$ are stiffness-like material parameters. Furthermore, the exponents ${\beta_1}$, ${\beta_2}$, ${\gamma_1}$, ${\gamma_2}$, ${\delta_1}$, ${\delta_2}$ and ${\xi}$ are dimensionless quantities. These parameters will be later identified in Section \ref{sec:4} using experimental data. 

\section{Numerical implementation}
\label{sec:3}

The collagen density evolution is computed by implicitly integrating Eq.\ (\ref{eq:2-10}). A standard Backward-Euler scheme is implemented to compute the collagen density ${\rho}^{0}_{\mathrm{co}}$. The collagen density ${\rho}^{0}_{\mathrm{co}}$ is multiplied by the mass-specific Helmholtz free energy $\mathit{\psi}_{\mathrm{co, m}}$ to obtain the corresponding energy for the collagen fibers per unit reference volume $\mathit{\psi}_{\mathrm{co}}$ as described by Eq.\ (\ref{eq:2-20}). 

\subsection{Time integration}
To apply the time-discretization of the collagen evolution equation, we define the time step as $\Delta t = (t_{\mathrm{n+1}} - t_{\mathrm{n}})$ where $t_{\mathrm{n+1}}$ refers to the current time-step and $t_{\mathrm{n}}$ refers to the previous time-step. Variables from time-step $t_{\mathrm{n}}$ are denoted by the subscript $\mathrm{n}$, and variables from the current time $t_{\mathrm{n+1}}$ are without a subscript. Consequently, the implicit Backward-Euler integration scheme can be expressed as following \begin{equation}
\label{eq:3-1}
{\rho}^{0}_{\mathrm{co}} =  ({\rho}^{0}_{\mathrm{co}})_{\mathrm{n}} + {\Delta t} \, \dot{\rho}^{0}_{\mathrm{co}},
\end{equation}
and the corresponding residual equation is 
\begin{equation}
\label{eq:3-2}
r_{\mathrm{\rho}} = {\rho}^{0}_{\mathrm{co}} - ({\rho}^{0}_{\mathrm{co}})_{\mathrm{n}} - {\Delta t} \, \dot{\rho}^{0}_{\mathrm{co}}  = 0.
\end{equation}

Solving the residual equation using the Newton-Raphson scheme gives us the densification rate $\dot{\rho}^{0}_{\mathrm{co}}$ at the current time-step $t_{\mathrm{n+1}}$. By substituting with $\dot{\rho}^{0}_{\mathrm{co}}$ in Eq.\ (\ref{eq:3-1}), we can find the corresponding collagen density ${\rho}^{0}_{\mathrm{co}}$. Then, ${\rho}^{0}_{\mathrm{co}}$ is inserted in Eq.\ (\ref{eq:2-20}) to compute the corresponding strain energy $\mathit{\psi}_{\mathrm{co}}$. The time-integration steps are explained in Algorithm \ref{densification_algorithm}.

\vspace{0.5cm}
\setlength\belowcaptionskip{1ex}
\begin{algorithm}[H]
	\begin{algorithmic}[1]
		\State Initialize Backward-Euler integration scheme
		\State Input $a_{1}$, $a_{2}$, $c_{\mathrm{cell}}$,  $h$, $\tau$, $t$, ${\psi}_{\mathrm{crit}}$
		\State Output $\rho^{0}_{\mathrm{co}}$ 
		%\For{$t \leq t_{end}$} 
		\State Compute $\dot{\alpha}_{\mathrm{bio}} \gets  \frac{h}{\tau}\, e^{-(t / \tau)^{h}}  \, (\frac{t}{\tau})^{h - 1}$  
		\State Compute $\dot{\rho}^{0}_{\mathrm{bio}} \gets a_{1} \, c_{\mathrm{cell}} \, \dot{\alpha}_{\mathrm{bio}}$
		\State Compute ${\psi}_{\mathrm{co, m}}$
		\If{${\psi}_{\mathrm{co, m}}$ $\geq {\psi}_{\mathrm{crit}} $}
		\State Compute $\dot{\alpha}_{\mathrm{mech}} \gets  e^{-({\rho}^{0}_{\mathrm{co}} / \rho_{\mathrm{th}})}$ 
		\State Compute $\dot{\rho}^{0}_{\mathrm{mech}} \gets a_{2} \, c_{\mathrm{cell}} \, \dot{\alpha}_{\mathrm{mech}} \, {\rho}^{0}_{\mathrm{co}} \, \frac{({\psi}_{\mathrm{co, m}} - {\psi}_{\mathrm{crit}})}{{\psi}_{\mathrm{crit}}} $
		\State $r_{\mathrm{\rho}} \gets {\rho}^{0}_{\mathrm{co}} - ({\rho}^{0}_{\mathrm{co}})_{\mathrm{n}} - {\Delta t} \, \dot{\rho}^{0}_{\mathrm{co}}  = 0 $
		\While  {$|r_{\mathrm{\rho}}| \leq tolerance$}
		\State Compute $\rho^{0}_{\mathrm{co}}$ using Newton-Raphson method
		\EndWhile
		\EndIf
	\end{algorithmic} 
	\caption{Computing collagen fiber density}
	\label{densification_algorithm}
\end{algorithm}

\subsection{Computing stresses and material tangents}
To construct a global finite element system, first we compute the second Piola-Kirchhoff stresses for each constituent and then sum up their individual contribution to get the total second Piola-Kirchhoff stress $\mathbf{S}$ as described by the Eq.\ \ref{eq:2-27}. The computation of second Piola-Kirchhoff stress of the textile scaffold $\mathbf{S}_{\mathrm{tex}}$ and the isotropic matrix $\mathbf{S}_{\mathrm{matrix}}$ can be performed in a straightforward way as already described by the Eq.\ \ref{eq:2-28} and Eq.\ \ref{eq:2-29} respectively. However, the computation of the second Piola-Kirchhoff stress for the collagen part $\mathbf{S}_{\mathrm{co}}$ requires taking into consideration the influence of the collagen evolution equations. The collagen density ${\rho}^{0}_{\mathrm{co}}$ depends on the mass-specific energy ${\psi}_{\mathrm{co, m}}$ and consequently on the Cauchy-Green tensor $\mathbf{C}$, which
introduces additional terms to compute the derivatives as shown in Eq.\ \ref{eq:2-9}. By summing up the contributions from each constituent, we end with an expression for the second Piola-Kirchhoff stress that reads  
\begin{equation}
\label{eq:3-3}
\mathbf{S} = 2 \, \left(\frac{\partial {\psi}_{\mathrm{tex}}}{\partial \mathbf{C}} + \frac{\partial {\psi}_{\mathrm{matrix}}}{\partial \mathbf{C}} + \frac{\partial {\rho}^{0}_{\mathrm{co}}}{\partial \mathbf{C}} \, {\psi}_{\mathrm{co, m}} + {\rho}^{0}_{\mathrm{co}} \, \frac{\partial {\psi}_{\mathrm{co, m}}}{\partial \mathbf{C}} \right).
\end{equation}

 The next step is to compute the material tangent operator ${\bf \mathbb{C}}$. By applying the same approach used to compute $\mathbf{S}$, ${\bf \mathbb{C}}$ is computed by summing up the tangents of each constituent as described by the following expression
\begin{equation}
\label{eq:3-4}
{ \mathbb{C}} = {\mathbb{C}_{\mathrm{tex}}  + { \mathbb{C}_{\mathrm{matrix}}} + { \mathbb{C}_{\mathrm{co}}}}.
\end{equation}

The tangent tensor for each constituent is computed by taking the partial derivative of the second Piola-Kirchhoff stress with respect to the right Cauchy-Green tensor. This gives us the following term for the textile part
\begin{equation}
\label{eq:3-5}
{ \mathbb{C}_{\mathrm{tex}}} = 2 \, \frac{\partial \mathbf{S}_{\mathrm{tex}}}{\partial \mathbf{C}} = 4 \, \frac{\partial^{2} {\psi}_{\mathrm{tex}}}{\partial \mathbf{C}^{2}}.
\end{equation}

Analogously, for the matrix part, we compute the following term
\begin{equation}
\label{eq:3-6}
{ \mathbb{C}_{\mathrm{matrix}}} = 4 \, \frac{\partial^{2} {\psi}_{\mathrm{matrix}}}{\partial \mathbf{C}^{2}}.
\end{equation}

For the collagen part, computing the tangent operator is more complex, because the Helmholtz free energy ${\psi}_{\mathrm{co}}$ depends on the collagen density ${\rho}^{0}_{\mathrm{co}}$. Similar to the other constituents, we describe the tangent operator by the following expression
\begin{equation}
\label{eq:3-7}
{ \mathbb{C}_{\mathrm{co}}} = 4 \, \frac{\partial^{2} {\psi}_{\mathrm{co}}}{\partial \mathbf{C}^{2}}.
\end{equation}

The derivatives in Eq.\ \ref{eq:3-3} and Eqs.\ \ref{eq:3-5}-\ref{eq:3-7} are computed using the code generated by the automatic differentiation software package AceGen \cite{Korelc_2002, Korelc_2009}.

\subsection{Finite element implementation}
In our structural computations, we used a special finite element technology with reduced integration, namely the solid-shell element Q1STs \cite{Reese_2007, Barfusz_2021b}. Q1STs is a low-order isoparametric element with eight nodes. Due to the application of the reduced integration concept, the element contains only one Gauss point within the shell plane. It is especially beneficial for modeling thin shell structures such as heart valves because we can use an arbitrary number of Gauss points through the element's thickness. Furthermore, Q1STs element formulation offers a remedy to volumetric and shear locking. Locking treatment and hourglass stabilization in Q1STs are achieved by applying the concepts of enhanced assumed strain and assumed natural strain. Using elements capable of treating such locking phenomena is essential for us to compute the examples presented in Section \ref{sec:5} because: (i) soft biological tissues are almost incompressible materials which makes them susceptible to volumetric locking, and (b) the structure presented in our work are under severe bending which would cause shear locking in case of using standard low-order element formulation. An additional advantage of using a solid-shell formulation is its ability to model the non-linear material behavior along the thickness direction using only one element. Consequently, we can drastically reduce the number of elements needed in our computations compared to using a standard continuum solid element.

\section{Identification of material parameters}
\label{sec:4}

To accurately model the mechanical behavior of the material, it is necessary to identify the material parameters of our constitutive equations. First, mechanical tests are performed on the textile scaffold and tissue-engineered material. Then, we chemically measure the collagen content of the ECM during the cultivation process. In this way, it is possible to characterize the strain energy function in Eq.\ (\ref{eq:2-20}), which introduces a linear correlation between collagen density and the collagen fibers strain energy ${\psi}_{\mathrm{co}}$. Later on, we identify the parameters of the Weibull cumulative distribution function introduced in Eq.\ (\ref{eq:2-11}), which describes the biologically-driven part of the collagen evolution.

\subsection{Electrospun textile scaffold}

\begin{figure}[h]
	\centering
	\subfloat[\centering \label{fig:3_1a}]{\includegraphics[scale=0.4]{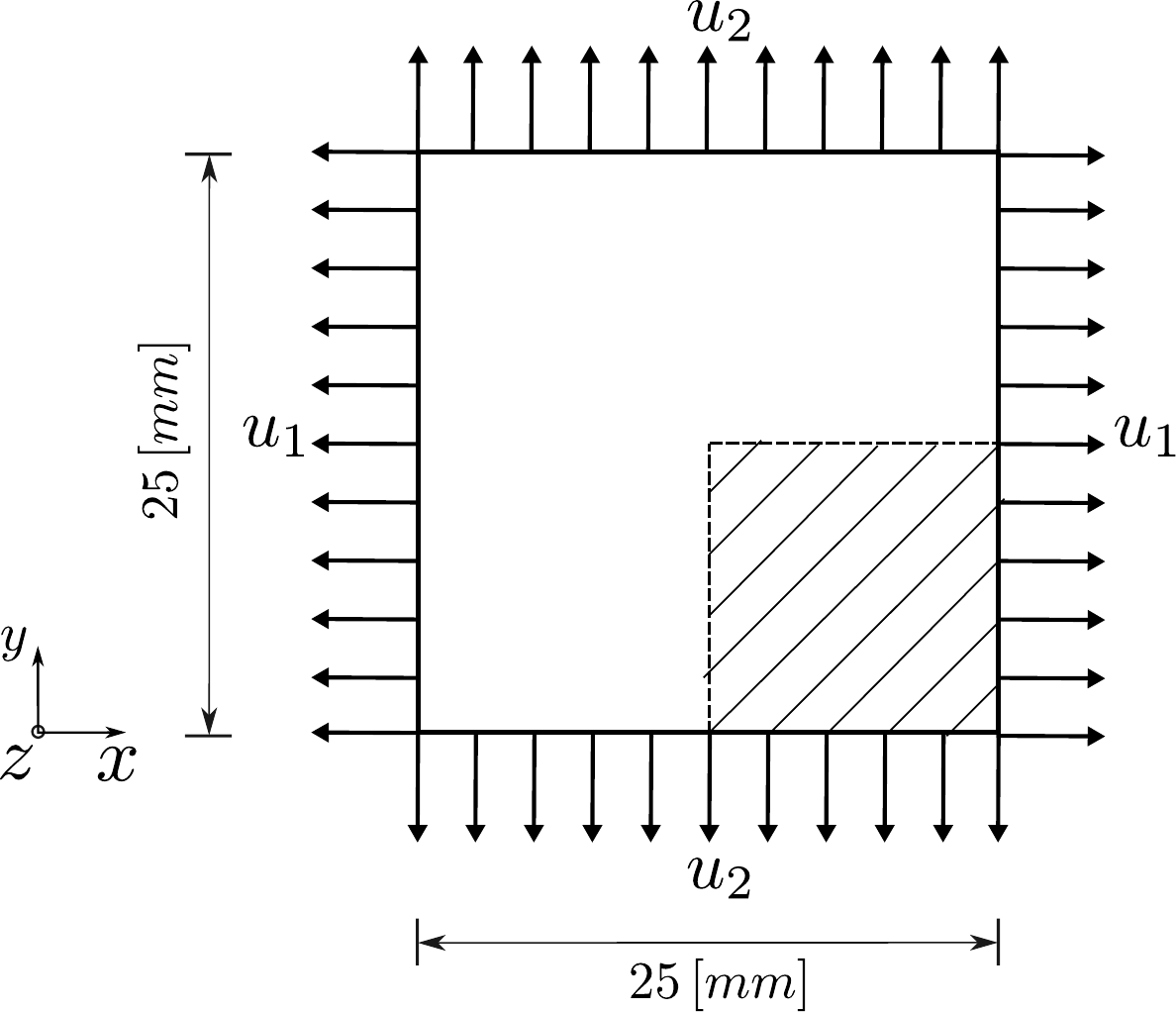}}
	\quad 
	\subfloat[\centering \label{fig:3_1b}]{\includegraphics[scale=0.4]{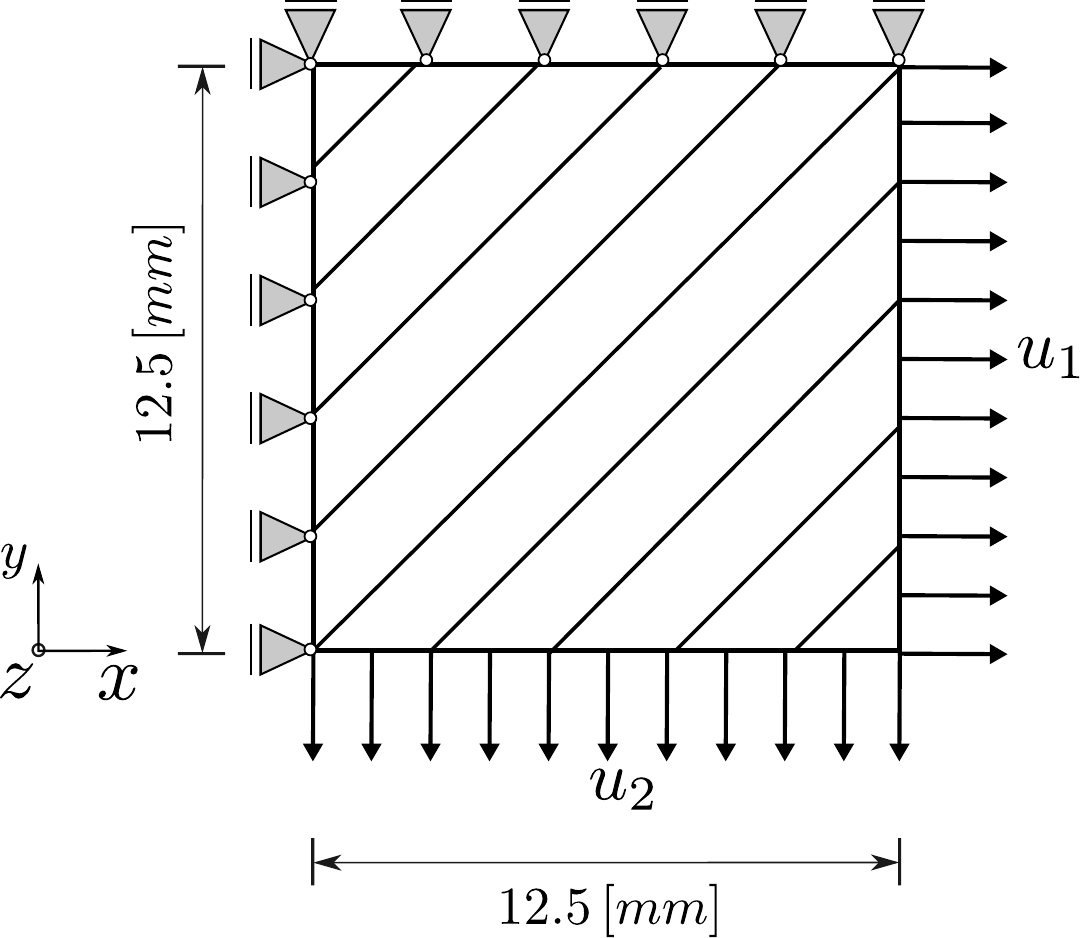}}
	%\vspace{-6pt}
	\caption{(a) Schematic representation for the entire structure of the biaxial tensile test experiment; (b) Boundary value problem for the symmetric part.}
	\label{fig:3_1}    
\end{figure}

The first material investigated here is the textile scaffold. Biaxial tensile testing experiments are performed to measure the mechanical behavior of the scaffold.  The choice for this specific experimental setup is motivated by the deformation behavior in heart valves, where the valve wall is under biaxial tension during the valve closure. In large segments of the heart valve, the strain along the radial direction is significantly higher than the strain along the circumferential direction. The stress-strain behavior of the scaffold is measured for two different setups. In the first setup, equal deformation was applied along the radial direction $u_{1}$ and circumferential directions $u_{2}$ as shown in Fig.\ \ref{fig:3_1}. In the second setup, we apply different displacement boundary conditions along the two directions, with $u_{1} = 3 \, u_{2}$.

%\textcolor{red}{[Add a photo for scaffold material]}

\pgfplotsset{%
	width=0.46\textwidth,
	height=0.45\textwidth
}
\begin{figure}[H]
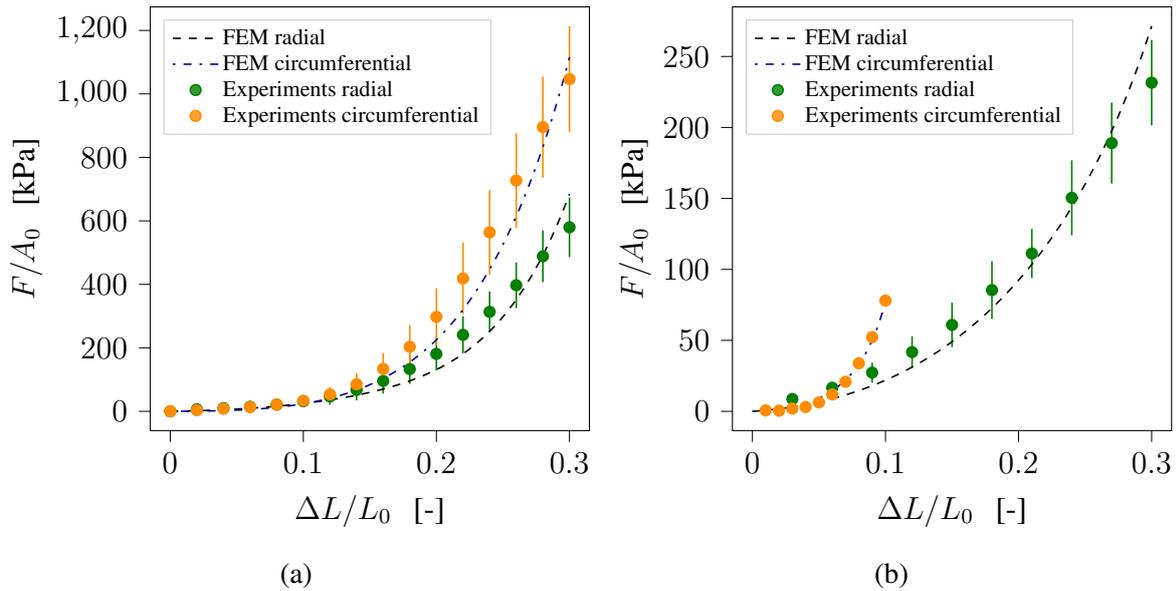

	\centering
	\subfloat[\centering]{\input{figures/equal_strain.tex}}
	\subfloat[\centering]{\input{figures/unequal_strain.tex}}
	%\vspace{-6pt}
	\caption{These plots for the engineering stress $(F/A_{0})$ against the engineering strain $(\Delta L/L_{0})$, where $F$ is the applied load, $A_{0}$ is the initial cross-sectional area, $\Delta L$ is the change in length and $L_{0}$ is the initial length. The relation between the displacement boundary conditions along the two directions is (a) $u_{1} = u_{2}$; and (b) $u_{1} = 3 u_{2}$.}
	\label{fig:3_2}    
\end{figure}

We set up a FEM (Finite Element Method) model for the structure. Displacement boundary conditions are defined as shown by the schematic in Fig.\ \ref{fig:3_1b}, and reaction forces are computed. To identify the material parameters, we implemented an optimization process in MATLAB \cite{MATLAB_2019b} that uses the derivative-free optimization function {\it fminsearch}. The Matlab code simultaneously fits the FEM results to the experimental data for the two experimental setups. The parameter values identified by the optimization routine are listed in Tab.\ \ref{table:3_1}. In Fig. \ref{fig:3_2}, we show the mean and standard deviation of the experimental results and the corresponding FEM results.

\begin{table}[H]
	\centering
	\setlength{\tabcolsep}{30pt}
	\begin{tabular}{c c c}
		\toprule
		Parameter & Value & Units \\
		\midrule
		$K^\mathrm{tex,1}_\mathrm{1} $  & $38.51$ & $\mathrm{[kPa]}$ \\
		$K^\mathrm{tex,1}_\mathrm{2} $ & $1.48$ & $\mathrm{[kPa]}$ \\
		$\beta_1 $ & $3$ &  $[-]$ \\
		$\beta_2 $ & $2$ & $[-]$ \\
		$K^\mathrm{tex,2}_\mathrm{1}$ & $214.39$ & $\mathrm{[kPa]}$ \\
		$K^\mathrm{tex,2}_\mathrm{2}$ & $0.0001$  & $\mathrm{[kPa]}$  \\
		$\gamma_1 $   & $4$ & $[-]$ \\
		$\gamma_2 $  & $2$ & $[-]$ \\
		$ K^\mathrm{tex}_\mathrm{coup,1}$  & $183.72$ & $\mathrm{[kPa]}$ \\
		$\delta_1$  & $2$ &$[-]$ \\
		$ K^\mathrm{tex}_\mathrm{coup,2}$ &  $58.71$ &  $\mathrm{[kPa]}$ \\
		$\delta_2$ & $3$ & $[-]$ \\
		$K^\mathrm{tex}_\mathrm{coup,ani}$  & $571.83$ & $\mathrm{[kPa]}$ \\
		 $\xi $  & $12$ & $[-]$ \\
		\bottomrule
	\end{tabular}
	\caption{Identified material parameters of the textile scaffold constitutive model.}
\label{table:3_1}
\end{table}

\subsection{Tissue-engineered construct}

\begin{figure}[H]
	\centering
	\subfloat[\centering \label{fig:3_3a}]{\includegraphics[height=0.15 \textheight]{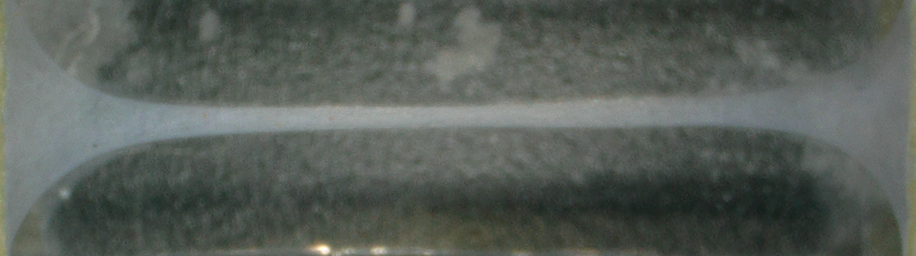}}
	
	\subfloat[\centering \label{fig:3_3b}]{\includegraphics[height=0.25 \textheight]{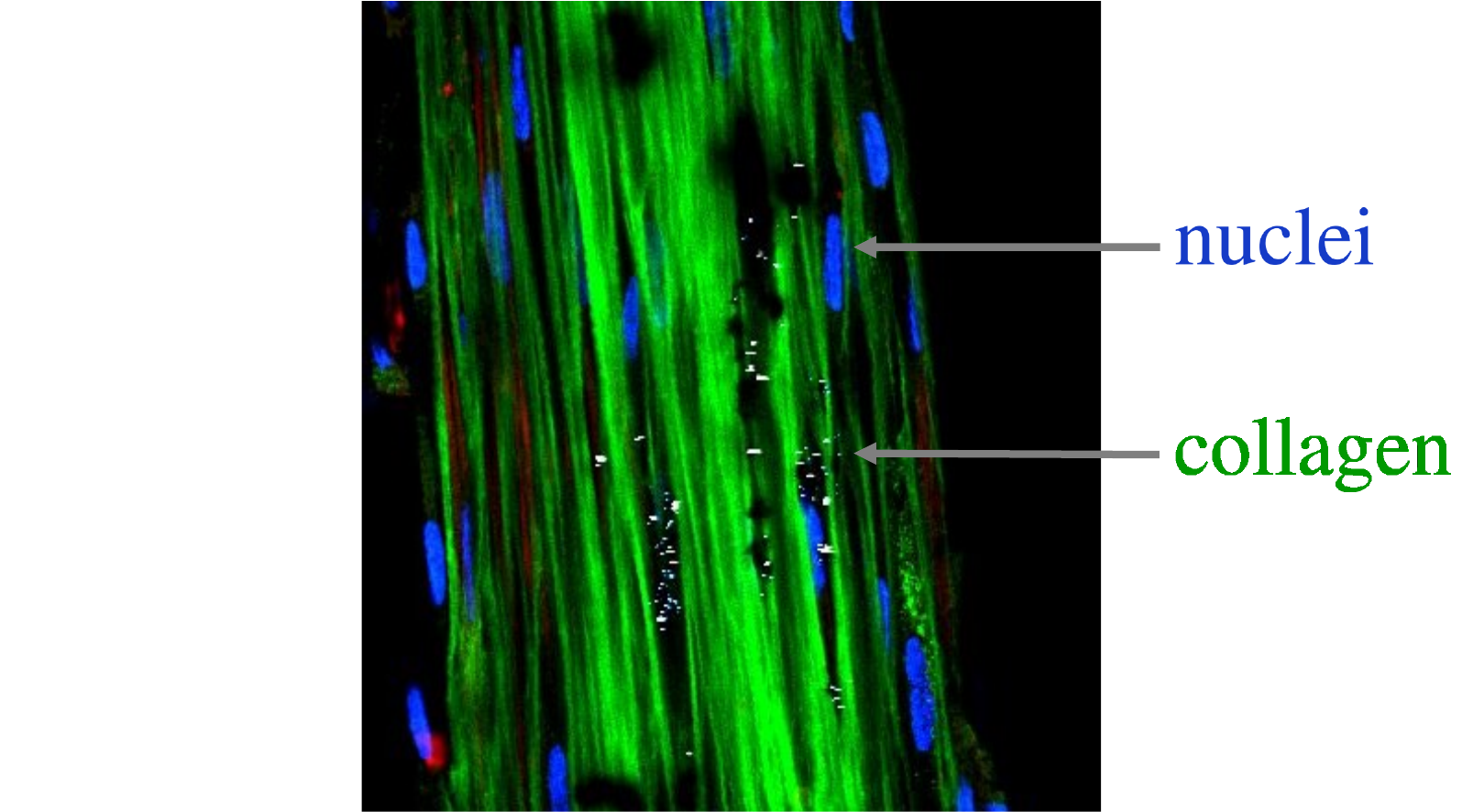}}
	
	\caption{(a) In-vitro cultivated tissue-engineered construct. (b) Image taken by 2-photon microscopy for in-vitro cultivated collagenous tissue. The collagen fibers are indicated by a green stain, and cell nuclei by a blue stain.}
	\label{fig:3_3}    
\end{figure}

The second constituent investigated here is the regenerative collagenous tissue. The tissue is produced by an in-vitro maturation process in a controlled bio-reactor. We fabricated rod-shaped samples where deformation is constrained from both ends of the rod, as shown in Fig.\ \ref{fig:3_3a}. Mechanical loads were not applied during the entire cultivation period. Images taken by 2-photon microscopy have shown that this experimental setup leads to fabricating a tissue with highly oriented collagen fibers. 
There are several approaches to measuring the change in collagen content during the maturation process. Among these approaches is to take microscopy images during the maturation process and then quantify the relative change in collagen content using image analysis techniques. Another approach is to chemically quantify the collagen content using a method called {\it hydroxyproline assay}. Chemical quantification is especially valuable since we can accurately measure the absolute collagen density in the tissue. Furthermore, to measure the changes in the tissue's mechanical behavior during the maturation process, we perform tensile test experiments after 14, 21 and 28 days of maturation. By quantifying the collagen density and measuring corresponding stress-stretch curves, we can assess the validity of our choice for the energy function $\mathit{\psi}_{\mathrm{co}}$ introduced in Eq.\ (\ref{eq:2-20}), where we assume that  $\mathit{\psi}_{\mathrm{co}}$ is linearly dependent on the relative collagen density $ \rho^{0}_{\mathrm{co}} /\rho_{\mathrm{co, f}}$. This requires finding a set of material parameters that accurately describes the tensile behavior of the tissue after 14, 21 and 28 days of maturation. 

\pgfplotsset{%
	width=0.45\textwidth,
	height=0.42\textwidth
}
\begin{figure}[H]
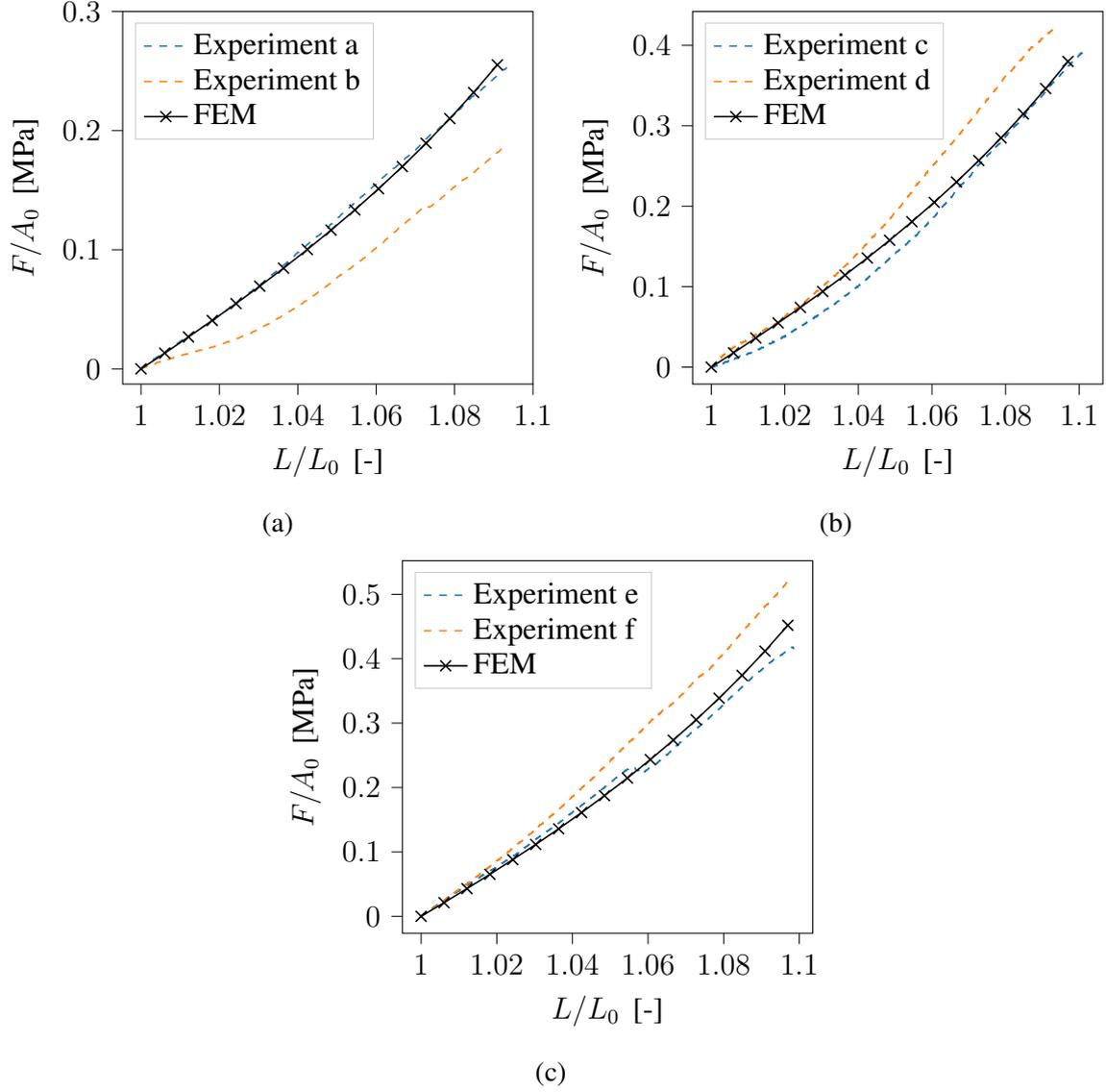

	\centering
	\subfloat[\centering \label{fig:3_4a}]{\input{figures/fitting_14_days.tex}}
	\subfloat[\centering \label{fig:3_4b}]{\input{figures/fitting_21_days.tex}}
	\qquad \qquad
	\subfloat[ \label{fig:3_4c}]{\input{figures/fitting_28_days.tex}}
	\caption{These plots for the engineering stress $(F/A_{0})$ against the stretch $(L/L_{0})$, where $F$ is the applied load, $A_{0}$ is the initial cross-sectional area, $L$ is the current length and $L_{0}$ is the initial length. The plots are for samples after (a) 14, (b) 21, and (c) 28 days of maturation.}
	\label{fig:3_4}    
\end{figure}

\begin{table}[H]
	\centering
	\setlength{\tabcolsep}{6pt}
	\begin{tabular}{c c c}
		\toprule
		Duration $\mathrm{[days]}$ & Average collagen density $\mathrm{[\SI{}{\micro\gram} / \SI{}{\micro l}]}$ & Relative collagen density $[-]$ \\
		\midrule
		$ 14 $  & $23.46$ & $0.6060$ \\
		$ 21 $ & $32.35$ & $0.8357$ \\
		$ 28 $ & $38.71$ &  $1.0$ \\
		\bottomrule
	\end{tabular}
	\caption{Collagen density values measured after different periods of tissue maturation using hydroxyproline assay.}
	\label{table:3_2}
\end{table}

We set up a FEM simulation of a soft collagenous rod to identify the material parameters. In our simulation, we use an idealized geometry with equal rod width and thickness. The rod width chosen in the simulation is equal to the average width of experimentally tested samples. The model is then fitted to the experimental data provided by tensile test specimens for the tissue after 14, 21 and 28 days of maturation. In the simulation, we only vary the relative collagen density  $\rho^{0}_{\mathrm{co}} /\rho_{\mathrm{co, f}}$ where the values are listed in Tab.\ \ref{table:3_2}; other material parameters listed in Tab.\ \ref{table:3_3} remain the same in all the simulations. The value of  $\kappa$ was chosen to be $\kappa = 0$ since the microscopy images show that collagen fibers are uni-axially oriented for rods produced in this specific setups.

 It is important to emphasize that these mechanical tests were performed for six different samples cultivated under the same conditions, with two samples tested for each time points. As a result the collagen content varies from one sample to another, leading to slightly different stress-strain behavior. Furthermore, due to the destructive nature of these tests, the collagen densities listed in Tab.\ \ref{table:3_2} were measured for a different set of samples cultivated under the same conditions. Then, we used the mean value of the collagen density in our FEM computations. Despite this uncertainty, we can see in Fig.\ \ref{fig:3_4} that the constitutive model and the material parameters listed in Tab.\ \ref{table:3_3} can reasonably describe the mechanical behavior of tissue-engineered soft material at different time points during the maturation process.

\begin{table}[H]
	\centering
	\setlength{\tabcolsep}{30pt}
	\begin{tabular}{c c c}
		\toprule
		Parameter & Value & Units \\
		\midrule
		$ \lambda $  & $10.0$ & $\mathrm{[MPa]}$ \\
		$ \mu  $ & $0.05$ & $\mathrm{[MPa]}$ \\
		$ k_1 $ & $0.825$ &  $\mathrm{[MPa]}$ \\
		$ k_2 $ & $4.0$ & $[-]$ \\
		$ \kappa $ & $0.0$ & $[-]$ \\
		\bottomrule
	\end{tabular}
	\caption{Identified material parameters of the collagenous tissue.}
	\label{table:3_3}
\end{table}

\pgfplotsset{%
	width=0.6\textwidth,
	height=0.5\textwidth
}
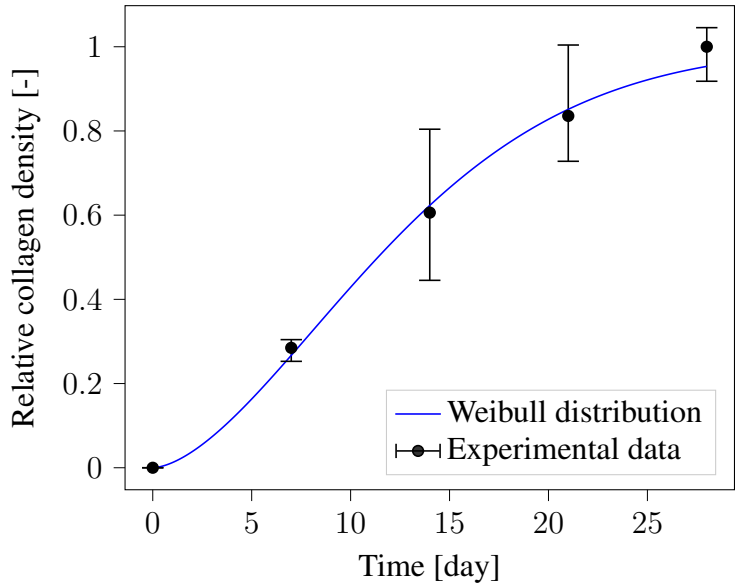
\begin{figure}[ht]
	\centering
	\subfloat[\centering]{% This file was created with tikzplotlib v0.10.1.
\begin{tikzpicture}

\definecolor{darkgray176}{RGB}{176,176,176}
\definecolor{lightgray204}{RGB}{204,204,204}

\begin{axis}[
legend cell align={left},
legend style={
  fill opacity=0.8,
  draw opacity=1,
  text opacity=1,
  at={(0.97,0.03)},
  anchor=south east,
  draw=lightgray204
},
tick align=outside,
tick pos=left,
x grid style={darkgray176},
xlabel={Time [day]},
xmin=-1.4, xmax=29.4,
xtick style={color=black},
y grid style={darkgray176},
ylabel={Relative collagen density [-]},
ymin=-0.0522664686127616, ymax=1.09759584086799,
ytick style={color=black}
]
\path [draw=black, semithick]
(axis cs:0,0)
--(axis cs:0,0);

\path [draw=black, semithick]
(axis cs:7,0.252741684319297)
--(axis cs:7,0.304312337897184);

\path [draw=black, semithick]
(axis cs:14,0.445152673727719)
--(axis cs:14,0.804140015499871);

\path [draw=black, semithick]
(axis cs:21,0.727950064582795)
--(axis cs:21,1.00410609661586);

\path [draw=black, semithick]
(axis cs:28,0.918057349522087)
--(axis cs:28,1.04532937225523);

\addplot [semithick, blue]
table {%
0 0
0.1 0.000278557798793488
0.2 0.000874872482071742
0.3 0.00170837391717771
0.4 0.00274597653943254
0.5 0.00396716574655931
0.6 0.00535731544343621
0.7 0.0069051604130328
0.8 0.00860158001697875
0.9 0.0104389257302074
1 0.0124106133693292
1.1 0.0145108588371266
1.2 0.0167344980388572
1.3 0.0190768592192057
1.4 0.0215336695344619
1.5 0.024100984863505
1.6 0.0267751359067735
1.7 0.0295526860141629
1.8 0.0324303976563994
1.9 0.0354052053947923
2 0.0384741938227781
2.1 0.041634579370332
2.2 0.0448836951509939
2.3 0.0482189782348923
2.4 0.0516379588774761
2.5 0.0551382513405442
2.6 0.0587175460214135
2.7 0.0623736026656182
2.8 0.0661042444838462
2.9 0.0699073530286909
3 0.0737808637139022
3.1 0.0777227618801112
3.2 0.0817310793278543
3.3 0.085803891252192
3.4 0.0899393135240446
3.5 0.0941355002721533
3.6 0.0983906417267411
3.7 0.102702962291826
3.8 0.10707071881801
3.9 0.11149219905159
4 0.115965720239218
4.1 0.120489627870172
4.2 0.125062294540659
4.3 0.129682118926615
4.4 0.134347524853187
4.5 0.13905696045051
4.6 0.143808897386708
4.7 0.148601830170068
4.8 0.153434275513321
4.9 0.158304771753729
5 0.163211878323417
5.1 0.168154175264971
5.2 0.173130262787872
5.3 0.178138760861808
5.4 0.183178308843308
5.5 0.188247565132506
5.6 0.193345206857173
5.7 0.198469929581436
5.8 0.203620447036838
5.9 0.208795490873644
6 0.213993810430477
6.1 0.21921417252056
6.2 0.224455361232982
6.3 0.229716177747575
6.4 0.234995440162091
6.5 0.240291983330503
6.6 0.245604658711342
6.7 0.250932334225098
6.8 0.256273894119774
6.9 0.261628238843774
7 0.266994284925374
7.1 0.272370964858086
7.2 0.27775722699128
7.3 0.283152035425498
7.4 0.288554369911909
7.5 0.293963225755438
7.6 0.299377613721119
7.7 0.30479655994325
7.8 0.310219105836992
7.9 0.315644308012057
8 0.321071238188175
8.1 0.326498983112042
8.2 0.331926644475494
8.3 0.33735333883466
8.4 0.342778197529864
8.5 0.348200366606085
8.6 0.353619006733783
8.7 0.359033293129915
8.8 0.364442415478998
8.9 0.36984557785407
9 0.375241998637424
9.1 0.380630910440995
9.2 0.386011560026305
9.3 0.391383208223858
9.4 0.396745129851911
9.5 0.402096613634538
9.6 0.407436962118925
9.7 0.412765491591832
9.8 0.418081531995168
9.9 0.423384426840635
10 0.42867353312341
10.1 0.433948221234807
10.2 0.439207874873916
10.3 0.444451890958183
10.4 0.449679679532913
10.5 0.45489066367968
10.6 0.46008427942364
10.7 0.465259975639733
10.8 0.470417213957775
10.9 0.475555468666439
11 0.48067422661613
11.1 0.485772987120753
11.2 0.490851261858397
11.3 0.49590857477093
11.4 0.500944461962529
11.5 0.505958471597156
11.6 0.510950163795004
11.7 0.515919110527915
11.8 0.52086489551382
11.9 0.525787114110197
12 0.53068537320658
12.1 0.53555929111615
12.2 0.540408497466431
12.3 0.545232633089111
12.4 0.550031349909027
12.5 0.554804310832345
12.6 0.559551189633948
12.7 0.564271670844085
12.8 0.568965449634303
12.9 0.573632231702681
13 0.578271733158433
13.1 0.582883680405874
13.2 0.587467810027806
13.3 0.592023868668362
13.4 0.596551612915316
13.5 0.601050809181929
13.6 0.605521233588329
13.7 0.609962671842494
13.8 0.614374919120844
13.9 0.618757779948498
14 0.623111068079219
14.1 0.627434606375082
14.2 0.631728226685907
14.3 0.635991769728488
14.4 0.640225084965647
14.5 0.644428030485156
14.6 0.648600472878554
14.7 0.652742287119899
14.8 0.65685335644448
14.9 0.660933572227535
15 0.664982833862991
15.1 0.669001048642278
15.2 0.672988131633228
15.3 0.676944005559112
15.4 0.68086860067783
15.5 0.684761854661295
15.6 0.688623712475033
15.7 0.692454126258041
15.8 0.696253055202919
15.9 0.700020465436317
16 0.70375632989972
16.1 0.707460628230596
16.2 0.711133346643945
16.3 0.714774477814264
16.4 0.718384020757968
16.5 0.721961980716279
16.6 0.725508369038618
16.7 0.729023203066527
16.8 0.732506506018136
16.9 0.735958306873207
17 0.739378640258776
17.1 0.742767546335416
17.2 0.746125070684142
17.3 0.749451264193985
17.4 0.752746182950248
17.5 0.756009888123468
17.6 0.759242445859115
17.7 0.762443927168022
17.8 0.7656144078176
17.9 0.768753968223814
18 0.771862693343986
18.1 0.774940672570394
18.2 0.777987999624724
18.3 0.781004772453364
18.4 0.783991093123573
18.5 0.78694706772052
18.6 0.789872806245238
18.7 0.792768422513471
18.8 0.795634034055459
18.9 0.798469762016647
19 0.801275731059358
19.1 0.804052069265411
19.2 0.806798908039726
19.3 0.809516382014894
19.4 0.812204628956762
19.5 0.814863789670996
19.6 0.817494007910676
19.7 0.820095430284893
19.8 0.822668206168391
19.9 0.825212487612227
20 0.827728429255489
20.1 0.830216188238051
20.2 0.832675924114385
20.3 0.835107798768443
20.4 0.837511976329589
20.5 0.839888623089615
20.6 0.842237907420821
20.7 0.844559999695179
20.8 0.846855072204579
20.9 0.849123299082152
21 0.851364856224687
21.1 0.853579921216135
21.2 0.855768673252199
21.3 0.85793129306602
21.4 0.860067962854952
21.5 0.862178866208433
21.6 0.864264188036945
21.7 0.866324114502064
21.8 0.868358832947611
21.9 0.870368531831883
22 0.872353400660975
22.1 0.874313629923194
22.2 0.876249411024551
22.3 0.878160936225341
22.4 0.880048398577801
22.5 0.881911991864837
22.6 0.883751910539837
22.7 0.885568349667545
22.8 0.887361504865998
22.9 0.88913157224953
23 0.890878748372831
23.1 0.892603230176054
23.2 0.894305214930968
23.3 0.895984900188158
23.4 0.897642483725257
23.5 0.8992781634962
23.6 0.900892137581514
23.7 0.902484604139617
23.8 0.904055761359128
23.9 0.905605807412182
24 0.907134940408742
24.1 0.908643358351899
24.2 0.910131259094153
24.3 0.911598840294673
24.4 0.913046299377519
24.5 0.914473833490825
24.6 0.915881639466934
24.7 0.917269913783473
24.8 0.918638852525363
24.9 0.919988651347756
25 0.921319505439889
25.1 0.922631609489845
25.2 0.923925157650213
25.3 0.925200343504643
25.4 0.926457360035272
25.5 0.927696399591039
25.6 0.928917653856843
25.7 0.930121313823567
25.8 0.931307569758945
25.9 0.932476611179258
26 0.933628626821857
26.1 0.934763804618502
26.2 0.935882331669503
26.3 0.936984394218661
26.4 0.938070177628985
26.5 0.939139866359195
26.6 0.940193643940983
26.7 0.941231692957029
26.8 0.942254195019766
26.9 0.943261330750875
27 0.944253279761509
27.1 0.945230220633228
27.2 0.946192330899637
27.3 0.947139787028723
27.4 0.948072764405872
27.5 0.948991437317559
27.6 0.949895978935703
27.7 0.950786561302674
27.8 0.951663355316944
27.9 0.952526530719367
28 0.953376256080085
};
\addlegendentry{Weibull distribution}
\addplot [semithick, black, mark=*, mark size=2, mark options={solid}, only marks, legend image post style={sharp plot,|-|}]
table {%
	0 0
	7 0.28486
	14 0.606
	21 0.8357
	28 1
};
\addlegendentry{Experimental data}
\addplot [semithick, black, mark=-, mark size=4, mark options={solid}, only marks]
table {%
	0 0
	7 0.252741684319297
	14 0.445152673727719
	21 0.727950064582795
	28 0.918057349522087
};
\addplot [semithick, black, mark=-, mark size=4, mark options={solid}, only marks]
table {%
	0 0
	7 0.304312337897184
	14 0.804140015499871
	21 1.00410609661586
	28 1.04532937225523
};
\end{axis}

\end{tikzpicture}}
	\caption{Relative collagen density measured after 7, 14, 21 and 28 days of maturation. The experimental data are fitted by a Weibull cumulative distribution curve.}
	\label{fig:3_5}    
\end{figure}

The increase in the relative collagen density during the maturation process is plotted in Fig.\ \ref{fig:3_5} where the black dot represents the average collagen density, and the black lines represent the ranges of density values measured experimentally by hydroxyproline assay. The cumulative increase in collagen density deposition can be accurately modeled by the Weibull cumulative distribution function introduced in Eq.\ (\ref{eq:2-11}). The parameters for the plotted Weibull cumulative distribution are listed in Tab.\ \ref{table:3_4}. 

\begin{table}[H]
	\centering
	\setlength{\tabcolsep}{30pt}
	\begin{tabular}{c c c}
		\toprule
		Parameter & Value & Units \\
		\midrule
		$ \tau $  & $14.21$ & $\mathrm{[days]}$ \\
		$ h  $ & $1.65$ & $[-]$ \\
		\bottomrule
	\end{tabular}
	\caption{Parameters identified for the Weibull distribution curve.}
	\label{table:3_4}
\end{table}
\section{Results and discussion}
\label{sec:5}

Our next step is to apply the material model and finite element implementation explained in Sections \ref{sec:2} and \ref{sec:3} to compute structural mechanics examples. To better understand our model's characteristics, we evaluate the maturation of a pressurized shell construct. Collagen growth in soft cardiovascular tissues under pressure loads, such as the formation of atheromatous fibrous caps or collagen accumulation due to aortic aneurysm, are well investigated phenomena. This makes it possible to qualitatively assess the validity of our model. In the next step, we compute the collagen evolution in a tubular-shaped biohybrid heart valve. As explained in Section \ref{sec:1}, the tubular design was introduced by Weber et al.\ \cite{Weber_etal_2014} to improve the reliability of the fabrication and suturing process of the valve. Since this special design is not widely investigated in the literature, applying our model can give us a better insight on the growth behavior during the tissue maturation process. Furthermore, from a numerical point of view, the tubular valve example is highly challenging, since the structure shows a snap-through behavior under quasi-static loading conditions. In addition to that, it is necessary to consider the contact conditions.

We perform the finite element computations using the software package FEAP \cite{Taylor_2020}. To visualize our results, we use ParaView \cite{Ahrens_2005} to generate the contour plots and Matplotlib \cite{Hunter_2007} for curve plotting. 

\subsection{Pressurized shell construct}

\begin{figure}[H]
	\centering
	\captionsetup{justification=raggedright,singlelinecheck=false}
	\begin{subfigure}{0.65\linewidth}
		\subcaption{}
		\label{}
		\includegraphics[scale=0.45]{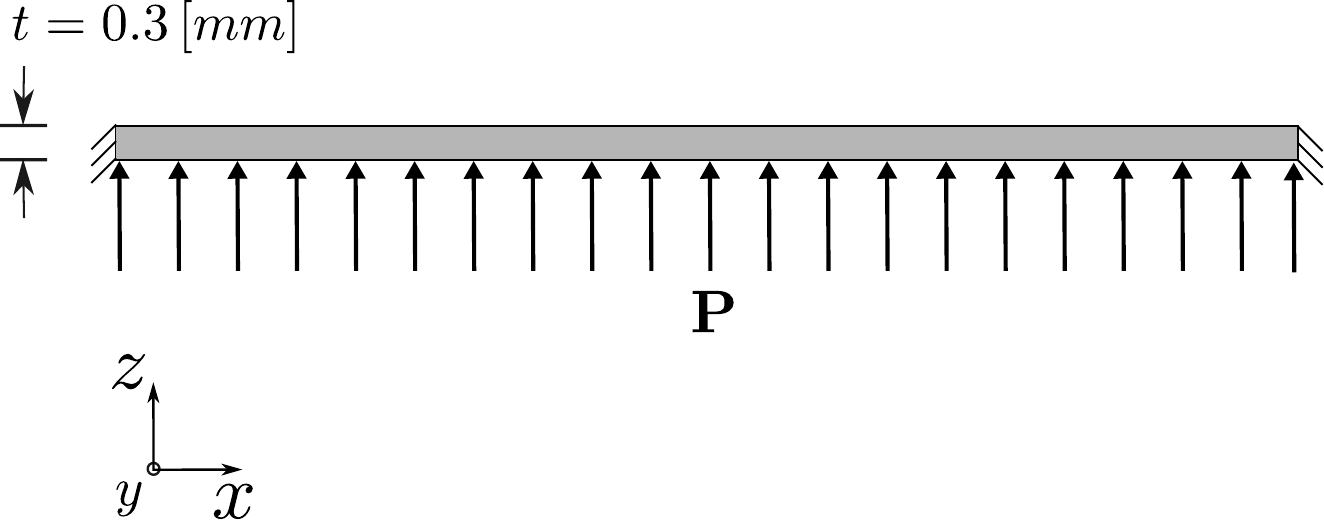}
	\end{subfigure}
	\begin{subfigure}{0.2\linewidth}
		\label{}
		\subcaption{}
		\includegraphics[scale=0.45]{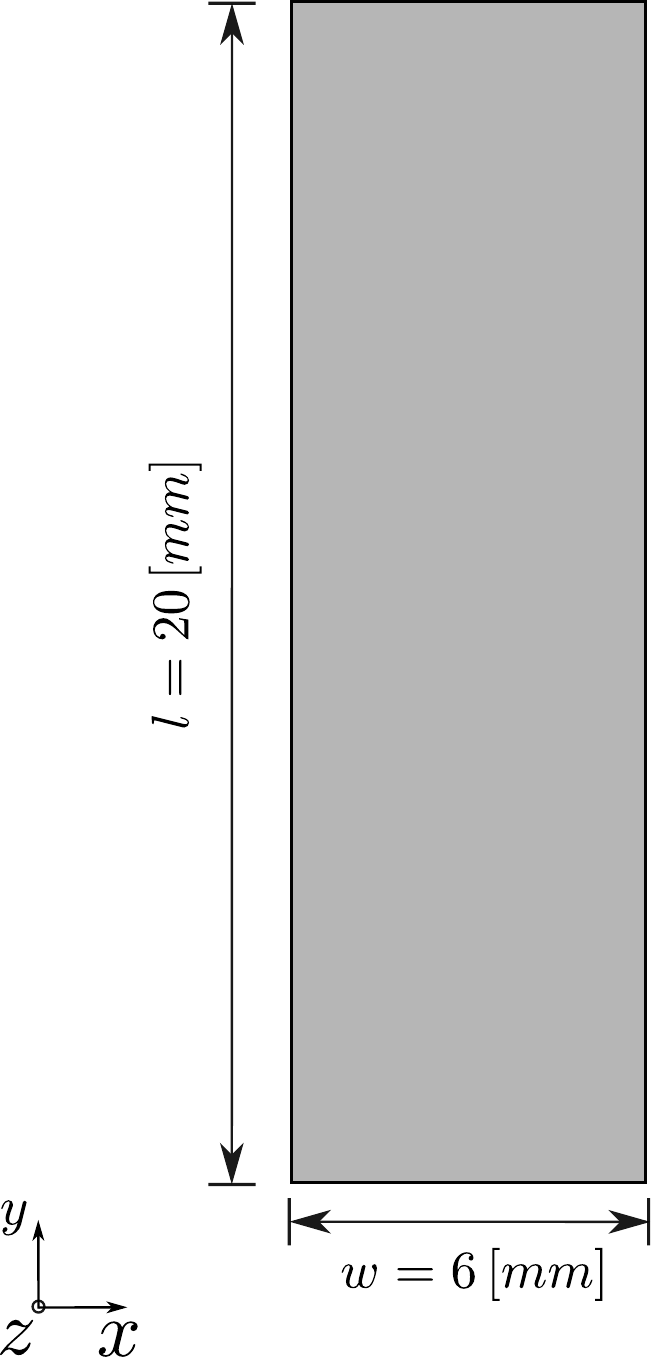}
	\end{subfigure}
	
	\caption{(a) Schematic representation of the boundary value problem for pressure-loaded soft biological tissue with thickness $t = 0.3$ mm ; (b) top view of the structure.}
	\label{fig:5_1}   
\end{figure}
\vspace{-0.8cm}

In this problem, we consider a tissue-engineered biohybrid shell construct. During the in-vitro maturation process a constant pressure load is applied, as illustrated in Fig.\ \ref{fig:5_1}. The construct length is $l = 20 \; \mathrm{mm}$ and width $w = 6 \; \mathrm{mm}$. The thickness is $t = 0.3 \; \mathrm{mm}$. The tissue is fixed at both ends, and a constant pressure load of $P = 2 \; \mathrm{kPa}$ is applied on the bottom surface.

Initially, the mechanical characteristics of the construct are dominated by the mechanical behavior of the scaffold. However, during the maturation process, the density of collagen increases which affects the overall behavior of the structure. We investigated the mechanical behavior of the scaffold and soft collagenous tissue in Section \ref{sec:3}. The identified material parameters are listed in Tab.\ \ref{table:3_1} and \ref{table:3_3}. These parameters are used to compute this example. However, to consider that collagen fibers are not fully uni-axially aligned within the scaffold, we change the value of the parameter $\kappa$ to $\kappa = 0.15$, with the mean orientation of the collagen fibers defined to be along the x-axis. 

The last set of parameters needed for the computations are parameters that define the collagen density evolution during the maturation process. For the parameters $\tau$ and $h$, which describe the Weibull function, we use the values listed in Tab.\ \ref{table:3_4}. The values used for the remaining parameters and the corresponding references are listed in Tab.\ \ref{table:5_1}.
\begin{table}[H]
	\centering
	\setlength{\tabcolsep}{30pt}
	\begin{tabular}{c c c c}
		\toprule
		Parameter & Value & Units & Reference \\
		\midrule
		$ a_{1} $  & $5 \times 10^{-4}$ & $\mathrm{[\SI{}{\micro\gram} / cells]}$ & own \\
		$ a_{2} $  & $5 \times 10^{-7}$ & $\mathrm{[mm^{3}/cells/day]}$ & own \\
		$ {\psi}_{\mathrm{crit}} $  & $2 \times 10^{-5}$ & $\mathrm{[J/\SI{}{\micro\gram}]}$ & own \\
		$ {\rho}_{\mathrm{th}} $ & $10.0$ & $\mathrm{[\SI{}{\micro\gram} / mm^{3}]}$ & own \\
		$ {\rho}_{\mathrm{co, f}} $ & $38.71$ & $\mathrm{[\SI{}{\micro\gram} / mm^{3}]}$ & Table \ref{table:3_2} \\
		$ c_{\mathrm{cell}}  $ & $15 \times 10^{3}$ & $\mathrm{[cells/mm^{3}]}$ & \cite{Hermans_etal_2022} \\
		\bottomrule
	\end{tabular}
	\caption{Values for additional modeling parameters.}
	\label{table:5_1}
\end{table}

\begin{figure}[H]
	\centering 
	\subfloat[\centering \label{fig:5_11a}]{\includegraphics[scale=0.6]{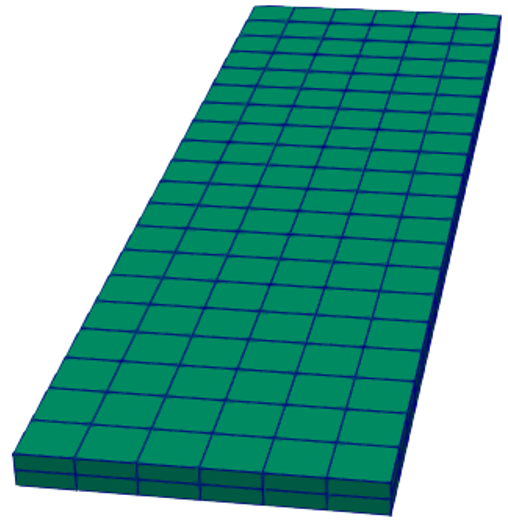}}
	\quad 
	\subfloat[\centering \label{fig:5_11b}]{\includegraphics[scale=0.6]{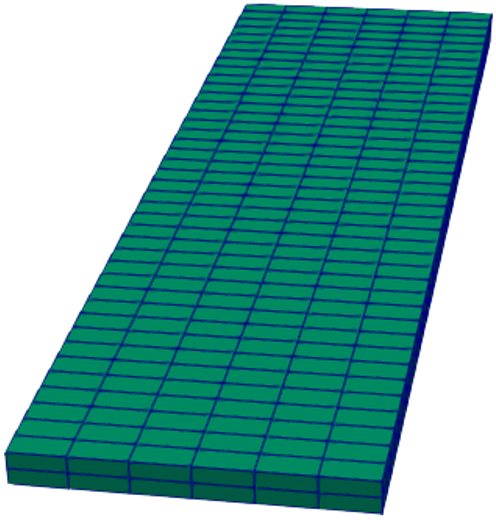}}
	\subfloat[\centering
	\label{fig:5_11c}]{\includegraphics[scale=0.6]{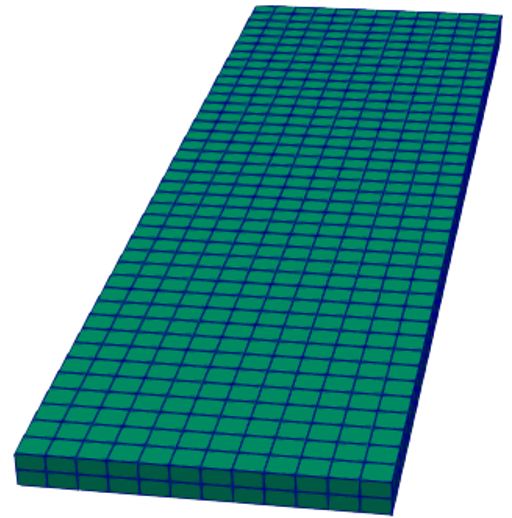}}
	\caption{Finite element mesh with three different mesh refinements. The total number of elements is: (a) 240 elements,  (b) 480 elements and (c) 960 elements. }
	\label{fig:5_11}    
\end{figure}

\begin{figure}[ht]
	\centering
	\input{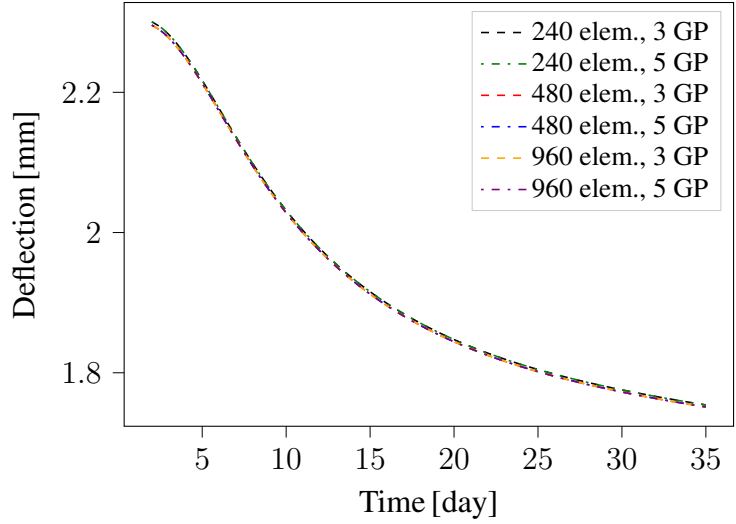}
	\caption{Maximum deflection of the tissue-engineered construct over time under constant pressure loading of $2 \; \mathrm{kPa}$. Results are computed using three refined meshes, using three and five Gauss points through the element's thickness. }
	\label{fig:5_3}    
\end{figure}

The computations are performed using the solid-shell element with reduced integration Q1STs \cite{Reese_2007, Barfusz_2021b}. A mesh convergence study is performed using three mesh refinement levels with 240, 480 and 960 elements, as shown in Fig.\ \ref{fig:5_11}. At each refinement level; the boundary value problem was computed using three and five Gauss points defined along the thickness of each element. By plotting the maximum deflection along the z-direction against the time, as shown in Fig.\ \ref{fig:5_3}, we observe an excellent mesh convergence behavior, even with a coarse mesh with only 240 elements, where each element contains three Gauss points. Therefore, we use for our simulations the mesh in Fig.\ \ref{fig:5_11a}, which consists 240 elements, with three Gauss points in each element.

\pgfplotsset{%
	width=0.6\textwidth,
	height=0.45\textwidth
}

\begin{figure}[H]
	\begin{minipage}{.7\columnwidth}
		\captionsetup{justification=raggedright,singlelinecheck=false}
		\centering
		\begin{subfigure}{0.5\linewidth}
			\subcaption{}
			\label{}
			\includegraphics[scale=0.35]{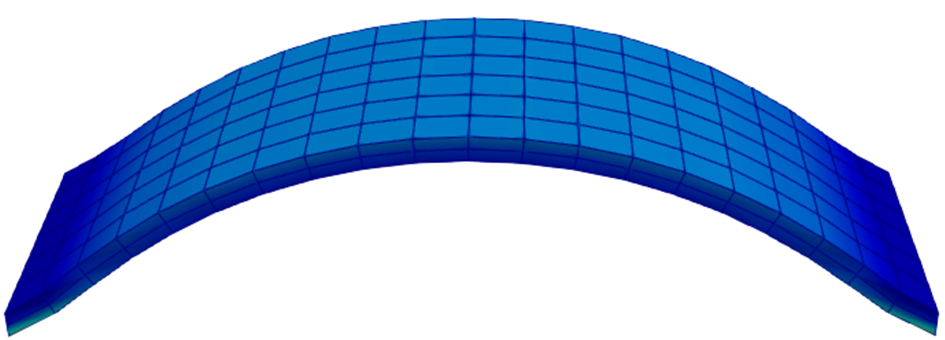}
		\end{subfigure}
		%\vspace{0.5cm}
		
		\begin{subfigure}{0.5\linewidth}
			\subcaption{}
			\label{}
			\includegraphics[scale=0.35]{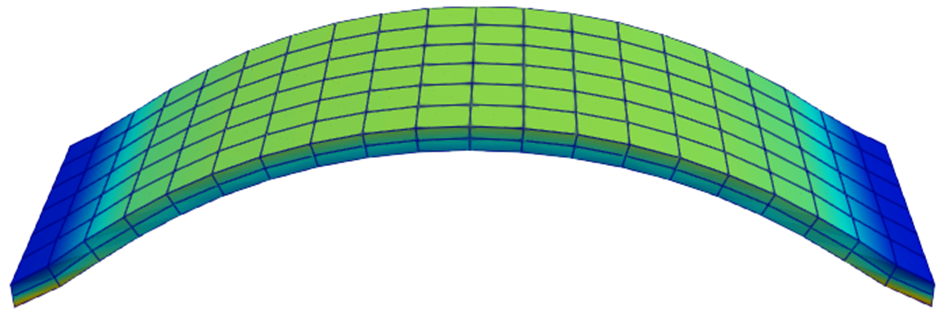}
		\end{subfigure}
		%\vspace{0.5cm}
		
		\begin{subfigure}{0.5\linewidth}
			\subcaption{}
			\label{}
			\includegraphics[scale=0.35]{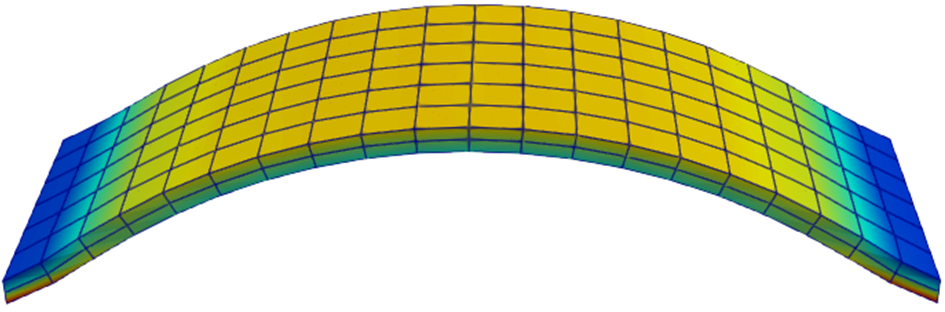}
		\end{subfigure}
		%\vspace{0.5cm}

		\begin{subfigure}{0.5\linewidth}
			\subcaption{}
			\label{}
			\includegraphics[scale=0.35]{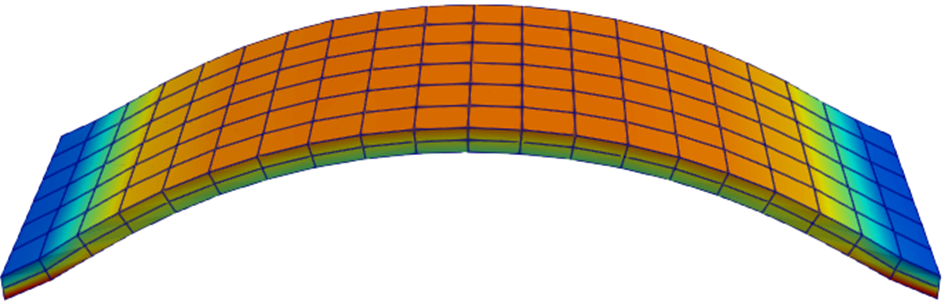}
		\end{subfigure}
	\end{minipage}\qquad \quad \qquad 
	\begin{minipage} {.15\columnwidth}
		\vspace{2cm} \hspace{5cm}
		\includegraphics[height=0.20\textheight]{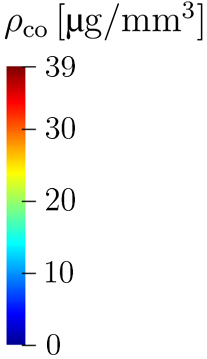}
	\end{minipage}
	\caption{Evolution of collagen density during the maturation process for the tissue-engineered construct after: (a) 7 days, (b) 14 days, (c) 21 days and (d) 28 days.}
	\label{fig:5_2}    
\end{figure}

The contour plots in Fig.\ \ref{fig:5_2} show the collagen density distribution at different time points during the maturation process. We can observe that collagen density gradually increases during the maturation process. Collagen fibers accumulate in regions which are highly strained along the fiber direction. The results here resemble the accumulation of collagen in abdominal aortic aneurysm \cite{Takahashi_2023}. The accumulation of collagen fibers increases the stiffness of the construct. As a result, the maximum deflection of the construct along the z-direction decreases over time, as shown in Fig.\ \ref{fig:5_3}.

\begin{figure}[H]
	\centering
	\includegraphics[scale=0.45]{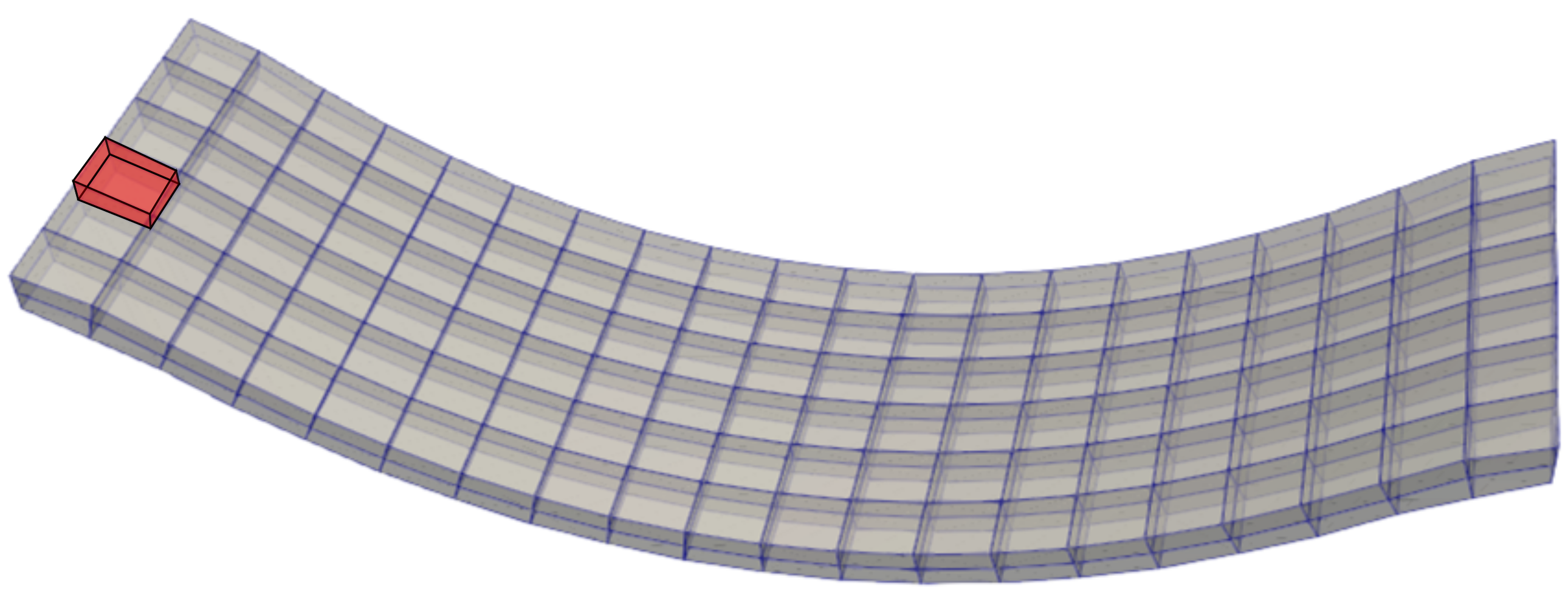}
	
	\caption{Bottom view of the deformed shell construct. The parameter study plots in Figs.\ \ref{fig:5_5} and \ref{fig:5_6} are for the element highlighted in red.}
	\label{fig:5_4}    
\end{figure}

To understand the influence of the parameters introduced in the collagen evolution model, we study the influence of various parameters on the collagen density evolution in the element highlighted in Fig.\ \ref{fig:5_4}. We choose this specific element because it shows the highest collagen concentration in structure. There, we output the average collagen density in the highlighted element. The parameters of interest are $a_{1}$, $a_{2}$, ${\psi}_{\mathrm{crit}}$ and ${\rho}_{\mathrm{th}}$, which are listed in Tab.\ \ref{table:5_1}. We vary the value of each parameter one at a time and plot the results for three different values. The results are plotted in Fig.\ \ref{fig:5_5}.  

%\vspace{-1cm}

\pgfplotsset{%
	width=0.495\textwidth,
	height=0.45\textwidth
}

\begin{figure}[H]
	\centering 
	\subfloat[\centering \label{fig:5_5a}]{% This file was created with tikzplotlib v0.10.1.
\begin{tikzpicture}

\definecolor{darkgray176}{RGB}{176,176,176}
\definecolor{darkorange25512714}{RGB}{255,127,14}
\definecolor{forestgreen4416044}{RGB}{44,160,44}
\definecolor{lightgray204}{RGB}{204,204,204}
\definecolor{steelblue31119180}{RGB}{31,119,180}

\begin{axis}[
legend cell align={left},
legend style={
  fill opacity=0.8,
  draw opacity=1,
  text opacity=1,
  at={(0.97,0.03)},
  anchor=south east,
  draw=lightgray204,
  font=\fontsize{10.5}{10.5}
},
tick align=outside,
tick pos=left,
x grid style={darkgray176},
x label style={at={(axis description cs:0.5,-0.125)},anchor=north},
y label style={at={(axis description cs:-0.115,.5)},anchor=south},
xlabel={Time\,[day]},
xmin=-1.7479, xmax=36.7499,
xtick style={color=black},
y grid style={darkgray176},
ylabel={${\rho}^{0}_{\mathrm{co}} \, \mathrm{[\SI{}{\micro\gram} / mm^{3}]}$},
ymin=0, ymax=40,
ytick style={color=black}
]
\addplot [semithick, steelblue31119180]
table {%
0.002 1.0927e-05
0.004 2.8073e-05
0.006 5.0389e-05
0.008 7.7294e-05
0.01 0.0001084
0.012 0.00014342
0.014 0.00018212
0.016 0.00022434
0.018 0.00026992
0.02 0.00031873
0.025 0.00045981
0.03 0.00061864
0.035 0.00079423
0.04 0.00098576
0.045 0.0011926
0.05 0.001414
0.055 0.0016497
0.06 0.0018991
0.065 0.0021619
0.07 0.0024378
0.075 0.0027263
0.08 0.0030273
0.085 0.0033405
0.09 0.0036656
0.095 0.0040025
0.1 0.0043509
0.11 0.0050929
0.12 0.0058786
0.13 0.0067071
0.14 0.0075772
0.15 0.0084883
0.16 0.0094394
0.17 0.01043
0.18 0.011459
0.19 0.012527
0.2 0.013632
0.21 0.014774
0.22 0.015954
0.23 0.017169
0.24 0.018421
0.25 0.019709
0.26 0.021032
0.27 0.022391
0.28 0.023785
0.29 0.025213
0.3 0.026677
0.35 0.034859
0.4 0.043899
0.45 0.053799
0.5 0.064569
0.55 0.076225
0.6 0.08879
0.65 0.10229
0.7 0.11677
0.75 0.13225
0.8 0.14879
0.85 0.16644
0.9 0.18525
0.95 0.20528
1 0.22661
1.05 0.2493
1.1 0.27344
1.15 0.29912
1.2 0.32643
1.25 0.35548
1.3 0.38638
1.35 0.41924
1.4 0.45421
1.45 0.4914
1.5 0.53098
1.55 0.57309
1.6 0.6179
1.65 0.66558
1.7 0.71631
1.75 0.77028
1.8 0.82768
1.85 0.88872
1.9 0.9536
1.95 1.0225
2 1.0957
2.05 1.1722
2.1 1.252
2.15 1.3351
2.2 1.4217
2.25 1.4658
2.5 2.0515
2.75 2.6664
3 3.3383
3.25 4.0487
3.5 4.7815
3.75 5.5239
4 6.2668
4.25 7.0035
4.5 7.7297
4.75 8.4426
5 9.1406
5.25 9.8227
5.5 10.488
5.75 11.138
6 11.771
6.25 12.388
6.5 12.989
6.75 13.576
7 14.148
7.25 14.706
7.5 15.25
7.75 15.781
8 16.299
8.25 16.805
8.5 17.3
8.75 17.783
9 18.254
9.25 18.716
9.5 19.166
9.75 19.607
10 20.038
10.25 20.459
10.5 20.872
10.75 21.275
11 21.67
11.25 22.056
11.5 22.433
11.75 22.803
12 23.165
12.25 23.519
12.5 23.865
12.75 24.204
13 24.536
13.25 24.861
13.5 25.179
13.75 25.49
14 25.794
14.25 26.092
14.5 26.383
14.75 26.668
15 26.947
15.25 27.22
15.5 27.487
15.75 27.749
16 28.004
16.25 28.254
16.5 28.499
16.75 28.738
17 28.972
17.25 29.201
17.5 29.425
17.75 29.644
18 29.858
18.25 30.067
18.5 30.272
18.75 30.472
19 30.667
19.25 30.859
19.5 31.046
19.75 31.228
20 31.407
20.25 31.582
20.5 31.753
20.75 31.92
21 32.083
21.25 32.243
21.5 32.399
21.75 32.552
22 32.701
22.25 32.847
22.5 32.99
22.75 33.129
23 33.266
23.25 33.399
23.5 33.53
23.75 33.658
24 33.783
24.25 33.905
24.5 34.025
24.75 34.142
25 34.256
25.25 34.369
25.5 34.478
25.75 34.586
26 34.691
26.25 34.795
26.5 34.896
26.75 34.995
27 35.092
27.25 35.187
27.5 35.28
27.75 35.372
28 35.461
28.25 35.549
28.5 35.636
28.75 35.72
29 35.803
29.25 35.885
29.5 35.965
29.75 36.044
30 36.121
30.25 36.197
30.5 36.272
30.75 36.346
31 36.418
31.25 36.489
31.5 36.559
31.75 36.628
32 36.696
32.25 36.763
32.5 36.828
32.75 36.893
33 36.957
33.25 37.02
33.5 37.082
33.75 37.143
34 37.204
34.25 37.264
34.5 37.322
34.75 37.381
35 37.438
};
\addlegendentry{${a}_{1} = 1 * 10^{-3}$}
\addplot [semithick, darkorange25512714]
table {%
0.002 5.4634e-06
0.004 1.4036e-05
0.006 2.5195e-05
0.008 3.8647e-05
0.01 5.4199e-05
0.012 7.1708e-05
0.014 9.1062e-05
0.016 0.00011217
0.018 0.00013496
0.02 0.00015937
0.025 0.0002299
0.03 0.00030932
0.035 0.00039711
0.04 0.00049288
0.045 0.00059628
0.05 0.00070702
0.055 0.00082485
0.06 0.00094957
0.065 0.001081
0.07 0.0012189
0.08 0.0015199
0.09 0.001845
0.1 0.0021935
0.11 0.0025644
0.12 0.0029573
0.13 0.0033716
0.14 0.0038067
0.15 0.0042622
0.16 0.0047378
0.17 0.0052331
0.18 0.0057477
0.19 0.0062815
0.2 0.0068341
0.25 0.010056
0.3 0.013719
0.35 0.017815
0.4 0.022341
0.45 0.027298
0.5 0.032691
0.55 0.03853
0.6 0.044827
0.65 0.051595
0.7 0.058853
0.75 0.066622
0.8 0.074926
0.85 0.08379
0.9 0.093247
0.95 0.10333
1 0.11407
1.05 0.12552
1.1 0.13771
1.15 0.15071
1.2 0.16455
1.25 0.17931
1.3 0.19505
1.35 0.21183
1.4 0.22974
1.45 0.24885
1.5 0.26927
1.55 0.29109
1.6 0.31442
1.65 0.33937
1.7 0.36608
1.75 0.39466
1.8 0.42529
1.85 0.4581
1.9 0.49326
1.95 0.53096
2 0.57138
2.05 0.61402
2.1 0.65899
2.25 0.81787
2.5 1.1601
2.75 1.5808
3 2.0734
3.25 2.6247
3.5 3.2192
3.75 3.8418
4 4.4798
4.25 5.1236
4.5 5.7658
4.75 6.4018
5 7.0281
5.25 7.6427
5.5 8.2442
5.75 8.8318
6 9.4054
6.25 9.9646
6.5 10.51
6.75 11.041
7 11.558
7.25 12.063
7.5 12.554
7.75 13.033
8 13.5
8.25 13.956
8.5 14.4
8.75 14.833
9 15.256
9.25 15.668
9.5 16.071
9.75 16.465
10 16.849
10.25 17.225
10.5 17.592
10.75 17.951
11 18.302
11.25 18.645
11.5 18.981
11.75 19.309
12 19.63
12.25 19.945
12.5 20.252
12.75 20.554
13 20.849
13.25 21.138
13.5 21.421
13.75 21.698
14 21.969
14.25 22.235
14.5 22.496
14.75 22.751
15 23.001
15.25 23.246
15.5 23.486
15.75 23.722
16 23.953
16.25 24.179
16.5 24.401
16.75 24.619
17 24.832
17.25 25.042
17.5 25.247
17.75 25.448
18 25.646
18.25 25.84
18.5 26.03
18.75 26.217
19 26.4
19.25 26.579
19.5 26.756
19.75 26.929
20 27.099
20.25 27.266
20.5 27.43
20.75 27.591
21 27.749
21.25 27.904
21.5 28.057
21.75 28.207
22 28.354
22.25 28.499
22.5 28.641
22.75 28.781
23 28.919
23.25 29.054
23.5 29.187
23.75 29.317
24 29.446
24.25 29.573
24.5 29.697
24.75 29.82
25 29.94
25.25 30.059
25.5 30.176
25.75 30.291
26 30.405
26.25 30.516
26.5 30.626
26.75 30.735
27 30.842
27.25 30.947
27.5 31.051
27.75 31.153
28 31.254
28.25 31.354
28.5 31.452
28.75 31.549
29 31.645
29.25 31.74
29.5 31.833
29.75 31.925
30 32.016
30.25 32.106
30.5 32.195
30.75 32.282
31 32.369
31.25 32.455
31.5 32.539
31.75 32.623
32 32.706
32.25 32.788
32.5 32.869
32.75 32.949
33 33.028
33.25 33.107
33.5 33.184
33.75 33.261
34 33.337
34.25 33.413
34.5 33.487
34.75 33.561
35 33.635
};
\addlegendentry{${a}_{1} = 5 * 10^{-4}$}
\addplot [semithick, forestgreen4416044]
table {%
0.002 1.0927e-06
0.004 2.8073e-06
0.006 5.0389e-06
0.008 7.7294e-06
0.01 1.084e-05
0.012 1.4342e-05
0.014 1.8212e-05
0.016 2.2434e-05
0.018 2.6992e-05
0.02 3.1873e-05
0.025 4.5981e-05
0.03 6.1864e-05
0.035 7.9423e-05
0.04 9.8576e-05
0.045 0.00011926
0.05 0.0001414
0.055 0.00016497
0.06 0.00018991
0.065 0.00021619
0.07 0.00024378
0.075 0.00027263
0.08 0.00030273
0.085 0.00033405
0.09 0.00036657
0.095 0.00040025
0.1 0.00043509
0.105 0.00047107
0.11 0.00050817
0.115 0.00054637
0.12 0.00058565
0.125 0.00062601
0.13 0.00066743
0.135 0.0007099
0.14 0.00075341
0.145 0.00079795
0.15 0.00084349
0.155 0.00089005
0.16 0.0009376
0.165 0.00098614
0.17 0.0010357
0.175 0.0010862
0.18 0.0011376
0.185 0.00119
0.19 0.0012434
0.195 0.0012977
0.2 0.001353
0.21 0.0014672
0.22 0.0015851
0.23 0.0017067
0.24 0.0018319
0.25 0.0019607
0.26 0.002093
0.27 0.0022288
0.28 0.0023682
0.29 0.0025111
0.3 0.0026574
0.35 0.0034758
0.4 0.0043801
0.45 0.0053705
0.5 0.0064483
0.55 0.0076153
0.6 0.0088738
0.65 0.010227
0.7 0.011679
0.75 0.013233
0.8 0.014895
0.85 0.01667
0.9 0.018565
0.95 0.020587
1 0.022743
1.05 0.025043
1.1 0.027496
1.15 0.030113
1.2 0.032907
1.25 0.035889
1.3 0.039076
1.35 0.042483
1.4 0.046127
1.45 0.05003
1.5 0.054212
1.55 0.058698
1.6 0.063515
1.65 0.068692
1.7 0.074262
1.75 0.080261
1.8 0.08673
1.85 0.093713
1.9 0.10126
1.95 0.10943
2 0.11827
2.05 0.1277
2.1 0.13776
2.15 0.1485
2.2 0.15995
2.25 0.17216
2.5 0.26105
2.75 0.38838
3 0.56575
3.25 0.80314
3.5 1.1056
3.75 1.4714
4 1.8921
4.25 2.3553
4.5 2.8478
4.75 3.3577
5 3.8758
5.25 4.3952
5.5 4.9113
5.75 5.4212
6 5.9231
6.25 6.416
6.5 6.8992
6.75 7.3725
7 7.8359
7.25 8.2893
7.5 8.7328
7.75 9.1664
8 9.5904
8.25 10.005
8.5 10.41
8.75 10.806
9 11.193
9.25 11.571
9.5 11.94
9.75 12.302
10 12.655
10.25 13.001
10.5 13.338
10.75 13.669
11 13.993
11.25 14.309
11.5 14.619
11.75 14.923
12 15.221
12.25 15.512
12.5 15.798
12.75 16.079
13 16.354
13.25 16.624
13.5 16.888
13.75 17.149
14 17.404
14.25 17.655
14.5 17.901
14.75 18.143
15 18.382
15.25 18.616
15.5 18.846
15.75 19.073
16 19.296
16.25 19.515
16.5 19.731
16.75 19.944
17 20.153
17.25 20.359
17.5 20.562
17.75 20.762
18 20.96
18.25 21.154
18.5 21.346
18.75 21.535
19 21.721
19.25 21.905
19.5 22.086
19.75 22.264
20 22.441
20.25 22.615
20.5 22.786
20.75 22.956
21 23.123
21.25 23.288
21.5 23.451
21.75 23.612
22 23.771
22.25 23.928
22.5 24.083
22.75 24.237
23 24.388
23.25 24.538
23.5 24.685
23.75 24.831
24 24.976
24.25 25.119
24.5 25.26
24.75 25.399
25 25.537
25.25 25.674
25.5 25.808
25.75 25.942
26 26.074
26.25 26.204
26.5 26.334
26.75 26.461
27 26.588
27.25 26.713
27.5 26.837
27.75 26.959
28 27.081
28.25 27.201
28.5 27.32
28.75 27.437
29 27.554
29.25 27.67
29.5 27.784
29.75 27.897
30 28.009
30.25 28.12
30.5 28.231
30.75 28.34
31 28.448
31.25 28.555
31.5 28.661
31.75 28.766
32 28.87
32.25 28.973
32.5 29.076
32.75 29.177
33 29.278
33.25 29.378
33.5 29.477
33.75 29.575
34 29.672
34.25 29.768
34.5 29.864
34.75 29.959
35 30.053
};
\addlegendentry{${a}_{1} = 1 * 10^{-4}$}
\end{axis}

\end{tikzpicture}}
	\
	\subfloat[\centering \label{fig:5_5b}]{% This file was created with tikzplotlib v0.10.1.
\begin{tikzpicture}

\definecolor{darkgray176}{RGB}{176,176,176}
\definecolor{darkorange25512714}{RGB}{255,127,14}
\definecolor{forestgreen4416044}{RGB}{44,160,44}
\definecolor{lightgray204}{RGB}{204,204,204}
\definecolor{steelblue31119180}{RGB}{31,119,180}

\begin{axis}[
legend cell align={left},
legend style={
  fill opacity=0.8,
  draw opacity=1,
  text opacity=1,
  at={(0.97,0.03)},
  anchor=south east,
  draw=lightgray204,
    font=\fontsize{10.5}{10.5}
},
tick align=outside,
tick pos=left,
x label style={at={(axis description cs:0.5,-0.125)},anchor=north},
y label style={at={(axis description cs:-0.115,.5)},anchor=south},
x grid style={darkgray176},
xlabel={Time\,[day]},
xmin=-1.7479, xmax=36.7499,
xtick style={color=black},
y grid style={darkgray176},
ylabel={${\rho}^{0}_{\mathrm{co}} \, \mathrm{[\SI{}{\micro\gram} / mm^{3}]}$},
ymin=0, ymax=45,
ytick style={color=black}
]
\addplot [semithick, steelblue31119180]
table {%
0.002 5.4634e-06
0.004 1.4036e-05
0.006 2.5195e-05
0.008 3.8647e-05
0.01 5.4199e-05
0.012 7.1708e-05
0.014 9.1063e-05
0.016 0.00011217
0.018 0.00013496
0.02 0.00015937
0.025 0.00022991
0.03 0.00030934
0.035 0.00039716
0.04 0.00049296
0.045 0.0005964
0.05 0.00070722
0.055 0.00082515
0.06 0.00094999
0.065 0.0010816
0.07 0.0012197
0.08 0.0015213
0.09 0.0018473
0.1 0.0021969
0.11 0.0025695
0.12 0.0029643
0.13 0.0033811
0.14 0.0038193
0.15 0.0042786
0.16 0.0047586
0.17 0.0052592
0.18 0.00578
0.19 0.006321
0.2 0.0068818
0.25 0.010182
0.3 0.013974
0.35 0.018268
0.4 0.02308
0.45 0.028437
0.5 0.03437
0.55 0.040921
0.6 0.048139
0.65 0.056081
0.7 0.064817
0.75 0.074426
0.8 0.084999
0.85 0.096644
0.9 0.10948
0.95 0.12366
1 0.13934
1.05 0.15671
1.1 0.17598
1.15 0.1974
1.2 0.22127
1.25 0.24789
1.3 0.27766
1.35 0.31098
1.4 0.34833
1.45 0.39026
1.5 0.43735
1.55 0.49026
1.6 0.54973
1.65 0.6165
1.7 0.69142
1.75 0.77531
1.8 0.86904
1.85 0.97344
1.9 1.0893
1.95 1.2173
2 1.3579
2.05 1.5085
2.1 1.6688
2.15 1.8384
2.2 2.0165
2.25 2.2025
2.5 3.2994
2.75 4.4461
3 5.5852
3.25 6.6904
3.5 7.7509
3.75 8.7633
4 9.7276
4.25 10.646
4.5 11.52
4.75 12.353
5 13.149
5.25 13.909
5.5 14.637
5.75 15.336
6 16.006
6.25 16.651
6.5 17.273
6.75 17.872
7 18.451
7.25 19.01
7.5 19.552
7.75 20.076
8 20.585
8.25 21.078
8.5 21.557
8.75 22.022
9 22.475
9.25 22.915
9.5 23.343
9.75 23.76
10 24.166
10.25 24.562
10.5 24.948
10.75 25.324
11 25.691
11.25 26.049
11.5 26.399
11.75 26.741
12 27.074
12.25 27.4
12.5 27.719
12.75 28.03
13 28.334
13.25 28.632
13.5 28.923
13.75 29.208
14 29.486
14.25 29.758
14.5 30.025
14.75 30.286
15 30.541
15.25 30.792
15.5 31.036
15.75 31.276
16 31.511
16.25 31.741
16.5 31.966
16.75 32.187
17 32.404
17.25 32.616
17.5 32.823
17.75 33.027
18 33.227
18.25 33.422
18.5 33.614
18.75 33.802
19 33.987
19.25 34.168
19.5 34.346
19.75 34.52
20 34.691
20.25 34.858
20.5 35.023
20.75 35.185
21 35.343
21.25 35.499
21.5 35.652
21.75 35.802
22 35.949
22.25 36.094
22.5 36.236
22.75 36.376
23 36.514
23.25 36.649
23.5 36.781
23.75 36.912
24 37.04
24.25 37.166
24.5 37.291
24.75 37.413
25 37.533
25.25 37.651
25.5 37.767
25.75 37.882
26 37.995
26.25 38.106
26.5 38.215
26.75 38.323
27 38.429
27.25 38.533
27.5 38.637
27.75 38.738
28 38.838
28.25 38.937
28.5 39.034
28.75 39.131
29 39.225
29.25 39.319
29.5 39.411
29.75 39.502
30 39.592
30.25 39.681
30.5 39.769
30.75 39.856
31 39.941
31.25 40.026
31.5 40.11
31.75 40.192
32 40.274
32.25 40.355
32.5 40.435
32.75 40.514
33 40.592
33.25 40.669
33.5 40.746
33.75 40.822
34 40.897
34.25 40.971
34.5 41.044
34.75 41.117
35 41.189
};
\addlegendentry{${a}_{2} = 1 * 10^{-6}$}
\addplot [semithick, darkorange25512714]
table {%
0.002 5.4634e-06
0.004 1.4036e-05
0.006 2.5195e-05
0.008 3.8647e-05
0.01 5.4199e-05
0.012 7.1708e-05
0.014 9.1062e-05
0.016 0.00011217
0.018 0.00013496
0.02 0.00015937
0.025 0.0002299
0.03 0.00030932
0.035 0.00039711
0.04 0.00049288
0.045 0.00059628
0.05 0.00070702
0.055 0.00082485
0.06 0.00094957
0.065 0.001081
0.07 0.0012189
0.08 0.0015199
0.09 0.001845
0.1 0.0021935
0.11 0.0025644
0.12 0.0029573
0.13 0.0033716
0.14 0.0038067
0.15 0.0042622
0.16 0.0047378
0.17 0.0052331
0.18 0.0057477
0.19 0.0062815
0.2 0.0068341
0.25 0.010056
0.3 0.013719
0.35 0.017815
0.4 0.022341
0.45 0.027298
0.5 0.032691
0.55 0.03853
0.6 0.044827
0.65 0.051595
0.7 0.058853
0.75 0.066622
0.8 0.074926
0.85 0.08379
0.9 0.093247
0.95 0.10333
1 0.11407
1.05 0.12552
1.1 0.13771
1.15 0.15071
1.2 0.16455
1.25 0.17931
1.3 0.19505
1.35 0.21183
1.4 0.22974
1.45 0.24885
1.5 0.26927
1.55 0.29109
1.6 0.31442
1.65 0.33937
1.7 0.36608
1.75 0.39466
1.8 0.42529
1.85 0.4581
1.9 0.49326
1.95 0.53096
2 0.57138
2.05 0.61402
2.1 0.65899
2.25 0.81787
2.5 1.1601
2.75 1.5808
3 2.0734
3.25 2.6247
3.5 3.2192
3.75 3.8418
4 4.4798
4.25 5.1236
4.5 5.7658
4.75 6.4018
5 7.0281
5.25 7.6427
5.5 8.2442
5.75 8.8318
6 9.4054
6.25 9.9646
6.5 10.51
6.75 11.041
7 11.558
7.25 12.063
7.5 12.554
7.75 13.033
8 13.5
8.25 13.956
8.5 14.4
8.75 14.833
9 15.256
9.25 15.668
9.5 16.071
9.75 16.465
10 16.849
10.25 17.225
10.5 17.592
10.75 17.951
11 18.302
11.25 18.645
11.5 18.981
11.75 19.309
12 19.63
12.25 19.945
12.5 20.252
12.75 20.554
13 20.849
13.25 21.138
13.5 21.421
13.75 21.698
14 21.969
14.25 22.235
14.5 22.496
14.75 22.751
15 23.001
15.25 23.246
15.5 23.486
15.75 23.722
16 23.953
16.25 24.179
16.5 24.401
16.75 24.619
17 24.832
17.25 25.042
17.5 25.247
17.75 25.448
18 25.646
18.25 25.84
18.5 26.03
18.75 26.217
19 26.4
19.25 26.579
19.5 26.756
19.75 26.929
20 27.099
20.25 27.266
20.5 27.43
20.75 27.591
21 27.749
21.25 27.904
21.5 28.057
21.75 28.207
22 28.354
22.25 28.499
22.5 28.641
22.75 28.781
23 28.919
23.25 29.054
23.5 29.187
23.75 29.317
24 29.446
24.25 29.573
24.5 29.697
24.75 29.82
25 29.94
25.25 30.059
25.5 30.176
25.75 30.291
26 30.405
26.25 30.516
26.5 30.626
26.75 30.735
27 30.842
27.25 30.947
27.5 31.051
27.75 31.153
28 31.254
28.25 31.354
28.5 31.452
28.75 31.549
29 31.645
29.25 31.74
29.5 31.833
29.75 31.925
30 32.016
30.25 32.106
30.5 32.195
30.75 32.282
31 32.369
31.25 32.455
31.5 32.539
31.75 32.623
32 32.706
32.25 32.788
32.5 32.869
32.75 32.949
33 33.028
33.25 33.107
33.5 33.184
33.75 33.261
34 33.337
34.25 33.413
34.5 33.487
34.75 33.561
35 33.635
};
\addlegendentry{${a}_{2} = 5 * 10^{-7}$}
\addplot [semithick, forestgreen4416044]
table {%
0.002 5.4634e-06
0.004 1.4036e-05
0.006 2.5195e-05
0.008 3.8647e-05
0.01 5.4199e-05
0.012 7.1708e-05
0.014 9.1062e-05
0.016 0.00011217
0.018 0.00013496
0.02 0.00015936
0.025 0.0002299
0.03 0.00030931
0.035 0.00039709
0.04 0.00049284
0.045 0.00059621
0.05 0.00070692
0.055 0.0008247
0.06 0.00094936
0.065 0.0010807
0.07 0.0012185
0.075 0.0013626
0.08 0.001513
0.085 0.0016694
0.09 0.0018318
0.095 0.0019999
0.1 0.0021739
0.11 0.0025441
0.12 0.0029359
0.13 0.0033489
0.14 0.0037825
0.15 0.0042362
0.16 0.0047095
0.17 0.0052022
0.18 0.0057138
0.19 0.006244
0.2 0.0065182
0.25 0.0099764
0.3 0.013578
0.35 0.01758
0.4 0.021972
0.45 0.026746
0.5 0.031896
0.55 0.037421
0.6 0.043319
0.65 0.049592
0.7 0.05624
0.75 0.063269
0.8 0.070683
0.85 0.078489
0.9 0.086694
0.95 0.095305
1 0.10433
1.05 0.11379
1.1 0.12369
1.15 0.13403
1.2 0.14485
1.25 0.15614
1.3 0.16793
1.35 0.18024
1.4 0.19308
1.45 0.20647
1.5 0.22044
1.55 0.235
1.6 0.25018
1.65 0.26601
1.7 0.28251
1.75 0.29971
1.8 0.31764
1.85 0.33634
1.9 0.35582
1.95 0.37613
2 0.39731
2.05 0.41917
2.1 0.44172
2.15 0.46499
2.2 0.48898
2.25 0.51371
2.5 0.65787
2.75 0.82396
3 1.0131
3.25 1.226
3.5 1.4629
3.75 1.7232
4 2.0058
4.25 2.3093
4.5 2.6316
4.75 2.9702
5 3.3228
5.25 3.6868
5.5 4.0599
5.75 4.4398
6 4.8244
6.25 5.212
6.5 5.601
6.75 5.99
7 6.378
7.25 6.7639
7.5 7.147
7.75 7.5267
8 7.9024
8.25 8.2737
8.5 8.6403
8.75 9.002
9 9.3585
9.25 9.7096
9.5 10.055
9.75 10.396
10 10.73
10.25 11.06
10.5 11.383
10.75 11.701
11 12.014
11.25 12.321
11.5 12.623
11.75 12.919
12 13.21
12.25 13.496
12.5 13.777
12.75 14.052
13 14.323
13.25 14.588
13.5 14.849
13.75 15.105
14 15.357
14.25 15.603
14.5 15.846
14.75 16.083
15 16.317
15.25 16.546
15.5 16.771
15.75 16.992
16 17.208
16.25 17.421
16.5 17.63
16.75 17.835
17 18.036
17.25 18.234
17.5 18.428
17.75 18.618
18 18.805
18.25 18.989
18.5 19.169
18.75 19.347
19 19.52
19.25 19.691
19.5 19.859
19.75 20.024
20 20.186
20.25 20.345
20.5 20.501
20.75 20.655
21 20.806
21.25 20.954
21.5 21.1
21.75 21.243
22 21.384
22.25 21.522
22.5 21.659
22.75 21.792
23 21.924
23.25 22.054
23.5 22.181
23.75 22.307
24 22.43
24.25 22.551
24.5 22.671
24.75 22.789
25 22.905
25.25 23.019
25.5 23.131
25.75 23.242
26 23.351
26.25 23.458
26.5 23.564
26.75 23.669
27 23.772
27.25 23.873
27.5 23.973
27.75 24.072
28 24.169
28.25 24.265
28.5 24.36
28.75 24.453
29 24.546
29.25 24.637
29.5 24.727
29.75 24.816
30 24.904
30.25 24.991
30.5 25.076
30.75 25.161
31 25.245
31.25 25.328
31.5 25.41
31.75 25.491
32 25.571
32.25 25.65
32.5 25.729
32.75 25.806
33 25.883
33.25 25.959
33.5 26.034
33.75 26.109
34 26.183
34.25 26.256
34.5 26.328
34.75 26.4
35 26.471
};
\addlegendentry{${a}_{2} = 2.5 * 10^{-7}$}
\end{axis}

\end{tikzpicture}}
	
	\subfloat[\centering \label{fig:5_5c}]{% This file was created with tikzplotlib v0.10.1.
\begin{tikzpicture}

\definecolor{darkgray176}{RGB}{176,176,176}
\definecolor{darkorange25512714}{RGB}{255,127,14}
\definecolor{forestgreen4416044}{RGB}{44,160,44}
\definecolor{lightgray204}{RGB}{204,204,204}
\definecolor{steelblue31119180}{RGB}{31,119,180}

\begin{axis}[
legend cell align={left},
legend style={
  fill opacity=0.8,
  draw opacity=1,
  text opacity=1,
  at={(0.97,0.03)},
  anchor=south east,
  draw=lightgray204,
    font=\fontsize{10.5}{10.5}
},
tick align=outside,
tick pos=left,
x label style={at={(axis description cs:0.5,-0.125)},anchor=north},
y label style={at={(axis description cs:-0.115,.5)},anchor=south},
x grid style={darkgray176},
xlabel={Time\,[day]},
xmin=-1.7479, xmax=36.7499,
xtick style={color=black},
y grid style={darkgray176},
ylabel={${\rho}^{0}_{\mathrm{co}} \, \mathrm{[\SI{}{\micro\gram} / mm^{3}]}$},
ymin=-1.97509426343, ymax=41.47709972683,
ytick style={color=black}
]
\addplot [semithick, steelblue31119180]
table {%
0.002 5.4634e-06
0.004 1.4036e-05
0.006 2.5195e-05
0.008 3.8647e-05
0.01 5.4199e-05
0.012 7.1709e-05
0.014 9.1064e-05
0.016 0.00011217
0.018 0.00013496
0.02 0.00015937
0.025 0.00022992
0.03 0.00030936
0.035 0.00039718
0.04 0.000493
0.045 0.00059646
0.05 0.00070729
0.055 0.00082525
0.06 0.00095013
0.065 0.0010817
0.07 0.0012199
0.08 0.0015216
0.09 0.0018478
0.1 0.0021976
0.15 0.0044981
0.2 0.0073114
0.25 0.010622
0.3 0.014428
0.35 0.018738
0.4 0.02357
0.45 0.028949
0.5 0.034909
0.55 0.041488
0.6 0.048737
0.65 0.056712
0.7 0.065481
0.75 0.075121
0.8 0.085721
0.85 0.097386
0.9 0.11023
0.95 0.12439
1 0.14002
1.05 0.15729
1.1 0.1764
1.15 0.19757
1.2 0.22104
1.25 0.2471
1.3 0.27604
1.35 0.3082
1.4 0.34395
1.45 0.38368
1.5 0.4278
1.55 0.47672
1.6 0.5309
1.65 0.59074
1.7 0.65665
1.75 0.72901
1.8 0.80813
1.85 0.89425
1.9 0.98755
1.95 1.0881
2 1.1959
2.05 1.3085
2.1 1.4256
2.15 1.5467
2.2 1.6714
2.25 1.7991
2.3 1.9294
2.35 2.0619
2.4 2.1961
2.45 2.3318
2.5 2.4684
2.55 2.6058
2.6 2.7437
2.65 2.8818
2.7 3.02
2.75 3.158
2.8 3.2958
2.85 3.4331
2.9 3.5699
2.95 3.706
3 3.8414
3.05 3.976
3.1 4.1098
3.15 4.2426
3.2 4.3745
3.25 4.5054
3.3 4.6353
3.35 4.7642
3.4 4.892
3.45 5.0188
3.5 5.1444
3.55 5.269
3.6 5.3925
3.65 5.5148
3.7 5.6361
3.75 5.7562
3.8 5.8753
3.85 5.9933
3.9 6.1102
3.95 6.2259
4 6.3407
4.05 6.4543
4.1 6.5669
4.15 6.6785
4.2 6.789
4.25 6.8985
4.3 7.007
4.35 7.1144
4.4 7.2209
4.45 7.3264
4.5 7.431
4.55 7.5346
4.6 7.6372
4.65 7.739
4.7 7.8398
4.75 7.9397
4.8 8.0388
4.85 8.1369
4.9 8.2342
4.95 8.3307
5 8.4263
5.05 8.5211
5.1 8.615
5.15 8.7082
5.2 8.8006
5.25 8.8922
5.3 8.9831
5.35 9.0731
5.4 9.1625
5.45 9.2511
5.5 9.339
5.55 9.4262
5.6 9.5127
5.65 9.5985
5.7 9.6836
5.75 9.768
5.8 9.8518
5.85 9.9349
5.9 10.017
5.95 10.099
6 10.18
6.05 10.261
6.1 10.341
6.15 10.42
6.2 10.499
6.25 10.578
6.3 10.655
6.35 10.732
6.4 10.809
6.45 10.885
6.5 10.96
6.55 11.035
6.6 11.109
6.65 11.183
6.7 11.257
6.75 11.329
6.8 11.402
6.85 11.473
6.9 11.545
6.95 11.616
7 11.686
7.25 12.027
7.5 12.358
7.75 12.678
8 12.989
8.25 13.29
8.5 13.583
8.75 13.867
9 14.143
9.25 14.412
9.5 14.674
9.75 14.928
10 15.176
10.25 15.418
10.5 15.653
10.75 15.883
11 16.107
11.25 16.325
11.5 16.538
11.75 16.746
12 16.949
12.25 17.147
12.5 17.341
12.75 17.53
13 17.714
13.25 17.894
13.5 18.071
13.75 18.243
14 18.411
14.25 18.575
14.5 18.736
14.75 18.893
15 19.046
15.25 19.196
15.5 19.343
15.75 19.486
16 19.626
16.25 19.763
16.5 19.897
16.75 20.028
17 20.155
17.25 20.28
17.5 20.402
17.75 20.522
18 20.639
18.25 20.753
18.5 20.864
18.75 20.973
19 21.08
19.25 21.184
19.5 21.286
19.75 21.386
20 21.483
20.25 21.578
20.5 21.671
20.75 21.762
21 21.851
21.25 21.938
21.5 22.024
21.75 22.107
22 22.188
22.25 22.268
22.5 22.346
22.75 22.422
23 22.497
23.25 22.569
23.5 22.641
23.75 22.711
24 22.779
24.25 22.846
24.5 22.912
24.75 22.976
25 23.039
25.25 23.1
25.5 23.16
25.75 23.219
26 23.277
26.25 23.334
26.5 23.39
26.75 23.444
27 23.498
27.25 23.55
27.5 23.601
27.75 23.652
28 23.701
28.25 23.75
28.5 23.798
28.75 23.845
29 23.891
29.25 23.936
29.5 23.98
29.75 24.024
30 24.067
30.25 24.109
30.5 24.15
30.75 24.191
31 24.231
31.25 24.271
31.5 24.31
31.75 24.348
32 24.386
32.25 24.423
32.5 24.46
32.75 24.496
33 24.532
33.25 24.567
33.5 24.602
33.75 24.636
34 24.67
34.25 24.703
34.5 24.736
34.75 24.769
35 24.801
};
\addlegendentry{${\rho}_{\mathrm{th}} = 5$}
\addplot [semithick, darkorange25512714]
table {%
0.002 5.4634e-06
0.004 1.4036e-05
0.006 2.5195e-05
0.008 3.8647e-05
0.01 5.4199e-05
0.012 7.1708e-05
0.014 9.1062e-05
0.016 0.00011217
0.018 0.00013496
0.02 0.00015937
0.025 0.0002299
0.03 0.00030932
0.035 0.00039711
0.04 0.00049288
0.045 0.00059628
0.05 0.00070702
0.055 0.00082485
0.06 0.00094957
0.065 0.001081
0.07 0.0012189
0.12 0.0031843
0.15 0.0045517
0.2 0.0073183
0.25 0.010544
0.3 0.014211
0.35 0.018313
0.4 0.022844
0.45 0.027808
0.5 0.033209
0.55 0.039057
0.6 0.045362
0.65 0.052141
0.7 0.05941
0.75 0.067191
0.8 0.075508
0.85 0.084387
0.9 0.093859
0.95 0.10396
1 0.11472
1.05 0.12619
1.1 0.1384
1.15 0.15142
1.2 0.16529
1.25 0.18007
1.3 0.19584
1.35 0.21265
1.4 0.23059
1.45 0.24974
1.5 0.2702
1.55 0.29206
1.6 0.31543
1.65 0.34043
1.7 0.36718
1.75 0.39582
1.8 0.4265
1.85 0.45937
1.9 0.4946
1.95 0.53237
2 0.57286
2.05 0.61557
2.2 0.76701
2.25 0.81998
2.5 1.1627
2.75 1.584
3 2.0769
3.25 2.6286
3.5 3.2233
3.75 3.846
4 4.4841
4.25 5.1278
4.5 5.77
4.75 6.4059
5 7.0321
5.25 7.6466
5.5 8.248
5.75 8.8355
6 9.4089
6.25 9.9681
6.5 10.513
6.75 11.044
7 11.562
7.25 12.066
7.5 12.557
7.75 13.036
8 13.503
8.25 13.958
8.5 14.402
8.75 14.835
9 15.258
9.25 15.671
9.5 16.074
9.75 16.467
10 16.852
10.25 17.227
10.5 17.594
10.75 17.953
11 18.304
11.25 18.647
11.5 18.983
11.75 19.311
12 19.632
12.25 19.947
12.5 20.254
12.75 20.556
13 20.851
13.25 21.139
13.5 21.422
13.75 21.699
14 21.971
14.25 22.237
14.5 22.497
14.75 22.752
15 23.003
15.25 23.248
15.5 23.488
15.75 23.723
16 23.954
16.25 24.181
16.5 24.403
16.75 24.62
17 24.834
17.25 25.043
17.5 25.248
17.75 25.45
18 25.647
18.25 25.841
18.5 26.031
18.75 26.218
19 26.401
19.25 26.581
19.5 26.757
19.75 26.93
20 27.1
20.25 27.267
20.5 27.431
20.75 27.592
21 27.75
21.25 27.905
21.5 28.058
21.75 28.208
22 28.355
22.25 28.5
22.5 28.642
22.75 28.782
23 28.92
23.25 29.055
23.5 29.188
23.75 29.319
24 29.447
24.25 29.574
24.5 29.698
24.75 29.821
25 29.941
25.25 30.06
25.5 30.177
25.75 30.292
26 30.406
26.25 30.517
26.5 30.627
26.75 30.736
27 30.843
27.25 30.948
27.5 31.052
27.75 31.154
28 31.255
28.25 31.355
28.5 31.453
28.75 31.55
29 31.646
29.25 31.74
29.5 31.834
29.75 31.926
30 32.017
30.25 32.107
30.5 32.195
30.75 32.283
31 32.37
31.25 32.455
31.5 32.54
31.75 32.624
32 32.707
32.25 32.788
32.5 32.869
32.75 32.95
33 33.029
33.25 33.107
33.5 33.185
33.75 33.262
34 33.338
34.25 33.414
34.5 33.488
34.75 33.562
35 33.635
};
\addlegendentry{${\rho}_{\mathrm{th}} = 10$}
\addplot [semithick, forestgreen4416044]
table {%
0.002 5.4634e-06
0.004 1.4036e-05
0.006 2.5195e-05
0.008 3.8647e-05
0.01 5.4199e-05
0.012 7.1708e-05
0.014 9.1062e-05
0.016 0.00011217
0.018 0.00013496
0.02 0.00015936
0.025 0.0002299
0.03 0.0003093
0.035 0.00039708
0.04 0.00049282
0.045 0.00059619
0.05 0.00070688
0.055 0.00082466
0.06 0.0009493
0.065 0.0010806
0.07 0.0012184
0.08 0.0015191
0.09 0.0018437
0.1 0.0021915
0.11 0.0025616
0.12 0.0029534
0.13 0.0033662
0.14 0.0037997
0.15 0.0042532
0.16 0.0047264
0.17 0.0052189
0.18 0.0057303
0.19 0.0062603
0.2 0.0068086
0.25 0.0099905
0.3 0.013589
0.35 0.017588
0.4 0.021976
0.45 0.026746
0.5 0.031891
0.55 0.037409
0.6 0.043299
0.65 0.049563
0.7 0.056203
0.75 0.063221
0.8 0.070625
0.85 0.078418
0.9 0.08661
0.95 0.095209
1 0.10422
1.05 0.11367
1.1 0.12355
1.15 0.13388
1.2 0.14468
1.25 0.15597
1.3 0.16775
1.35 0.18005
1.4 0.19289
1.45 0.20628
1.5 0.22025
1.55 0.23483
1.6 0.25003
1.65 0.26589
1.7 0.28243
1.75 0.29968
1.8 0.31768
1.85 0.33646
1.9 0.35605
1.95 0.3765
2 0.39783
2.05 0.41988
2.1 0.44265
2.15 0.46617
2.2 0.49046
2.25 0.51552
2.5 0.6628
2.75 0.83442
3 1.0327
3.25 1.2597
3.5 1.5171
3.75 1.8064
4 2.1279
4.25 2.4816
4.5 2.8665
4.75 3.2809
5 3.7225
5.25 4.1886
5.5 4.6759
5.75 5.1811
6 5.7012
6.25 6.2327
6.5 6.773
6.75 7.3192
7 7.869
7.25 8.4203
7.5 8.9715
7.75 9.5208
8 10.067
8.25 10.609
8.5 11.147
8.75 11.679
9 12.204
9.25 12.724
9.5 13.236
9.75 13.741
10 14.239
10.25 14.729
10.5 15.212
10.75 15.687
11 16.155
11.25 16.615
11.5 17.068
11.75 17.513
12 17.951
12.25 18.382
12.5 18.806
12.75 19.222
13 19.632
13.25 20.035
13.5 20.431
13.75 20.82
14 21.203
14.25 21.579
14.5 21.949
14.75 22.313
15 22.671
15.25 23.023
15.5 23.37
15.75 23.71
16 24.045
16.25 24.375
16.5 24.699
16.75 25.018
17 25.331
17.25 25.64
17.5 25.944
17.75 26.243
18 26.537
18.25 26.826
18.5 27.111
18.75 27.392
19 27.668
19.25 27.94
19.5 28.208
19.75 28.471
20 28.731
20.25 28.987
20.5 29.239
20.75 29.487
21 29.731
21.25 29.972
21.5 30.21
21.75 30.444
22 30.675
22.25 30.902
22.5 31.126
22.75 31.347
23 31.566
23.25 31.781
23.5 31.993
23.75 32.202
24 32.409
24.25 32.612
24.5 32.814
24.75 33.012
25 33.208
25.25 33.402
25.5 33.593
25.75 33.782
26 33.968
26.25 34.152
26.5 34.334
26.75 34.514
27 34.691
27.25 34.867
27.5 35.04
27.75 35.212
28 35.382
28.25 35.55
28.5 35.715
28.75 35.88
29 36.042
29.25 36.203
29.5 36.362
29.75 36.519
30 36.675
30.25 36.829
30.5 36.982
30.75 37.133
31 37.282
31.25 37.431
31.5 37.577
31.75 37.723
32 37.867
32.25 38.01
32.5 38.151
32.75 38.292
33 38.431
33.25 38.568
33.5 38.705
33.75 38.841
34 38.975
34.25 39.108
34.5 39.241
34.75 39.372
35 39.502
};
\addlegendentry{${\rho}_{\mathrm{th}} = 20$}
\end{axis}

\end{tikzpicture}}
	\
	\subfloat[\centering \label{fig:5_5d}]{% This file was created with tikzplotlib v0.10.1.
\begin{tikzpicture}

\definecolor{darkgray176}{RGB}{176,176,176}
\definecolor{darkorange25512714}{RGB}{255,127,14}
\definecolor{forestgreen4416044}{RGB}{44,160,44}
\definecolor{lightgray204}{RGB}{204,204,204}
\definecolor{steelblue31119180}{RGB}{31,119,180}

\begin{axis}[
legend cell align={left},
legend style={
  fill opacity=0.8,
  draw opacity=1,
  text opacity=1,
  at={(0.97,0.03)},
  anchor=south east,
  draw=lightgray204,
  font=\fontsize{10.5}{10.5}
},
tick align=outside,
tick pos=left,
x label style={at={(axis description cs:0.5,-0.125)},anchor=north},
y label style={at={(axis description cs:-0.115,.5)},anchor=south},
x grid style={darkgray176},
xlabel={Time\,[day]},
xmin=-1.7479, xmax=36.7499,
xtick style={color=black},
y grid style={darkgray176},
ylabel={${\rho}^{0}_{\mathrm{co}} \, \mathrm{[\SI{}{\micro\gram} / mm^{3}]}$},
ymin=0, ymax=45,
ytick style={color=black}
]
\addplot [semithick, steelblue31119180]
table {%
0.002 5.4634e-06
0.004 1.4036e-05
0.006 2.5195e-05
0.008 3.8647e-05
0.01 5.4199e-05
0.012 7.1709e-05
0.014 9.1064e-05
0.016 0.00011217
0.018 0.00013496
0.02 0.00015937
0.025 0.00022992
0.03 0.00030936
0.035 0.00039718
0.04 0.000493
0.045 0.00059646
0.05 0.00070729
0.055 0.00082525
0.06 0.00095013
0.065 0.0010817
0.07 0.0012199
0.08 0.0015216
0.09 0.0018478
0.1 0.0021976
0.15 0.0044982
0.2 0.0073115
0.25 0.010622
0.3 0.014428
0.35 0.01874
0.4 0.023573
0.45 0.028955
0.5 0.034918
0.55 0.041504
0.6 0.048763
0.65 0.056754
0.7 0.065545
0.75 0.075218
0.8 0.085866
0.85 0.097596
0.9 0.11053
0.95 0.12483
1 0.14063
1.05 0.15815
1.1 0.17759
1.15 0.19922
1.2 0.22331
1.25 0.2502
1.3 0.28027
1.35 0.31393
1.4 0.35168
1.45 0.39406
1.5 0.44166
1.55 0.49516
1.6 0.55529
1.65 0.62281
1.7 0.69856
1.75 0.78338
1.8 0.87813
1.85 0.98364
1.9 1.1007
1.95 1.23
2 1.372
2.05 1.524
2.1 1.6857
2.25 2.2567
2.5 3.3627
2.75 4.5147
3 5.6575
3.25 6.7654
3.5 7.8282
3.75 8.8427
4 9.8089
4.25 10.729
4.5 11.605
4.75 12.44
5 13.238
5.25 14
5.5 14.73
5.75 15.431
6 16.103
6.25 16.751
6.5 17.374
6.75 17.975
7 18.556
7.25 19.118
7.5 19.661
7.75 20.188
8 20.698
8.25 21.193
8.5 21.674
8.75 22.141
9 22.595
9.25 23.037
9.5 23.467
9.75 23.886
10 24.293
10.25 24.691
10.5 25.078
10.75 25.456
11 25.824
11.25 26.184
11.5 26.535
11.75 26.878
12 27.213
12.25 27.54
12.5 27.86
12.75 28.172
13 28.477
13.25 28.776
13.5 29.068
13.75 29.354
14 29.633
14.25 29.907
14.5 30.174
14.75 30.436
15 30.692
15.25 30.943
15.5 31.189
15.75 31.43
16 31.665
16.25 31.896
16.5 32.122
16.75 32.344
17 32.561
17.25 32.774
17.5 32.982
17.75 33.186
18 33.387
18.25 33.583
18.5 33.776
18.75 33.964
19 34.15
19.25 34.331
19.5 34.509
19.75 34.684
20 34.856
20.25 35.024
20.5 35.189
20.75 35.351
21 35.51
21.25 35.666
21.5 35.82
21.75 35.97
22 36.118
22.25 36.264
22.5 36.406
22.75 36.547
23 36.685
23.25 36.82
23.5 36.953
23.75 37.084
24 37.213
24.25 37.34
24.5 37.464
24.75 37.587
25 37.707
25.25 37.826
25.5 37.943
25.75 38.058
26 38.171
26.25 38.282
26.5 38.392
26.75 38.5
27 38.606
27.25 38.711
27.5 38.815
27.75 38.917
28 39.017
28.25 39.116
28.5 39.214
28.75 39.311
29 39.406
29.25 39.5
29.5 39.592
29.75 39.684
30 39.774
30.25 39.863
30.5 39.951
30.75 40.038
31 40.124
31.25 40.209
31.5 40.293
31.75 40.376
32 40.458
32.25 40.539
32.5 40.619
32.75 40.699
33 40.777
33.25 40.855
33.5 40.932
33.75 41.008
34 41.083
34.25 41.157
34.5 41.231
34.75 41.304
35 41.377
};
\addlegendentry{${\psi}_{\mathrm{crit}} = 1 * 10^{-5}$}
\addplot [semithick, darkorange25512714]
table {%
0.002 5.4634e-06
0.004 1.4036e-05
0.006 2.5195e-05
0.008 3.8647e-05
0.01 5.4199e-05
0.012 7.1708e-05
0.014 9.1062e-05
0.016 0.00011217
0.018 0.00013496
0.02 0.00015937
0.025 0.0002299
0.03 0.00030932
0.035 0.00039711
0.04 0.00049288
0.045 0.00059628
0.05 0.00070702
0.055 0.00082485
0.06 0.00094957
0.065 0.001081
0.07 0.0012189
0.12 0.0031843
0.15 0.0045517
0.2 0.0073183
0.25 0.010544
0.3 0.014211
0.35 0.018313
0.4 0.022844
0.45 0.027808
0.5 0.033209
0.55 0.039057
0.6 0.045362
0.65 0.052141
0.7 0.05941
0.75 0.067191
0.8 0.075508
0.85 0.084387
0.9 0.093859
0.95 0.10396
1 0.11472
1.05 0.12619
1.1 0.1384
1.15 0.15142
1.2 0.16529
1.25 0.18007
1.3 0.19584
1.35 0.21265
1.4 0.23059
1.45 0.24974
1.5 0.2702
1.55 0.29206
1.6 0.31543
1.65 0.34043
1.7 0.36718
1.75 0.39582
1.8 0.4265
1.85 0.45937
1.9 0.4946
1.95 0.53237
2 0.57286
2.05 0.61557
2.2 0.76701
2.25 0.81998
2.5 1.1627
2.75 1.584
3 2.0769
3.25 2.6286
3.5 3.2233
3.75 3.846
4 4.4841
4.25 5.1278
4.5 5.77
4.75 6.4059
5 7.0321
5.25 7.6466
5.5 8.248
5.75 8.8355
6 9.4089
6.25 9.9681
6.5 10.513
6.75 11.044
7 11.562
7.25 12.066
7.5 12.557
7.75 13.036
8 13.503
8.25 13.958
8.5 14.402
8.75 14.835
9 15.258
9.25 15.671
9.5 16.074
9.75 16.467
10 16.852
10.25 17.227
10.5 17.594
10.75 17.953
11 18.304
11.25 18.647
11.5 18.983
11.75 19.311
12 19.632
12.25 19.947
12.5 20.254
12.75 20.556
13 20.851
13.25 21.139
13.5 21.422
13.75 21.699
14 21.971
14.25 22.237
14.5 22.497
14.75 22.752
15 23.003
15.25 23.248
15.5 23.488
15.75 23.723
16 23.954
16.25 24.181
16.5 24.403
16.75 24.62
17 24.834
17.25 25.043
17.5 25.248
17.75 25.45
18 25.647
18.25 25.841
18.5 26.031
18.75 26.218
19 26.401
19.25 26.581
19.5 26.757
19.75 26.93
20 27.1
20.25 27.267
20.5 27.431
20.75 27.592
21 27.75
21.25 27.905
21.5 28.058
21.75 28.208
22 28.355
22.25 28.5
22.5 28.642
22.75 28.782
23 28.92
23.25 29.055
23.5 29.188
23.75 29.319
24 29.447
24.25 29.574
24.5 29.698
24.75 29.821
25 29.941
25.25 30.06
25.5 30.177
25.75 30.292
26 30.406
26.25 30.517
26.5 30.627
26.75 30.736
27 30.843
27.25 30.948
27.5 31.052
27.75 31.154
28 31.255
28.25 31.355
28.5 31.453
28.75 31.55
29 31.646
29.25 31.74
29.5 31.834
29.75 31.926
30 32.017
30.25 32.107
30.5 32.195
30.75 32.283
31 32.37
31.25 32.455
31.5 32.54
31.75 32.624
32 32.707
32.25 32.788
32.5 32.869
32.75 32.95
33 33.029
33.25 33.107
33.5 33.185
33.75 33.262
34 33.338
34.25 33.414
34.5 33.488
34.75 33.562
35 33.635
};
\addlegendentry{${\psi}_{\mathrm{crit}} = 2 * 10^{-5}$}
\addplot [semithick, forestgreen4416044]
table {%
0.002 5.4634e-06
0.004 1.4036e-05
0.006 2.5195e-05
0.008 3.8647e-05
0.01 5.4199e-05
0.012 7.1708e-05
0.014 9.1062e-05
0.016 0.00011217
0.018 0.00013496
0.02 0.00015936
0.025 0.0002299
0.03 0.0003093
0.035 0.00039708
0.04 0.00049282
0.045 0.00059619
0.05 0.00070688
0.055 0.00082466
0.06 0.0009493
0.065 0.0010806
0.07 0.0012184
0.08 0.0015191
0.09 0.0018437
0.1 0.0021915
0.11 0.0025616
0.12 0.0029534
0.13 0.0033662
0.14 0.0037997
0.15 0.0042532
0.16 0.0047264
0.17 0.0052189
0.18 0.0057303
0.19 0.0062603
0.2 0.0068086
0.25 0.0099905
0.3 0.013589
0.35 0.017588
0.4 0.021976
0.45 0.026745
0.5 0.03189
0.55 0.037407
0.6 0.043297
0.65 0.04956
0.7 0.056198
0.75 0.063214
0.8 0.070615
0.85 0.078405
0.9 0.086592
0.95 0.095185
1 0.10419
1.05 0.11362
1.1 0.12349
1.15 0.13381
1.2 0.1446
1.25 0.15586
1.3 0.16761
1.35 0.17988
1.4 0.19268
1.45 0.20603
1.5 0.21994
1.55 0.23445
1.6 0.24958
1.65 0.26535
1.7 0.28178
1.75 0.29891
1.8 0.31676
1.85 0.33537
1.9 0.35477
1.95 0.37499
2 0.39607
2.05 0.41782
2.1 0.44026
2.25 0.51426
2.5 0.6579
2.75 0.82325
3 1.0114
3.25 1.2232
3.5 1.4585
3.75 1.7171
4 1.9978
4.25 2.2992
4.5 2.619
4.75 2.9552
5 3.3051
5.25 3.6665
5.5 4.0368
5.75 4.4138
6 4.7956
6.25 5.1804
6.5 5.5665
6.75 5.9527
7 6.3379
7.25 6.7211
7.5 7.1015
7.75 7.4785
8 7.8516
8.25 8.2203
8.5 8.5844
8.75 8.9435
9 9.2975
9.25 9.6462
9.5 9.9895
9.75 10.327
10 10.66
10.25 10.987
10.5 11.308
10.75 11.624
11 11.934
11.25 12.239
11.5 12.539
11.75 12.833
12 13.122
12.25 13.406
12.5 13.684
12.75 13.958
13 14.227
13.25 14.49
13.5 14.749
13.75 15.004
14 15.253
14.25 15.498
14.5 15.738
14.75 15.974
15 16.206
15.25 16.433
15.5 16.657
15.75 16.876
16 17.091
16.25 17.302
16.5 17.509
16.75 17.713
17 17.912
17.25 18.108
17.5 18.301
17.75 18.49
18 18.675
18.25 18.857
18.5 19.036
18.75 19.212
19 19.384
19.25 19.554
19.5 19.72
19.75 19.884
20 20.044
20.25 20.202
20.5 20.357
20.75 20.509
21 20.659
21.25 20.806
21.5 20.95
21.75 21.092
22 21.232
22.25 21.369
22.5 21.504
22.75 21.637
23 21.767
23.25 21.895
23.5 22.022
23.75 22.146
24 22.268
24.25 22.388
24.5 22.507
24.75 22.623
25 22.738
25.25 22.851
25.5 22.962
25.75 23.072
26 23.18
26.25 23.286
26.5 23.391
26.75 23.494
27 23.596
27.25 23.697
27.5 23.796
27.75 23.893
28 23.99
28.25 24.085
28.5 24.179
28.75 24.271
29 24.363
29.25 24.453
29.5 24.542
29.75 24.63
30 24.717
30.25 24.802
30.5 24.887
30.75 24.971
31 25.054
31.25 25.136
31.5 25.217
31.75 25.297
32 25.376
32.25 25.455
32.5 25.532
32.75 25.609
33 25.685
33.25 25.76
33.5 25.834
33.75 25.908
34 25.981
34.25 26.053
34.5 26.125
34.75 26.196
35 26.266
};
\addlegendentry{${\psi}_{\mathrm{crit}} = 4 * 10^{-5}$}
\end{axis}

\end{tikzpicture}}
	
	\caption{Parameter study for the collagen density evolution during the cultivation process. The modeling parameters investigated are: (a) $a_{1}$, (b) $a_{2}$, (c) ${\rho}_{\mathrm{th}}$ and (d) ${\psi}_{\mathrm{crit}}$.}
	\label{fig:5_5}    
\end{figure}
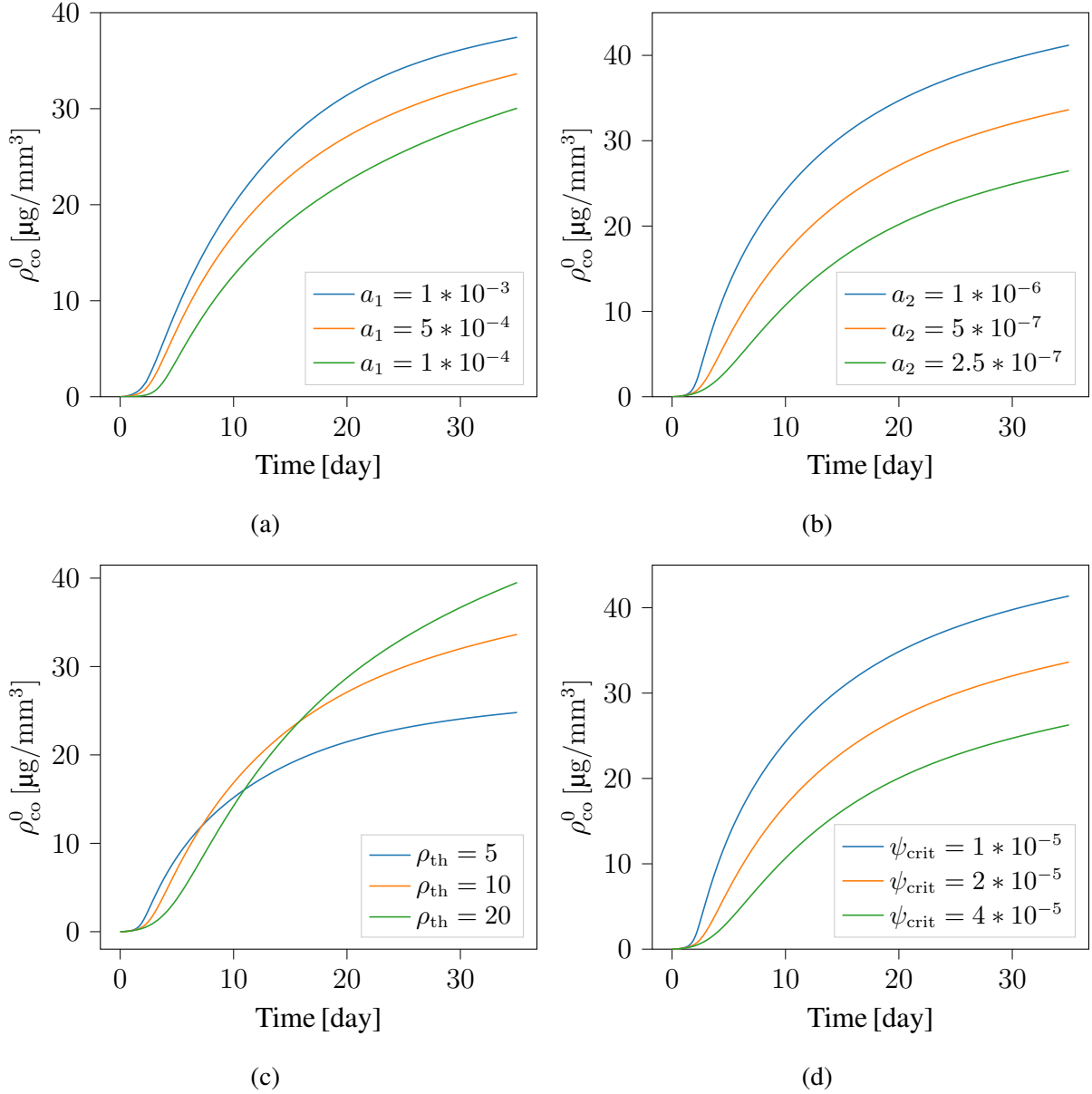

The parameters $a_{1}$ and $a_{2}$ influence the biologically and mechanically-driven part of the collagen growth, respectively. We can see in Fig.\ \ref{fig:5_5a} that a higher value for $a_{1}$ leads to a steeper increase in collagen density at the initial period of the process and, consequently to a higher collagen density at the end of the process. Similarly, in Fig.\ \ref{fig:5_5b} we see a similar pattern with higher $a_{2}$ values leading to a higher collagen densification rate. However, it is evident from both figures that in highly strained regions, our results are more sensitive to $a_{2}$ than $a_{1}$ since mechanical stimulation is the dominant factor for collagen growth in these regions. In Fig.\ \ref{fig:5_5c} we see the influence of ${\rho}_{\mathrm{th}}$ on the S-shaped curve and and the collagen density's saturation level. As expected, a lower value for ${\rho}_{\mathrm{th}}$ shifts the steep part of the curve to the left and leads to reaching the saturation level at a lower collagen density. We also looked at the influence of ${\psi}_{\mathrm{crit}}$ on the results in Fig.\ \ref{fig:5_5d}. There, we can also see that a lower value of ${\psi}_{\mathrm{crit}}$ leads to higher collagen density. Such an outcome is reasonable, as we can see in Eq.\ (\ref{eq:2-14}) that the densification rate is a function of the difference between collagen fiber strain energy ${\psi}_{\mathrm{co, m}}$ and the threshold value ${\psi}_{\mathrm{crit}}$. Furthermore, ${\psi}_{\mathrm{crit}}$ shows up in the denominator of Eq.\ (\ref{eq:2-14}), making our results highly sensitive to its value.

The material parameters of the constitutive model can also influence the collagen evolution behavior. Here we only look at the influence of collagen fiber alignment on the results. As we can see in Eq.\ \ref{eq:2-19}, the collagen fiber stretching $\lambda_{\mathrm{co}}$ is a function of the structural tensor $\mathbf{H}$. A lower value of $\kappa$ means collagen fibers are highly oriented along the mean orientation vector ${\mathbf a}$, which leads to a higher value for the fiber stretch $\lambda_{\mathrm{co}}$ and the fiber strain $E_{\mathrm{co}}$. This leads to a higher rate of collagen accumulation due to mechanical stimulation, as shown in Fig.\ \ref{fig:5_6}.

\pgfplotsset{%
	width=0.55\textwidth,
	height=0.45\textwidth
}
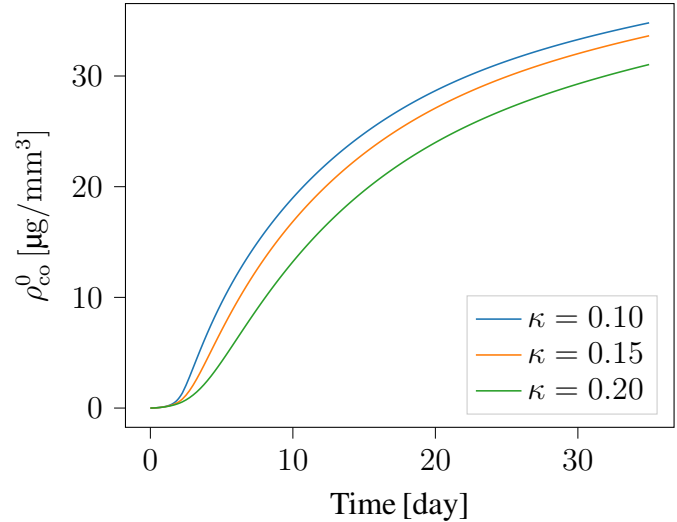
\begin{figure}[ht]
	\centering
	% This file was created with tikzplotlib v0.10.1.
\begin{tikzpicture}

\definecolor{darkgray176}{RGB}{176,176,176}
\definecolor{darkorange25512714}{RGB}{255,127,14}
\definecolor{forestgreen4416044}{RGB}{44,160,44}
\definecolor{lightgray204}{RGB}{204,204,204}
\definecolor{steelblue31119180}{RGB}{31,119,180}

\begin{axis}[
legend cell align={left},
legend style={
  fill opacity=0.8,
  draw opacity=1,
  text opacity=1,
  at={(0.97,0.03)},
  anchor=south east,
  draw=lightgray204
},
tick align=outside,
tick pos=left,
x grid style={darkgray176},
xlabel={Time\,[day]},
xmin=-1.7479, xmax=36.7499,
xtick style={color=black},
y grid style={darkgray176},
ylabel={${\rho}^{0}_{\mathrm{co}} \, \mathrm{[\SI{}{\micro\gram} / mm^{3}]}$},
ymin=-1.74019426343, ymax=36.54419972683,
ytick style={color=black}
]
\addplot [semithick, steelblue31119180]
table {%
0.002 5.4634e-06
0.004 1.4036e-05
0.006 2.5195e-05
0.008 3.8647e-05
0.01 5.4199e-05
0.012 7.1708e-05
0.014 9.1063e-05
0.016 0.00011217
0.018 0.00013496
0.02 0.00015937
0.025 0.00022991
0.03 0.00030934
0.035 0.00039716
0.04 0.00049295
0.045 0.00059638
0.05 0.00070718
0.055 0.00082508
0.06 0.00094989
0.065 0.0010814
0.07 0.0012195
0.08 0.0015209
0.09 0.0018466
0.1 0.0021958
0.11 0.0025678
0.12 0.002962
0.13 0.0033778
0.14 0.0038148
0.15 0.0042726
0.16 0.0047509
0.17 0.0052494
0.18 0.0057678
0.19 0.0063059
0.2 0.0068634
0.25 0.010131
0.3 0.013868
0.35 0.018076
0.4 0.022761
0.45 0.027938
0.5 0.033625
0.55 0.039847
0.6 0.046633
0.65 0.054018
0.7 0.062044
0.75 0.070757
0.8 0.080213
0.85 0.090472
0.9 0.1016
0.95 0.11369
1 0.12682
1.05 0.14109
1.1 0.15661
1.15 0.17352
1.2 0.19195
1.25 0.21206
1.3 0.23403
1.35 0.25806
1.4 0.28437
1.45 0.31321
1.5 0.34483
1.55 0.37954
1.6 0.41767
1.65 0.45955
1.7 0.50557
1.75 0.55614
1.8 0.61166
1.85 0.6726
1.9 0.73938
1.95 0.81248
2 0.89234
2.05 0.97766
2.1 1.0686
2.15 1.1652
2.2 1.2673
2.25 1.3749
2.5 2.0581
2.75 2.8264
3 3.6325
3.25 4.4446
3.5 5.2443
3.75 6.0226
4 6.7753
4.25 7.5012
4.5 8.2005
4.75 8.8741
5 9.5231
5.25 10.149
5.5 10.753
5.75 11.336
6 11.9
6.25 12.445
6.5 12.974
6.75 13.486
7 13.983
7.25 14.466
7.5 14.934
7.75 15.39
8 15.834
8.25 16.265
8.5 16.686
8.75 17.096
9 17.495
9.25 17.885
9.5 18.266
9.75 18.637
10 19
10.25 19.354
10.5 19.7
10.75 20.039
11 20.37
11.25 20.694
11.5 21.01
11.75 21.32
12 21.623
12.25 21.919
12.5 22.21
12.75 22.494
13 22.772
13.25 23.045
13.5 23.312
13.75 23.573
14 23.829
14.25 24.08
14.5 24.326
14.75 24.567
15 24.803
15.25 25.035
15.5 25.262
15.75 25.484
16 25.702
16.25 25.916
16.5 26.125
16.75 26.33
17 26.532
17.25 26.729
17.5 26.923
17.75 27.113
18 27.3
18.25 27.483
18.5 27.662
18.75 27.838
19 28.011
19.25 28.18
19.5 28.347
19.75 28.51
20 28.67
20.25 28.828
20.5 28.982
20.75 29.134
21 29.283
21.25 29.429
21.5 29.573
21.75 29.714
22 29.853
22.25 29.989
22.5 30.123
22.75 30.254
23 30.384
23.25 30.511
23.5 30.636
23.75 30.759
24 30.88
24.25 30.999
24.5 31.116
24.75 31.231
25 31.344
25.25 31.456
25.5 31.566
25.75 31.674
26 31.78
26.25 31.885
26.5 31.988
26.75 32.09
27 32.19
27.25 32.289
27.5 32.386
27.75 32.482
28 32.577
28.25 32.671
28.5 32.763
28.75 32.853
29 32.943
29.25 33.032
29.5 33.119
29.75 33.205
30 33.29
30.25 33.374
30.5 33.458
30.75 33.54
31 33.621
31.25 33.701
31.5 33.78
31.75 33.858
32 33.936
32.25 34.012
32.5 34.088
32.75 34.163
33 34.237
33.25 34.311
33.5 34.383
33.75 34.455
34 34.526
34.25 34.597
34.5 34.667
34.75 34.736
35 34.804
};
\addlegendentry{${\kappa} = 0.10$}
\addplot [semithick, darkorange25512714]
table {%
0.002 5.4634e-06
0.004 1.4036e-05
0.006 2.5195e-05
0.008 3.8647e-05
0.01 5.4199e-05
0.012 7.1708e-05
0.014 9.1062e-05
0.016 0.00011217
0.018 0.00013496
0.02 0.00015937
0.025 0.0002299
0.03 0.00030932
0.035 0.00039711
0.04 0.00049288
0.045 0.00059628
0.05 0.00070702
0.055 0.00082485
0.06 0.00094957
0.065 0.001081
0.07 0.0012189
0.08 0.0015199
0.09 0.001845
0.1 0.0021935
0.11 0.0025644
0.12 0.0029573
0.13 0.0033716
0.14 0.0038067
0.15 0.0042622
0.16 0.0047378
0.17 0.0052331
0.18 0.0057477
0.19 0.0062815
0.2 0.0068341
0.25 0.010056
0.3 0.013719
0.35 0.017815
0.4 0.022341
0.45 0.027298
0.5 0.032691
0.55 0.03853
0.6 0.044827
0.65 0.051595
0.7 0.058853
0.75 0.066622
0.8 0.074926
0.85 0.08379
0.9 0.093247
0.95 0.10333
1 0.11407
1.05 0.12552
1.1 0.13771
1.15 0.15071
1.2 0.16455
1.25 0.17931
1.3 0.19505
1.35 0.21183
1.4 0.22974
1.45 0.24885
1.5 0.26927
1.55 0.29109
1.6 0.31442
1.65 0.33937
1.7 0.36608
1.75 0.39466
1.8 0.42529
1.85 0.4581
1.9 0.49326
1.95 0.53096
2 0.57138
2.05 0.61402
2.1 0.65899
2.25 0.81787
2.5 1.1601
2.75 1.5808
3 2.0734
3.25 2.6247
3.5 3.2192
3.75 3.8418
4 4.4798
4.25 5.1236
4.5 5.7658
4.75 6.4018
5 7.0281
5.25 7.6427
5.5 8.2442
5.75 8.8318
6 9.4054
6.25 9.9646
6.5 10.51
6.75 11.041
7 11.558
7.25 12.063
7.5 12.554
7.75 13.033
8 13.5
8.25 13.956
8.5 14.4
8.75 14.833
9 15.256
9.25 15.668
9.5 16.071
9.75 16.465
10 16.849
10.25 17.225
10.5 17.592
10.75 17.951
11 18.302
11.25 18.645
11.5 18.981
11.75 19.309
12 19.63
12.25 19.945
12.5 20.252
12.75 20.554
13 20.849
13.25 21.138
13.5 21.421
13.75 21.698
14 21.969
14.25 22.235
14.5 22.496
14.75 22.751
15 23.001
15.25 23.246
15.5 23.486
15.75 23.722
16 23.953
16.25 24.179
16.5 24.401
16.75 24.619
17 24.832
17.25 25.042
17.5 25.247
17.75 25.448
18 25.646
18.25 25.84
18.5 26.03
18.75 26.217
19 26.4
19.25 26.579
19.5 26.756
19.75 26.929
20 27.099
20.25 27.266
20.5 27.43
20.75 27.591
21 27.749
21.25 27.904
21.5 28.057
21.75 28.207
22 28.354
22.25 28.499
22.5 28.641
22.75 28.781
23 28.919
23.25 29.054
23.5 29.187
23.75 29.317
24 29.446
24.25 29.573
24.5 29.697
24.75 29.82
25 29.94
25.25 30.059
25.5 30.176
25.75 30.291
26 30.405
26.25 30.516
26.5 30.626
26.75 30.735
27 30.842
27.25 30.947
27.5 31.051
27.75 31.153
28 31.254
28.25 31.354
28.5 31.452
28.75 31.549
29 31.645
29.25 31.74
29.5 31.833
29.75 31.925
30 32.016
30.25 32.106
30.5 32.195
30.75 32.282
31 32.369
31.25 32.455
31.5 32.539
31.75 32.623
32 32.706
32.25 32.788
32.5 32.869
32.75 32.949
33 33.028
33.25 33.107
33.5 33.184
33.75 33.261
34 33.337
34.25 33.413
34.5 33.487
34.75 33.561
35 33.635
};
\addlegendentry{${\kappa} = 0.15$}
\addplot [semithick, forestgreen4416044]
table {%
0.002 5.4634e-06
0.004 1.4036e-05
0.006 2.5195e-05
0.008 3.8647e-05
0.01 5.4199e-05
0.012 7.1708e-05
0.014 9.1062e-05
0.016 0.00011217
0.018 0.00013496
0.02 0.00015936
0.025 0.0002299
0.03 0.00030931
0.035 0.00039709
0.04 0.00049283
0.045 0.0005962
0.05 0.0007069
0.055 0.00082468
0.06 0.00094933
0.065 0.0010806
0.07 0.0012185
0.08 0.0015192
0.09 0.0018439
0.1 0.0021917
0.11 0.0025619
0.12 0.0029538
0.13 0.0033669
0.14 0.0038005
0.15 0.0042543
0.16 0.0047278
0.17 0.0052206
0.18 0.0057325
0.19 0.006263
0.2 0.0068119
0.25 0.0099995
0.3 0.013608
0.35 0.017621
0.4 0.022031
0.45 0.02683
0.5 0.032015
0.55 0.037586
0.6 0.043544
0.65 0.049893
0.7 0.056638
0.75 0.063785
0.8 0.071344
0.85 0.079323
0.9 0.087735
0.95 0.096592
1 0.10591
1.05 0.1157
1.1 0.12598
1.15 0.13678
1.2 0.14811
1.25 0.15999
1.3 0.17245
1.35 0.18551
1.4 0.19921
1.45 0.21356
1.5 0.22861
1.55 0.24438
1.6 0.26092
1.65 0.27825
1.7 0.29643
1.75 0.31549
1.8 0.33548
1.85 0.35645
1.9 0.37845
1.95 0.40154
2 0.42577
2.05 0.45089
2.1 0.47695
2.15 0.50397
2.2 0.53195
2.25 0.56094
2.5 0.73456
2.75 0.93959
3 1.1782
3.25 1.4515
3.5 1.7594
3.75 2.1008
4 2.4731
4.25 2.8729
4.5 3.2964
4.75 3.7392
5 4.1974
5.25 4.6669
5.5 5.1444
5.75 5.6267
6 6.1113
6.25 6.596
6.5 7.079
6.75 7.5589
7 8.0346
7.25 8.5052
7.5 8.9698
7.75 9.428
8 9.8794
8.25 10.324
8.5 10.761
8.75 11.19
9 11.612
9.25 12.026
9.5 12.432
9.75 12.831
10 13.223
10.25 13.606
10.5 13.983
10.75 14.352
11 14.714
11.25 15.069
11.5 15.417
11.75 15.759
12 16.093
12.25 16.421
12.5 16.743
12.75 17.058
13 17.367
13.25 17.67
13.5 17.967
13.75 18.259
14 18.545
14.25 18.825
14.5 19.1
14.75 19.369
15 19.634
15.25 19.893
15.5 20.147
15.75 20.396
16 20.641
16.25 20.881
16.5 21.117
16.75 21.348
17 21.574
17.25 21.797
17.5 22.015
17.75 22.23
18 22.44
18.25 22.646
18.5 22.849
18.75 23.048
19 23.243
19.25 23.435
19.5 23.623
19.75 23.808
20 23.99
20.25 24.168
20.5 24.344
20.75 24.516
21 24.685
21.25 24.851
21.5 25.015
21.75 25.175
22 25.333
22.25 25.488
22.5 25.641
22.75 25.791
23 25.939
23.25 26.084
23.5 26.227
23.75 26.367
24 26.506
24.25 26.642
24.5 26.776
24.75 26.908
25 27.038
25.25 27.166
25.5 27.292
25.75 27.416
26 27.538
26.25 27.659
26.5 27.777
26.75 27.894
27 28.01
27.25 28.124
27.5 28.236
27.75 28.347
28 28.456
28.25 28.564
28.5 28.67
28.75 28.775
29 28.878
29.25 28.981
29.5 29.082
29.75 29.181
30 29.28
30.25 29.377
30.5 29.473
30.75 29.569
31 29.663
31.25 29.755
31.5 29.847
31.75 29.938
32 30.028
32.25 30.117
32.5 30.205
32.75 30.292
33 30.378
33.25 30.463
33.5 30.547
33.75 30.631
34 30.714
34.25 30.796
34.5 30.877
34.75 30.957
35 31.037
};
\addlegendentry{${\kappa} = 0.20$}
\end{axis}

\end{tikzpicture}
	\caption{Studying the influence of fiber dispersion parameter $\kappa$ on the collagen density.}
	\label{fig:5_6}    
\end{figure}

\subsection{Tubular-shaped heart valve}

The tubular design was proposed to avoid the complications of synthesizing and suturing semilunar-shaped heart valves. Similar to the previous example, we study the in-vitro cultivation of textile reinforced heart valves. 

\begin{figure}[H]
	\centering
	\subfloat[\centering \label{fig:5_7a}]{\includegraphics[height=0.25 \textheight]{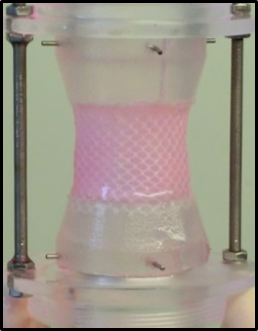}}
	\qquad \qquad \qquad
	\subfloat[\centering \label{fig:5_7b}]{\includegraphics[trim={0 {.10\textheight} {.25\textwidth} 0}, clip, height=0.25 \textheight]{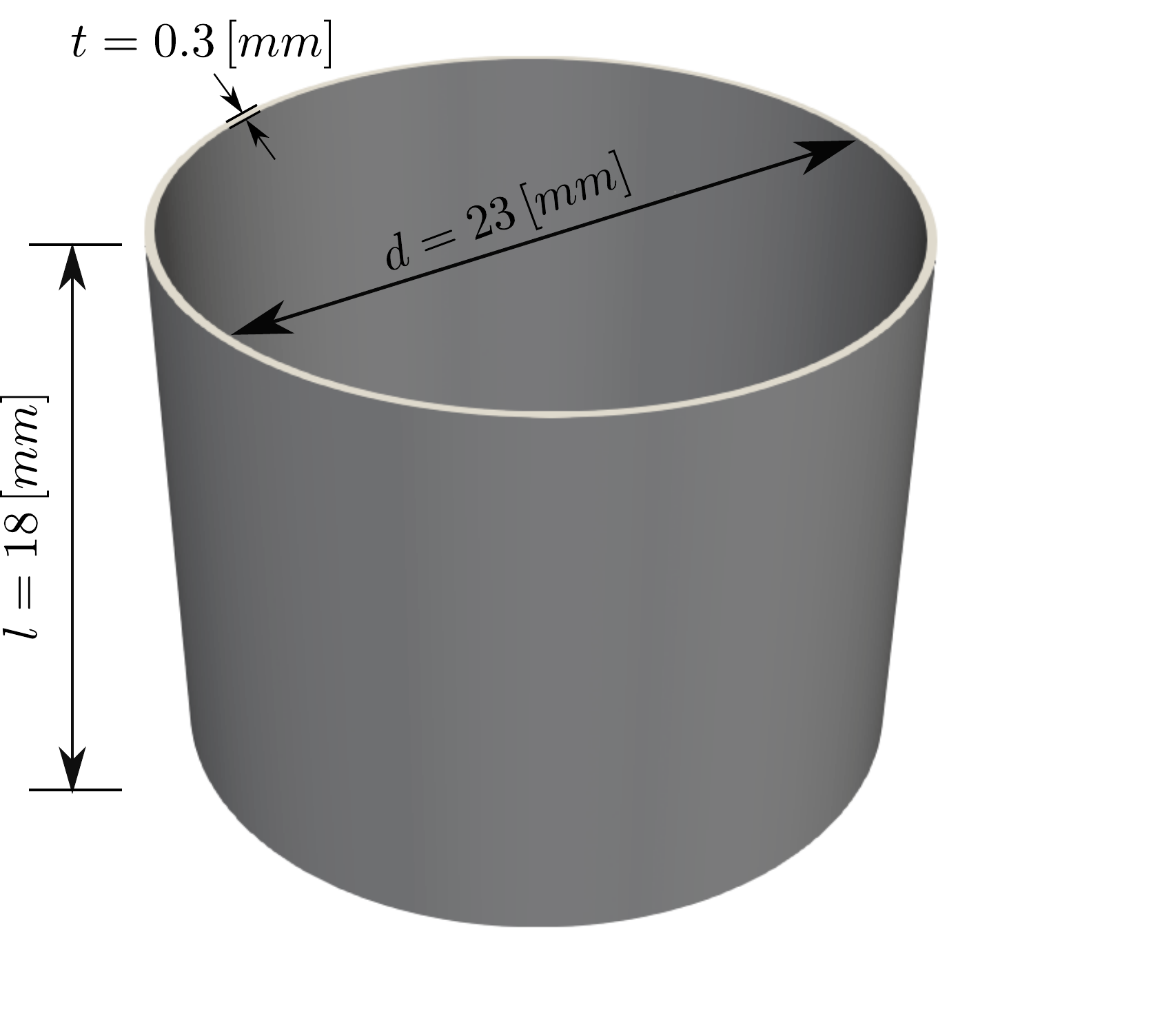}}
	
	\caption{(a) Tubular tissue-engineered heart valve design [Moreira et al. \cite{Moreira_etal_2015}]; (b) schematic illustrating the dimensions of exemplary tubular heart valve.}
	\label{fig:5_7}    
\end{figure}

\begin{figure}[H]
	\centering
	\subfloat[\centering \label{fig:5_8a}]{\includegraphics[height=0.25 \textheight]{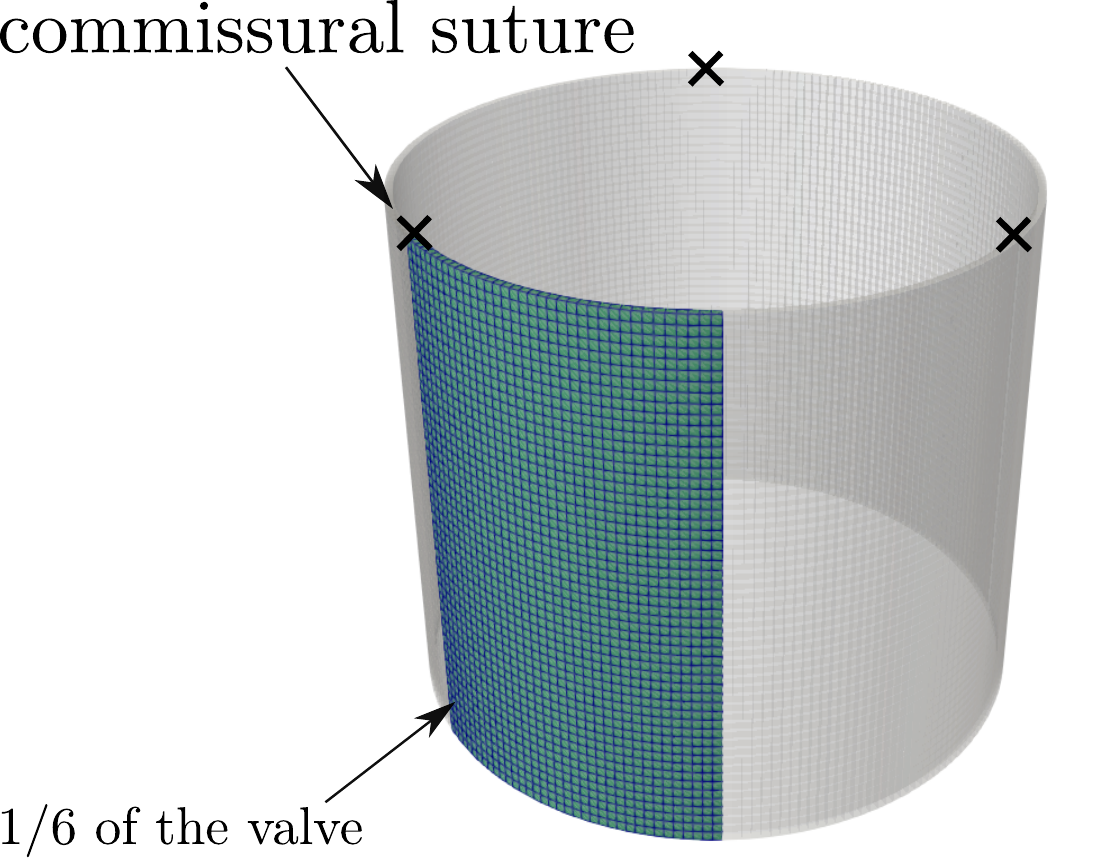}}
	\quad
	\subfloat[\centering \label{fig:5_8b}]{\includegraphics[height=0.25 \textheight]{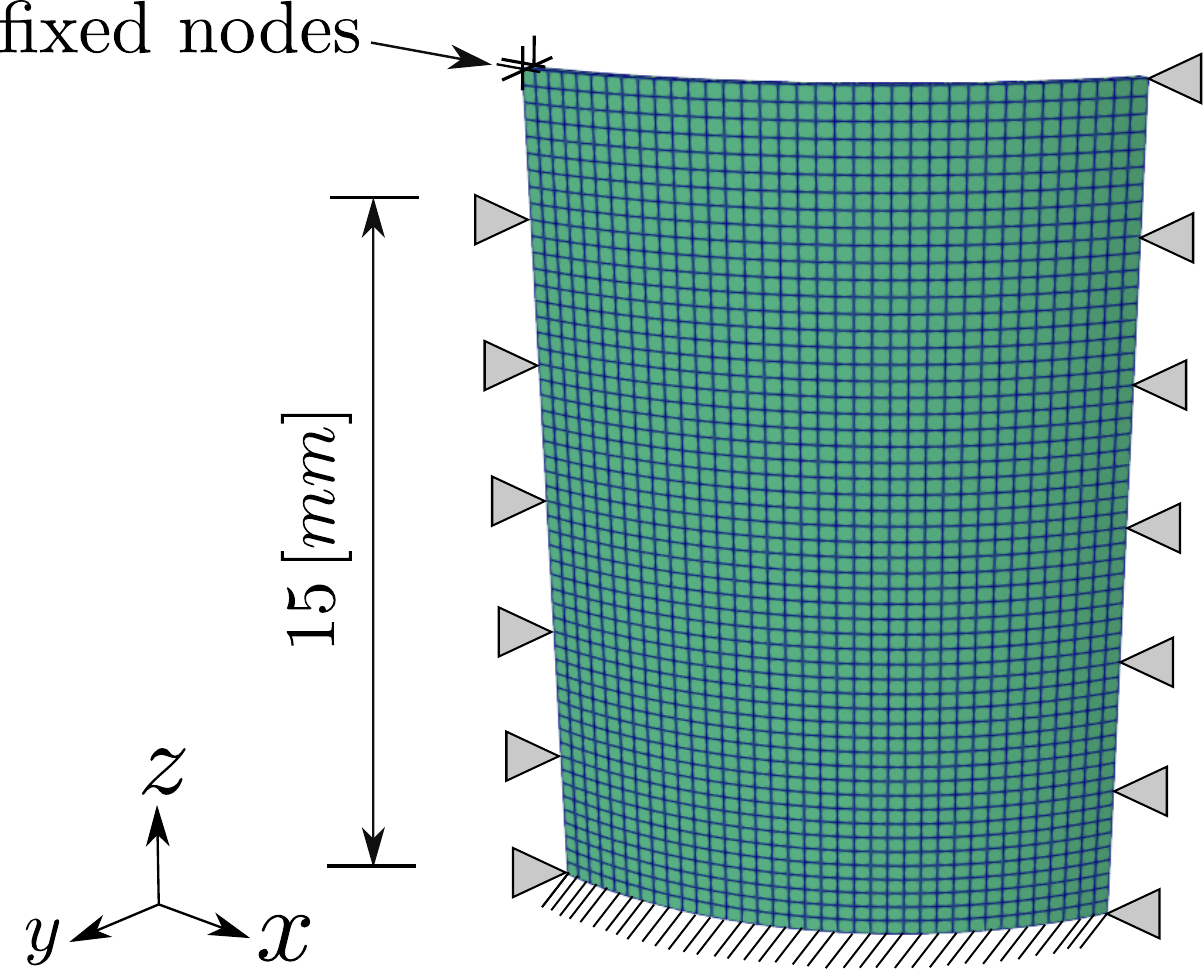}}
	
	\caption{(a) Schematic for a tubular heart valve showing the symmetry conditions. (b) Boundary valve problem for one-sixth of the heart valve.}
	\label{fig:5_8}    
\end{figure}

The valve's shape is shown in Fig.\ \ref{fig:5_7a} and its dimensions are indicated in Fig.\ \ref{fig:5_7b}, with diameter $d = 23 \; \, \mathrm{mm}$, length $l = 18 \; \mathrm{mm}$ and thickness $t = 0.3 \; \mathrm{mm}$. The valve has three  leaflets which are tied using a commissural suture as illustrated in Fig.\ \ref{fig:5_8a}. Due to the symmetry of the leaflets, it is sufficient to compute only one-half leaflet (one-sixth of the structure) as indicated in Fig.\ \ref{fig:5_8a}. We consider suture points in our boundary value problem by defining two fixed nodes, as indicated in Fig.\ \ref{fig:5_8b}. To avoid excessive distortion of the elements close to the commissural suture, we do not apply the symmetry boundary conditions on the nodes at the top $3 \; \mathrm{mm}$ of the left side. Therefore, symmetry boundary conditions are applied only on $15 \; \mathrm{mm}$ of the left side, as illustrated in Fig.\ \ref{fig:5_8b} \cite{Sesa_2023}. A similar approach was applied in the work of Stapleton et al.\ \cite{Stapleton_etal_2015} and Sodhani et al.\ \cite{Sodhani_etal_2017,Sodhani_etal_2018a, Sodhani_etal_2018b}. We discretize the structure using 1944 solid-shell elements Q1STs, with 54 elements along the longitudinal direction and 36 elements along the circumferential direction. Each element uses three Gauss points through the thickness. 

The leaflet contact is modeled in the same way as in \cite{Stapleton_etal_2015}, where we defined a rigid wall along the symmetry surface of the cylinder. Then, we define the contact surface to be between the valve leaflet's inner-surface and the rigid wall. We use the standard penalty method in FEAP \cite{Taylor_2020}, where the contact surface of the rigid wall is defined as the master surface and the valve's inner-surface is the slave.

\begin{figure}[H]
	\begin{minipage}{.78\columnwidth}
		\subfloat[\centering \label{fig:5_9a}]{\includegraphics[height=0.3 \textheight]{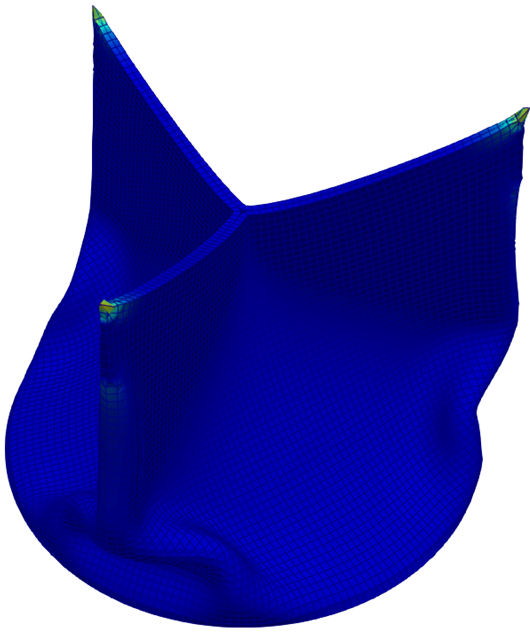}}
		\qquad
		\subfloat[\centering \label{fig:5_9b}]{\includegraphics[height=0.3 \textheight]{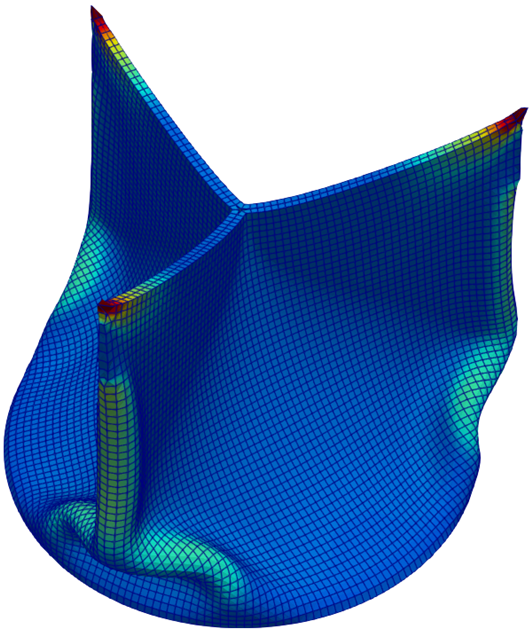}}
		
		\subfloat[\label{fig:5_9c}]{\includegraphics[height=0.3 \textheight]{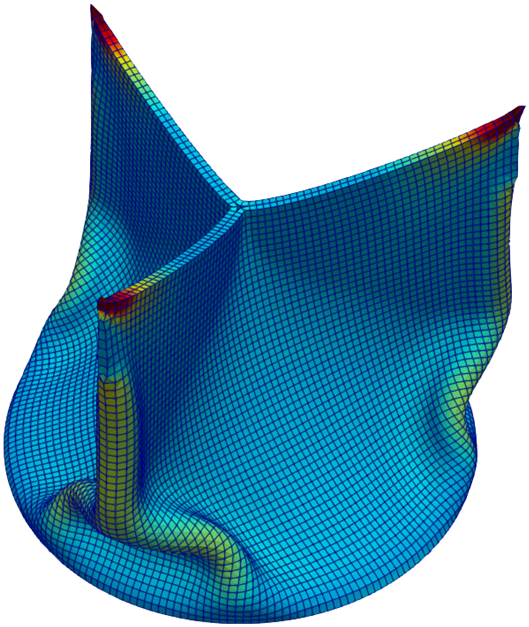}} \qquad
		\subfloat[\label{fig:5_9d}]{\includegraphics[height=0.3 \textheight]{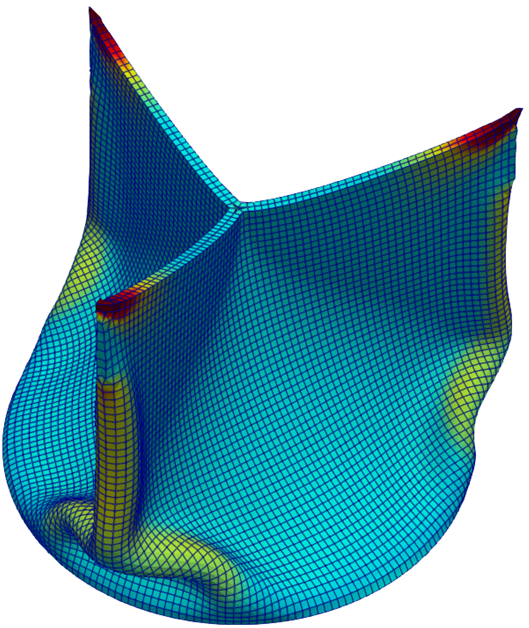}} 
	\end{minipage}
	\begin{minipage} {.21\columnwidth}
		\vspace{0.5cm} \hspace{0.5cm}
		\includegraphics[height=0.20\textheight]{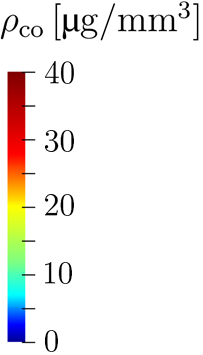}
	\end{minipage}
	
	\caption{Evolution of collagen density during the maturation process of the tubular heart valve. The figures show the density distribution after: (a) 7 days, (b) 14 days, (c) 21 days and (d) 28 days.}
	\label{fig:5_9}    
\end{figure}

The mechanical deformation of the heart valve is a highly complex fluid-structure interaction problem. However, in this example, we simplify it by modeling the blood pressure as a quasi-static pressure load. The computation is subdivided into two phases. In the first phase, we model the closure of the valve, and then in the second phase, we apply a constant pressure load. The heart valve closure is a physical instability problem, where the structure shows a snap-through behavior. Therefore, we perform the finite element computations using the arc-length method. In the second phase, we perform the computations using a classical force-controlled Newton-Raphson method, where we apply a constant pressure load of $P = 2 \; \, \mathrm{kPa}$. The modeling parameters are the same as in the previous example. However, to avoid numerical instability, we increase the shear modulus value to $\mu = 0.15 \, \mathrm{MPa}$, and use $\lambda = 1.35 \, \mathrm{MPa}$. In addition, we define the mean orientation of the collagen fibers to be along the circumferential direction and the fiber dispersion parameter as $\kappa = 0.15$. 

\begin{figure}[H]
	\centering 
	\subfloat[\centering \label{fig:5_10a}]{\includegraphics[scale=0.43]{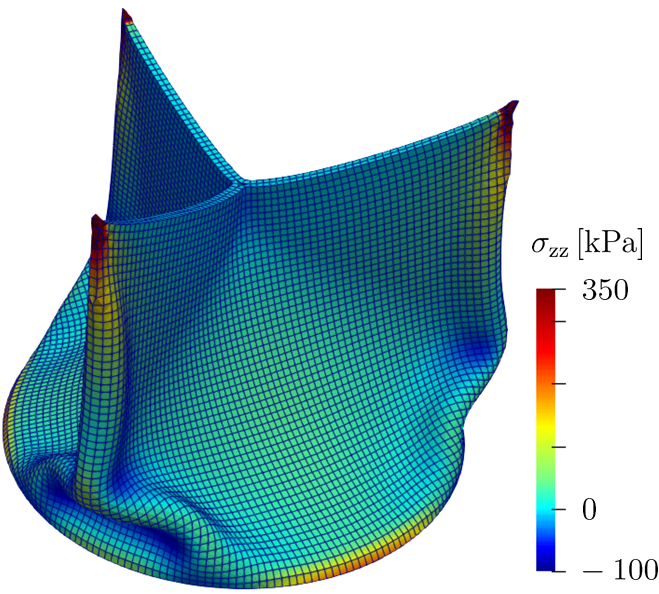}}
	\quad \quad
	\subfloat[\centering \label{fig:5_10b}]{\includegraphics[scale=0.43]{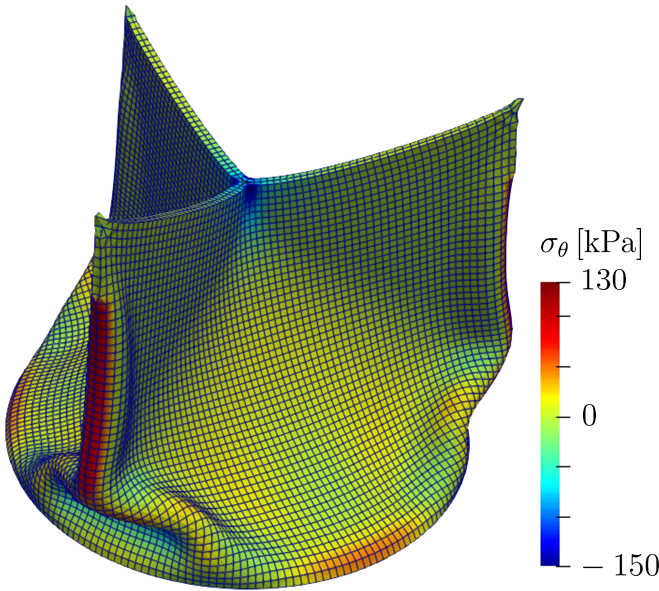}}
	\caption{Heart valve cauchy stress contours at $t = 28 \, \mathrm{days}$, with (a) stress along the circumferential direction and (b) stress along the longitudinal direction.}
	\label{fig:5_10}    
\end{figure}

In the contour plots for the collagen distribution in Fig.\ \ref{fig:5_9}, we can observe that collagen density is higher along the inner edge of the leaflets and regions surrounding the suture points. This can be explained by the fact that this structure experiences severe bending along the circumferential direction in these regions. In these results, we find a higher collagen density along the boundaries between the leaflets which is similar to the architecture of collagen fibers in native aortic heart valve leaflets shown by Peskin \& McQueen \cite{Peskin_1994}. However, our results does not show high concentration of collagen fibers along the lower edge of the leaflets. It is also important to point out that the bending behavior in tubular valves differs significantly from the standard semilunar-shaped valves. Furthermore, in biohybrid valves, the textile scaffold contributes considerably to the material's stiffness, while in native heart valves, collagen fibers are the main structural constituent of the tissue. We can also see in Fig.\ \ref{fig:5_10} the contours of the Cauchy stresses along the circumferential direction $\sigma_{\mathrm{\theta}}$ and the longitudinal direction $\sigma_{\mathrm{zz}}$. The stress contour shows higher circumferential stresses on the boundaries between the leaflets, due to the severe bending. We also see very high longitudinal stresses close to the commissural suture and the on bottom edge of the valve.

\section{Conclusion and outlook}
\label{sec:6}

 In this work, we developed a finite element framework that can predict the evolution of collagen density during the maturation process of textile-reinforced biohybrid implants. The model successfully applies an energy based approach to tissue-engineered materials, where the collagen density evolution is driven by the mass specific Helmholtz free energy of the collagen fibers. We also show that this energy based approach satisfies the second law of thermodynamics. During the cultivation of tissue-engineered materials, the initial collagen density is zero and accordingly the volume specific Helmholtz free energy is also zero. This characteristic poses a unique challenge to apply collagen densification models previously developed for other applications such as bone or cardiovascular tissue remodeling. This was solved by using the assumption that collagen evolution can be split into a part driven by biological factors and a second part driven by mechanobiological stimulation. In this way, our approach satisfies experimental results which showed collagen evolution for unloaded tissues and it allows us to model collagen evolution in a thermodynamically consistent manner. Later we used experimental data to identify most of the material parameters in our constitutive model. On the other hand, due to the lack of experimental data on collagen evolution caused by mechanical stimulation, the corresponding parameters are chosen to give physiologically reasonable results. The results produced with such a simple material model are qualitatively in accordance with collagen distribution observed in similarly loaded native tissues. 

To fully validate the model, we plan to perform a set of experiments to quantify collagen evolution under mechanical loading under in-vitro and in-situ experimental settings. Furthermore, the coupling effect between the ECM and the textile scaffold shall be quantified in the future. Another aspect that needs to be investigated is the influence of hemodynamics on tissue growth. This is especially challenging because it requires bridging two different time scales: the short-term scale for cardiac cycles (seconds) and the long-term time scale for tissue maturation (days to weeks). Work on fluid-solid-growth models can be a good starting point \cite{Figueroa_2009}.

The model can be extended by introducing additional material invariants to consider fiber-reinforced scaffolds. The extended model would allow us to study how the fiber reinforcement architecture influences collagen growth. With such a model we can optimize the fiber layout to achieve a mechanical behavior comparable to native human tissues. Other design variables such as the implant's geometry or the loading conditions during the maturation process can also be interesting to study. In our framework, we implemented the material and element routines using automatic differentiation techniques. This can be beneficial to perform sensitivity analysis studies since automatic differentiation can potentially improve the efficiency and reduce the computational errors that arise from the standard semi-analytic approaches often used in structural mechanics \cite{Korelc_2016}.

In this paper, the focus is on mass growth since inelastic volumetric deformation in textile-reinforced tissues is negligible. However, unreinforced biological tissues experience mass, volumetric growth, and fiber reorientation. The previous work on volumetric growth from Lamm et al.\ \cite{Lamm_2021, Lamm_2022} and fiber reorientation from Holthusen et al.\ \cite{Holthusen_2023} considered biological tissues with constant collagen content. A future goal would be to formulate an overarching model capable of describing the simultaneous processes of collagen deposition, fiber reorientation, and volumetric growth based on the work on these publications and the model proposed here.

\section{Acknowledgment}
S.\ Jockenhövel, S.\ Reese, and T.\ Brepols gratefully acknowledge the financial support provided by the German Research Foundation (DFG) for the subproject "Experimental investigations and modeling of biohybrid heart valves including tissue maturation – from in vitro to in situ tissue engineering" (Project number 403471716, RE 1057/45-1 and RE 1057/45-2) of DFG PAK-961 consortium "Towards a model based control of biohybrid implant maturation". Furthermore, S.\ Reese acknowledges the support granted by the DFG for the project "In-stent restenosis in coronary arteries - in silico investigations based on patient-specific data and meta modeling" (Project number 465213526, RE 1057/53-1). S.\ Reese and T.\ Brepols acknowledge the financial support for the project "Experimental and numerical investigations of laminated, fibre reininforced plastics under crash loading" (Project number 404502442, RE 1057/46-1).

%\input{sections/appendix.tex}

%%%%%%%%%%%%%%%%%%%%%%%%%%%%%%%%%%%%%%%%%%%%%%%%%%%%%%%%%%%%%%%%%%%%%%%%%%%%%%%%%%%%%%%%%%%%%%%%%%%%%%%%%
\bibliographystyle{IEEEtran.bst}
\bibliography{literature}

\end{document}